\documentclass[aps,prl,twocolumn,superscriptaddress]{revtex4-1}
\usepackage{bm}
\usepackage{graphicx}
\usepackage{color}
\usepackage{braket}
\usepackage{amsmath,amssymb,amsfonts,amsthm,mathtools}
\usepackage{enumerate}
\usepackage{enumitem}
\usepackage[colorlinks=true,linkcolor=blue,anchorcolor=red,citecolor=blue,urlcolor=blue]{hyperref}
\usepackage{titlesec}
\usepackage{tikz-cd}
\usepackage{float}
\usepackage{multirow}
\usepackage[title]{appendix}
\usepackage{pdfpages}
\usepackage{changes}
\usepackage{stackengine}
\usepackage{makecell}

\def \bM {\mathsf{M}}
\def \bL {\mathsf{L}}
\def \bR {\mathsf{R}}
\def \bg {\mathsf{g}}
\def \bG {\mathsf{G}}
\makeatletter
\AtBeginDocument{\let\LS@rot\@undefined}
\makeatother

\makeatletter 
\renewcommand\NAT@biblabelnum[1]{#1.} 
\makeatother

\setcitestyle{super}

\begin{document}

\title{Classification of time-reversal-invariant crystals with gauge structures}
\author{Z. Y. Chen}
\affiliation{National Laboratory of Solid State Microstructures and Department of Physics, Nanjing University, Nanjing 210093, China}

\author{Zheng Zhang}
\affiliation{National Laboratory of Solid State Microstructures and Department of Physics, Nanjing University, Nanjing 210093, China}

\author{Shengyuan A. Yang}
\affiliation{Research Laboratory for Quantum Materials, Singapore University of Technology and Design, Singapore 487372, Singapore}

\author{Y. X. Zhao}
\email[]{zhaoyx@nju.edu.cn}
\affiliation{National Laboratory of Solid State Microstructures and Department of Physics, Nanjing University, Nanjing 210093, China}
\affiliation{Collaborative Innovation Center of Advanced Microstructures, Nanjing University, Nanjing 210093, China}

\begin{abstract}
A peculiar feature of quantum states is that they may embody so-called projective representations of symmetries rather than ordinary representations. Projective representations of space groups—the defining symmetry of crystals—remain largely unexplored. Despite recent advances in artificial crystals, whose intrinsic gauge structures necessarily require a projective description, a unified theory is yet to be established. Here, we establish such a unified theory by exhaustively classifying and representing all 458 projective symmetry algebras of time-reversal-invariant crystals from 17 wallpaper groups in two dimensions—189 of which are algebraically non-equivalent. We discover three physical signatures resulting from projective symmetry algebras, including the shift of high-symmetry momenta, an enforced nontrivial Zak phase, and a spinless eight-fold nodal point. Our work offers a theoretical foundation for the field of artificial crystals and opens the door to a wealth of topological states and phenomena beyond the existing paradigms.
\end{abstract}

\maketitle

\noindent \textbf{INTRODUCTION}\\
Symmetry groups and their representations are at the heart of physics. When going from classical to quantum physics, a classical symmetry group $G$ becomes represented in the Hilbert space, where it makes no physical difference if all states are multiplied by a global phase. It follows that the representation allows an extra phase factor, i.e., for $g_1,g_2\in G$, their representations $\rho(g_1)$ and $\rho(g_2)$ may satisfy $\rho(g_1)\rho(g_2)=\nu(g_1,g_2)\rho(g_1g_2)$ with $\nu(g_1,g_2)\in U(1)$. This is known as the projective representation of $G$, and the phase factors $\nu$ called the factor system for this representation. As a well-known example, classifying the projective representations of Poincar\'{e} group for elementary particles leads to the two types of particles, bosons and fermions, corresponding to two distinct factor systems~\cite{Wigner_Lorentz}.

\begin{figure*}[ht]
	\centering
	\includegraphics[width=\textwidth]{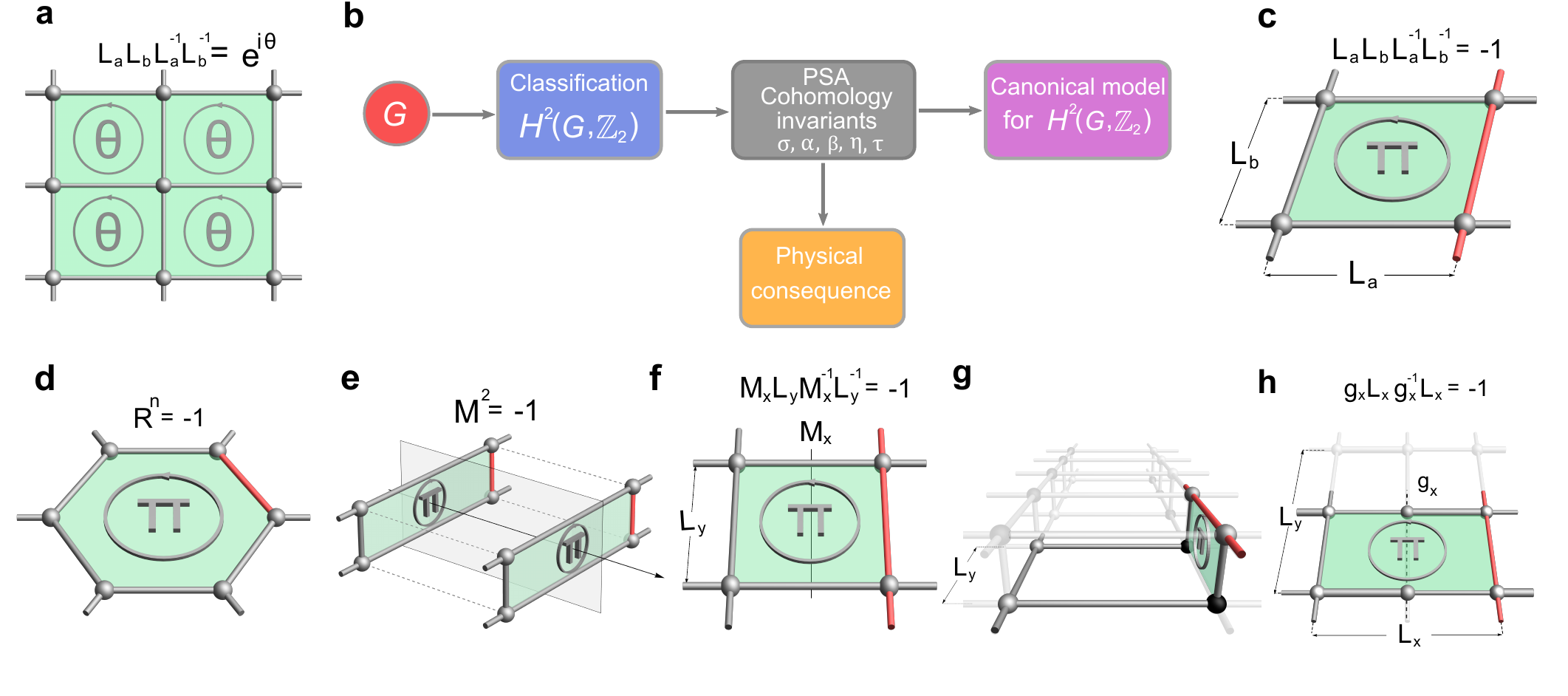}
	\caption{\textbf{Logic chart and flux realizations of projective algebraic relations.} \textbf{a}.~Gauge flux in a lattice requires the space group symmetries to be projectively represented. \textbf{b}.~Work flow diagram of this work. Given a space group $G$, the classification of projective representations are given by $H^2(G,\mathbb{Z}_2)$. These representations are captured by projective symmetry algebras (PSAs) with a complete set of cohomology invariants, from which we construct a canonical model that can realize all possible PSAs, and derive nontrivial physical consequences. \textbf{c}-\textbf{h} illustrate the construction method for realizing the five basic classes of PSAs (cohomology invariants). Specifically, \textbf{c}, \textbf{d}, \textbf{e}, \textbf{h} are for nontrivial $\sigma$, $\alpha$, $\beta$, $\tau$, respectively, and \textbf{f} and \textbf{g} are for nontrivial $\eta$. In \textbf{g}, the bond connecting the two black sites can be either present or absent, and the total flux through the vertical and horizontal rectangular plaquettes is required to be $\pi$. 
	Here, the $\pi$ flux in a plaquette is realized by a negative hopping amplitude (marked by red color) on an edge of that plaquette. }
	\label{fig:Flux_Invariant}
\end{figure*}

The defining symmetries for crystals are the space groups. What is the physical meaning of a projective representation in this context? Consider a spinless quantum particle on a two-dimensional (2D) lattice as shown in Fig.~\ref{fig:Flux_Invariant}\textbf{a}. A projective representation for the lattice translations allows (hereafter, we use bold letters to denote the represented symmetry operators) $\bL_a\bL_b\bL_a^{-1}\bL_b^{-1}=e^{i\theta}$, from which one observes that the extra phase factor corresponds to a gauge flux through the lattice.   This shows that projective representations of space groups are associated with gauge fluxes, and somewhat explains the previous ignorance of them in textbooks on solid-state physics~\cite{bradley_book}. Because most electronic crystals are free of gauge flux, one can show that their descriptions are restricted to ordinary representations. Nevertheless, it was recognized that rich gauge-flux configurations can emerge in certain strongly correlated spin systems, where projective representations of space groups are needed for their description~\cite{wen2002quantum,kitaev2006anyons,wang2006spin,essin2013classifying,messio2013time,bieri2016projective}.  

The rise of artificial crystals in recent years completely changes the situation. Artificial crystals cover a wide range of
systems, such as acoustic, photonic, mechanical, circuit, and cold-atom systems~\cite{Ozawa2019rmp,MaGuancong2019nrp,Lu2014,Yang2015,Acoustic_Crystal,Imhof2018,Yu2020,Prodan_Spring,Huber2016,Cooper2019,Optical_Lattice_RMP}.  Most artificial crystals intrinsically preserve time-reversal ($T$) symmetry, which allows fluxes $0$ and $\pi$ over the lattices. A salient feature is that these lattice gauge fluxes can be readily engineered. Recent works showed that these fluxes modify the physics in a fundamental way and projective representations are indispensable for understanding artificial crystals~\cite{Zhao_PRB_2020,Zhao_PT_Switch,Zhiyi_NC,Haoran_PRL_2022,Chunyin_PRL_2022,xue2022topological}. This urgently calls for a unified theory of projective representations of symmetries for $T$-invariant crystals, which constitutes the foundation of the whole field.

In this work, we develop such a theory and predict its distinguishing consequences. First, we characterize all possible projective symmetry algebras (PSAs) with time-reversal symmetry for any space group. This is demonstrated by 458 PSAs---189 of which are algebraically independent---for all 17 wallpaper groups in two dimensions. Then, we show all the $2$D PSAs can be systematically realized by lattice models with appropriate gauge fluxes. Finally, we present three signature results of PSAs, including the shift of high-symmetry momenta, an enforced nontrivial Zak phase, and a spinless eight-fold nodal point. 

\bigskip
\noindent \textbf{RESULTS}\\
\textbf{Projective symmetry algebras with time-reversal invariance}
We start by presenting a general result that reduces the problem for systems with time reversal symmetry $T$. Let $G$ be the space group, then the total symmetry group is $G\times \mathcal{Z}_2^T$, where $\mathcal{Z}_2^T=\{E,T\}$ is the two-element group generated by $T$. Mathematically, the classification of all possible factor systems for this group corresponds to
the second group cohomology $H^{2,c}(G\times \mathcal{Z}_2^T, U(1))$~\cite{witten1985current}. We have proven that due to the anti-unitary character of $T$, $H^{2,c}(G\times \mathcal{Z}_2^T, U(1))\cong H^2(G, \mathbb{Z}_2)\times H^2(\mathcal{Z}_2^T, \mathbb{Z}_2)$ with $\mathbb{Z}_2=\{\pm 1\}$ (see Methods). Hence, the computation is simplified to deriving $H^2(G,\mathbb{Z}_2)$, since it is known that $H^2(\mathcal{Z}_2^T, \mathbb{Z}_2)\cong \mathbb{Z}_2=\{\pm 1\}$ distinguishes integer and half integer spins, respectively. This means we only need to consider $G$ with factors restricted to $\mathbb{Z}_2=\{\pm 1\}$. Our discussion below focuses on spinless systems, which are pertinent to most artificial crystals. It can be directly extended to spinful systems, as we shall comment at the end.

The second group cohomology $H^2(G,\mathbb{Z}_2)$ can be derived from the abstract group cohomology theory, e.g., from the twisted tensor product of the cochain complex of the translation subgroup and that of the point group~\cite{brown_book}. Considering the 17 wallpaper groups for 2D, the results are listed in the second column of Table~\ref{coboundary-invariant}. Besides the classification, we also need to know the content of each $H^2(G,\mathbb{Z}_2)$, namely, the concrete algebraic relations satisfied by the symmetry operators, which are called the PSAs, because they are directly related to the physics of a system. We have worked out all PSAs in terms of generators of each group, as listed in the fourth column of Table~\ref{coboundary-invariant}. 
The technical details are given in Supplementary Note 2. The classification is complete, meaning that any $T$-invariant crystal system in two dimensions must belong to one of the PSAs listed here. 

Interestingly, each $H^2(G,\mathbb{Z}_2)$ is a product of $\mathbb{Z}_2$, i.e., $H^2(G,\mathbb{Z}_2)\cong \mathbb{Z}_2^n$. Meanwhile, we find that the corresponding PSA can be captured by a complete set of $n$ $\mathbb{Z}_2$-valued cohomology invariants, which are denoted by $\sigma$, $\alpha$, $\beta$, $\eta$, and $\tau$ in Table~\ref{coboundary-invariant}. The specific meaning of these symbols will be explained in a while. Here, one can easily check that they are indeed cohomology invariants, by noting that they are unchanged when multiplying symmetry operators by arbitrary  $\mathbb{Z}_2$  phases $\pm 1$. The ordinary representation just corresponds to the case with all invariants being $+1$.

It should be noted that for each $G$ the PSAs classified by $H^2(G,\mathbb{Z}_2)$ have redundancies for their abstract algebraic structures, because often $G$  has nontrivial automorphisms such that two nonequivalent factor systems lead to equivalent algebraic structures. We screen out all nonequivalent algebras, of which the numbers ($N_G$) are listed in the last column of Table~\ref{coboundary-invariant}. We find that there are 189 non-equivalent algebraic structures out of the 458 PSAs.

To illustrate our theory, we take group $P2$ as an example. The set of generators of $P2$ consists of two unit translations $L_a, L_b$ and the rotation $R$ by $\pi$ (along an out-of-plane axis). Its group algebras are expressed in terms of the four combinations, $R$, $L_aR$, $L_bR$ and $L_aL_bR$, each of which is squared to $1$. According to Table~\ref{coboundary-invariant}, $H^2(P2,\mathbb{Z}_2)\cong\mathbb{Z}_2^4$, so there are four cohomology invariants $\alpha_i$ ($i=1,2,3,4$), corresponding to the four PSA relations:

\begin{equation}\label{example-p2-relation-II}
	%\begin{split}
		\bR^2 =\alpha_1,~( \bL_a  \bR)^2  = \alpha_2,~( \bL_b \bR)^2  = \alpha_3,~( \bL_a \bL_b \bR)^2 =  \alpha_4.
	%\end{split}
\end{equation}

Since any permutation of the four twofold rotations above gives an isomorphic PSA, there are only five equivalence classes of PSAs, and each class is specified by how many $\alpha$'s equal $-1$.

Our result shows that the 17 wallpaper groups together with $T$ symmetry can generate $458\times 2$ PSAs (the additional factor $2$ is from spin), which is much richer than the case of Poincar\'{e} group with only twofold classification. Physically, this is due to the reduced symmetry which allows more gauge flux configurations.

\smallskip
	
\noindent\textbf{Flux realizations of projective symmetry algebras}
\begin{table*}[ht]
	\centering
	\begin{tabular}{ccclc}
		\addstackgap[5pt]{$\bm{G}$} & $\bm{H^2 (G,\mathbb{Z}_2)}$ & \textbf{Generators} &  \qquad \qquad \qquad \qquad \qquad \quad \textbf{Cohomology invariants} & $\bm{N_G}$ \\
		\Xhline{2\arrayrulewidth}
		\multirow{1}{*}{ $P1$}  & $\ \mathbb{Z}_2 $ & $\bL_a,\bL_b$ & $\begin{array}{c}\sigma = \bL_a  \bL_b \bL_a^{-1} \bL_b^{-1}. \end{array} $ & $2$  \\
		 
		\multirow{1}{*}{ $P2$}  &  \multirow{1}{*}{ $\mathbb{Z}_2^4 $} &\multirow{1}{*}{  $\bL_a,\bL_b,\bR$}  & $\begin{array}{c}
			\alpha_1  = \bR^2 ,\quad  \alpha_2   = (\bL_a \bR)^2,\quad  \alpha_3   = (\bL_b \bR)^2,\quad
			\alpha_4   = (\bL_a \bL_b \bR)^2.
		\end{array}$ & $5$  \\

		\multirow{1}{*}{ $Pm$}  &  \multirow{1}{*}{ $\mathbb{Z}_2^4 $} &\multirow{1}{*}{  $\bL_x,\bL_y,\bM_x$}  & $\begin{array}{c}
			\beta_1   =  \bM_x^2  ,\quad \beta_2   = (\bL_x \bM_x)^2,\quad 
			\eta_1 =  \bM_x \bL_y \bM_x^{-1} \bL_y^{-1} ,\quad \eta_2 = (\bL_x \bM_x)   \bL_y (\bL_x \bM_x)^{-1} \bL_y^{-1}. \\
		\end{array}$  & $10$ \\

		\multirow{1}{*}{ $Pg$}  & $\ \mathbb{Z}_2 $& $\ \bL_x,\bg_x$  & $\begin{array}{c} \tau = \bg_x \bL_x \bg_x^{-1} \bL_x.\end{array} $  & $2$ \\

		\multirow{1}{*}{ $Cm$}  &  \multirow{1}{*}{ $\mathbb{Z}_2^2 $} &\multirow{1}{*}{  $\bL_a,\bL_b,\bM$}  & $\begin{array}{c} \sigma = \bL_a \bL_b \bL_a^{-1}\bL_b^{-1},\quad\beta   = \bM^2,\quad 1 = \bM \bL_a \bM^{-1} \bL_b^{-1}. \end{array}$ & $4$  \\

		\multirow{1}{*}{ $Pmm$}  &  \multirow{1}{*}{ $\mathbb{Z}_2^8 $} &\multirow{1}{*}{  $\bL_x,\bL_y,\bM_x,\bM_y$}  & $\begin{array}{l}
			\alpha_1    = (\bM_x \bM_y)^2,\quad 
			\alpha_2     = (\bL_x \bM_x \bM_y)^2,\quad 
			\alpha_3   = (\bL_y \bM_x \bM_y)^2 ,\quad 
			\alpha_4   = (\bL_x \bL_y \bM_x \bM_y)^2,\\
			\beta_1   = \bM_x^2,\quad  \beta_2   = (\bL_x \bM_x)^2,\quad
			\beta_3   = \bM_y^2,\quad   \beta_4   = (\bL_y \bM_y)^2 .
			
		\end{array}
		$ & $51$  \\

		\multirow{1}{*}{ $Pmg$}  &  \multirow{1}{*}{ $\mathbb{Z}_2^4 $} &\multirow{1}{*}{  $\bg_y,\bL_y,\bM_x$}  & $\begin{array}{c}
			\alpha_1   = (\bM_x \bg_y)^2,\quad  \alpha_2   =  (\bL_y \bM_x \bg_y)^2,\quad 
			\beta   =\bM_x^2 ,\quad  \eta =  \bM_x \bL_y \bM_x^{-1} \bL_y^{-1} . 
		\end{array}$ & $12$  \\

		\multirow{1}{*}{ $Pgg$}  &  \multirow{1}{*}{ $\mathbb{Z}_2^2 $} &\multirow{1}{*}{  $\bg_x,\bg_y$}  & $ \begin{array}{c} \alpha_1   = ( \bg_x \bg_y)^2,\quad \alpha_2   = ( \bg_x \bg_y^{-1})^2. \end{array}$ & $3$  \\

		\multirow{1}{*}{ $Cmm$}  &  \multirow{1}{*}{ $\mathbb{Z}_2^5 $} &\multirow{1}{*}{  $\bL_a,\bM_x,\bM_y$}  & $\begin{array}{l}
			\alpha_1   = ( \bM_x \bM_y)^2 ,\quad  \alpha_2   = (\bL_a \bM_x \bM_y)^2,\quad 
			\alpha_3    = (\bL_a \bM_x \bL_a^{-1} \bM_y)^2 ,\quad
			\beta_1   = \bM_x^2,\quad \beta_2  =\bM_y^2.
		\end{array}$  & $18$ \\

		\multirow{1}{*}{ $P4$}  &  \multirow{1}{*}{ $\mathbb{Z}_2^3 $} &\multirow{1}{*}{  $\bL_x,\bR$}  & $\begin{array}{c}
			\alpha_1   =  \bR^4,\quad  \alpha_2   = ( \bL_x \bR^2)^2,\quad   \alpha_3   = ( \bL_x \bR)^4.
		\end{array}$ & $6$  \\

		\multirow{1}{*}{ $P4m$}  &  \multirow{1}{*}{ $\mathbb{Z}_2^6 $} &\multirow{1}{*}{  $\bL_x,\bR,\bM$} & $\begin{array}{l}
			\alpha_1   =  \bR^4,\quad  \alpha_2   = ( \bL_x \bR^2)^2,\quad \alpha_3    = ( \bL_x \bR)^4 ,\quad
			\beta_1  =  \bM^2,\quad \beta_2   = (  \bR\bM)^2,\quad \beta_3  = ( \bL_x \bM)^2.
		\end{array}$ & $40$  \\

		\multirow{1}{*}{ $P4g$}  &  \multirow{1}{*}{ $\mathbb{Z}_2^3 $} &\multirow{1}{*}{  $\bg_y,\bR$} & $\begin{array}{l} \alpha_1  = \bR^4,\quad  \alpha_2  = ( \bg_y^2 \bR^2)^2,\quad \beta_1   = ( \bg_y \bR)^2. \end{array}$  & $6$ \\

		\multirow{1}{*}{ $P3$}  &  \multirow{1}{*}{ $\mathbb{Z}_2 $} &\multirow{1}{*}{  $\bL_a,\bL_b,\bR$} & $\begin{array}{l} \sigma=\bL_a \bL_b \bL_a^{-1}  \bL_b^{-1},\quad 1 =\bR \bL_a \bR^{-1} \bL_b^{-1} \bL_a,\quad 1 = \bR \bL_b \bR^{-1}  \bL_a,\quad1 = \bR^3. \end{array} $ & $2$ \\

		\multirow{1}{*}{ $P3m1$}  &  \multirow{1}{*}{ $\mathbb{Z}_2^2 $} &\multirow{1}{*}{  $\bL_a,\bL_b,\bR,\bM$} & $\begin{array}{l}
			\sigma = \bL_a \bL_b \bL_a^{-1}  \bL_b^{-1},\quad   
			\beta   = \bM^2= (\bR \bM)^2=(\bL_a \bM)^2 ,\quad
			1 = \bR \bL_a \bR^{-1} \bL_b^{-1} \bL_a,\\
			1 =\bR \bL_b \bR^{-1}  \bL_a,\quad1 =\bR^3.
		\end{array}$ & $4$ \\

		\multirow{1}{*}{ $P31m$}  &  \multirow{1}{*}{ $\mathbb{Z}_2^2 $} &\multirow{1}{*}{  $\bL_a,\bL_b,\bR,\bM$}  & $\begin{array}{l}
			\sigma = \bL_a \bL_b \bL_a^{-1}  \bL_b^{-1} = \bL_a \bM \bL_a^{-1} \bM^{-1},\quad  
			\beta   = \bM^2= (\bR \bM)^2 ,\quad
			1 = \bR \bL_a \bR^{-1} \bL_b^{-1} \bL_a,\\
			1 =\bR \bL_b \bR^{-1}  \bL_a,\quad1 =\bR^3.
		\end{array}$ & $4$ \\

		\multirow{1}{*}{ $P6$}  &  \multirow{1}{*}{ $\mathbb{Z}_2^2 $} &\multirow{1}{*}{  $\bL_a,\bR$}  & $\begin{array}{c} \alpha_1   = \bR^6,\quad \alpha_2   = (\bL_a \bR^3)^2,\quad1 = (\bL_a \bR^2)^3. \end{array} $ & $4$  \\

		\multirow{1}{*}{ $P6m$}  &  \multirow{1}{*}{ $\mathbb{Z}_2^4 $} &\multirow{1}{*}{  $\bL_a,\bR,\bM$}  & $\begin{array}{c}
			\alpha_1   = \bR^6,\quad \alpha_2   = (\bL_a \bR^3)^2,\quad 
			\beta_1   = \bM^2 =(\bL_a \bM)^2 ,\quad\beta_2   = (\bR \bM)^2,\quad1 = (\bL_a \bR^2)^3.
		\end{array}$ & $16$ \\

	\end{tabular}
	\caption{\textbf{Projective symmetry algebras of 17 wallpaper groups}. The first column `$G$' lists the names of 17 wallpaper groups, and the second column `$H^2(G,\mathbb{Z}_2)$' presents the corresponding second cohomology groups. The chosen generators of each wallpaper group are given in the third column. The PSAs and the cohomology invariants are presented in the fourth column. $\bL$, $\bR$, $\bM$ and $\bg$ denote translation, rotation, mirror and glide reflection, respectively. $\sigma$, $\alpha$, $\beta$, $\eta$ and $\tau$ are the five classes of cohomology invariants valued in $\mathbb{Z}_2=\{\pm 1\}$. The last column `$N_G$' lists the number of equivalence classes of PSAs for each wallpaper group.}
	\label{coboundary-invariant}
\end{table*}
After completing the classification, our next task is to develop a construction method to realize each of the PSAs. This is important for two purposes. First, it serves as a validity check for our results in Table~\ref{coboundary-invariant}
and demonstrates that each PSA can indeed be realized in a physical system. Second, it provides guidance for the experimental realization of nontrivial PSAs in artificial crystals.

Our construction is via mapping each PSA in Table~\ref{coboundary-invariant} to a specific gauge flux pattern. 
In this process, we distinguish five classes of cohomology invariants in PSAs, corresponding to the five symbols $\sigma$, $\alpha$, $\beta$, $\eta$, and $\tau$ in Table~\ref{coboundary-invariant}. The flux lattices for them are illustrated in Fig.~\ref{fig:Flux_Invariant}\textbf{c}-\textbf{h}, and are elucidated below. The technical details for how these flux lattices represent PSAs can be found in Methods.

(\textit{i}) The first class refers to $\sigma = \bL_a \bL_b \bL_a^{-1} \bL_b^{-1}$ for the translation subgroup. $\sigma=\pm 1$ corresponds respectively to flux $0$ and $\pi$ through the plaquette spanned by $\bL_a$ and $\bL_b$, as illustrated in Fig.~\ref{fig:Flux_Invariant}\textbf{c}.

(\textit{ii}) The second class concerns cohomology invariants $\alpha$ of symmorphic rotational symmetries. That is, $\alpha = \bR^n $ for an $n$-fold rotation $R$, where $\alpha=\pm 1$ corresponds to flux $0$ or $\pi$ through the plaquette invariant under the rotation, as in Fig.~\ref{fig:Flux_Invariant}\textbf{d}.

(\textit{iii})  The third class corresponds to the square of a mirror reflection $M$, i.e., $\beta = \bM^2$. It turns out that $\bM^2=-1$ cannot be realized on lattices with only nearest neighbor hoppings within one layer. We propose to realize it either by second neighbor hopping as in Ref.~\cite{shao2022spinless} or on a bilayer lattice with $\pi$ flux through the interlayer plaquettes, as shown in Fig.~\ref{fig:Flux_Invariant}\textbf{e}.

(\textit{iv}) The fourth class ($\eta$ invariants) is on relations between translations and reflections. For example, $\eta = \bM_x \bL_y \bM_x^{-1} \bL_y^{-1} $, and $\eta=\pm 1$ corresponds to flux $0$ or $\pi$ through the plaquette in Fig.~1\textbf{f} that preserves $\bL_y$ and $\bM_x$. Moreover, if $\bM^2=\bM_x \bL_y \bM_x^{-1} \bL_y^{-1}=-1 $, we may design the flux pattern as in Fig.~\ref{fig:Flux_Invariant}\textbf{g}.

(\textit{v}) The fifth class consists of invariants $\tau$ that extend the algebraic relations between translations and glide reflections, e.g., $\tau = \bg_x \bL_x  \bg_x^{-1} \bL_x$. As illustrated in Fig.~1\textbf{h}, $\tau=\pm 1$ respectively corresponds to flux $0$ or $\pi$ through the area spanned by the translation and glide reflection, which is half of the plaquette spanned by unit translations. It appears only for $Pg$ group in Table~\ref{coboundary-invariant}.

With the above building blocks, we can systematically translate the cohomology invariants into fluxed lattices and obtain models realizing each of PSAs in Table~\ref{coboundary-invariant}. In this way, we have constructed a ``canonical'' lattice model for each wallpaper group $G$, in the sense that all PSAs for $G$ can be realized in this single model, by simply varying the $0/\pi$ flux distribution in the lattice. In Methods, we categorize the $17$ wallpaper groups into five classes to briefly introduce how the canonical lattice models are constructed. 

\begin{figure*}[ht]
	\centering
	\includegraphics[width=\textwidth]{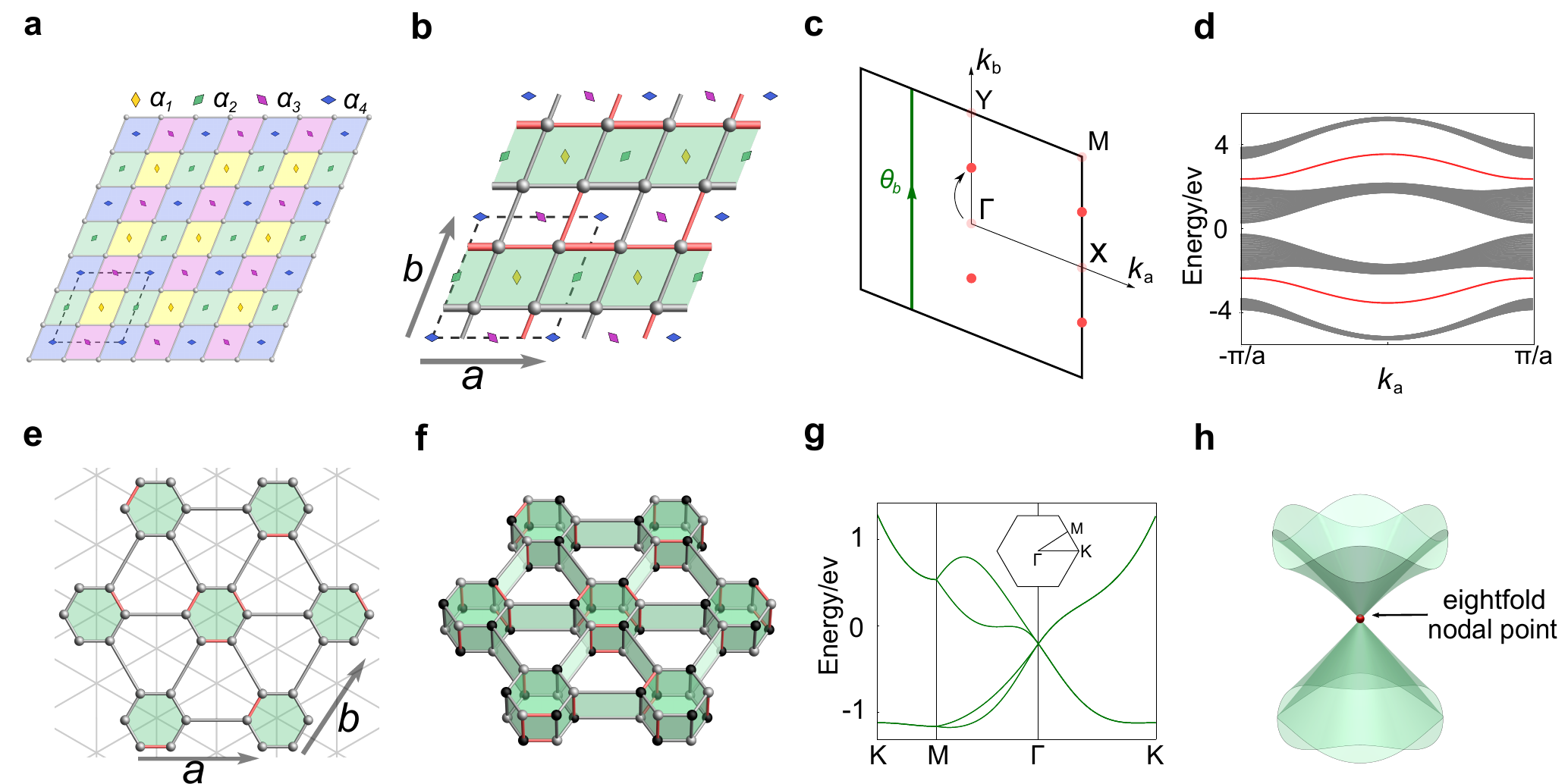}
	\caption{\textbf{Models and physical consequences of projective symmetry algebras.} \textbf{a}.~The canonical model for $P2$. The dashed line marks the unit cell. The four classes of translation-related rotation centers are colored in red, green, purple and blue, respectively. Each rotation center is the center of a plaquette shadowed with the same color of the rotation center. The four colors correspond to the four $\alpha$-invariants in Table~\ref{coboundary-invariant}. \textbf{b}.~The model of $P2$ that realizes the projective symmetry algebra (PSA) with $\alpha_1=\alpha_2=1$ and $\alpha_3=\alpha_4=-1$. The dashed line marks the unit cell, which contains four sites. The bonds with red color have a negative hopping amplitude, which makes the shaded plaquettes having a $\pi$ flux. $\bm{a}$ and $\bm{b}$ are lattice vectors. Here, we added a dimerization pattern in hopping to open spectral gaps, as in \textbf{d}.  \textbf{c}.~Due to the PSA in \textbf{b}, high-symmetry momenta are shifted and the Zak phase $\theta_b$ over a $\bm{G}_b$-periodic path must be nontrivial. Here, $k_{a,b}$ are the wave-vector components for the lattice vectors $\bm{a}$ and $\bm{b}$ in \textbf{b}. \textbf{d}.~Spectrum of the model in \textbf{b} on the slab geometry with the $b$ dimension confined. The spectrum is parametrized by $k_a$.  The in-gap edge states are colored in red, which arise from the nontrivial Zak phase. To construct a canonical model for $P3m1$, we first build a one-layer lattice model in \textbf{e} to accommodate the $\sigma$ invariant. Then, we double it into a bilayer model in \textbf{f} to further accommodate the $\beta$ invariant. In \textbf{f}, black and white colors mean the two sites are inequivalent, e.g., they may have different on-site energies. \textbf{g}.~Band structure of a $P3m1$ model in \textbf{f}, which exhibits an eightfold nodal point at $\Gamma$. Note that along $\Gamma$-$M$ ($\Gamma$-$K$), each band is twofold (fourfold) degenerate. \textbf{h}.~Dispersion in the vicinity of the eightfold degenerate nodal point. }
	\label{Fig-example-model}
\end{figure*}

As an example, consider $P2$ group with four $\alpha$ invariants. The algebraic relation for each $\alpha_i$ in Table~\ref{coboundary-invariant} corresponds to a twofold rotation center in the unit cell, as illustrated in Fig.~\ref{Fig-example-model}\textbf{a}. Under lattice translation, each $\alpha_i$ is associated with a class of translation-related rotation centers, which are distinguished by four colors in Fig.~\ref{Fig-example-model}\textbf{a}. Then, the canonical model can be constructed with each plaquette hosting a unique rotation center (see Fig.~\ref{Fig-example-model}\textbf{a}), corresponding to the dual lattice of the lattice of rotation centers.
%Accordingly, on the standard model, each $\alpha$ is associated with a class of translation-related plaquettes (shadowed by the color of the corresponding rotation center in Fig.~\ref{Fig-example-model}\textbf{a}).
Each $\alpha_i=\pm 1$ can then be realized by inserting flux $0$ or $\pi$ into the corresponding class of plaquettes.

The canonical models for all 17 wallpaper groups are illustrated in Fig.~\ref{fig:Standard_Models}, and are explicitly constructed in Supplementary Note 3. For each wallpaper group in Fig.~\ref{fig:Standard_Models}, the cohomology invariants correspond to independent fluxes in the lattice model, and we distinguish the fluxes by different colors. This is consistent with the number $2^n$ of PSAs, with $n$ the number of colors in each lattice model.

\smallskip
	
\noindent\textbf{Physical consequences of projective symmetry algebras} 
Our revealed PSAs
can lead to a wealth of new physics, beyond conventional systems based on ordinary representations. Below, we present three remarkable consequences for demonstration.
			
(\textit{1}) \emph{Shift of high-symmetry points.}
In ordinary band structures, high-symmetry points are located either at the center ($\Gamma$ point) or on the boundary of Brillouin zone (BZ)~\cite{bradley_book}. In contrast, with PSAs, the high-symmetry points are redistributed, and they can be at non-central points in the interior of BZ.

For instance, continue with the example of $P2$ group. Let's consider the PSA with $\alpha_1=\alpha_2=1$ and $\alpha_3=\alpha_4=-1$ (see the canonical model realization in Fig.~\ref{Fig-example-model}\textbf{b}). Clearly, in this case
the two translations $\bL_a$ and $\bL_b$ commute as usual, and therefore the BZ is unchanged. However, since $RL_bR^{-1}=L_b^{-1}$ is modified to $\bR\bL_b\bR^{-1}=-\bL_b^{-1}$ by the fluxes, the $R$-invariant momenta are transformed from $(0,0)$, $(0,\pi)$, $(\pi,0)$ and $(\pi,\pi)$ to $(0,\pm\pi/2)$ and $(\pi,\pm\pi/2)$, as illustrated in Fig.~\ref{Fig-example-model}\textbf{c} (see discussion in Methods).
			
(\textit{2}) \textit{Enforced nontrivial Zak phase.}
While ordinary crystal symmetries may protect topological structures of energy bands, we discover that some PSAs can even \emph{enforce} nontrivial topological structures. That is, once the PSA is realized, certain topological invariant is \emph{guaranteed} to be nontrivial.
%We propose the symmetry enforcement of topological structures ubiquitously exists for projective symmetry algebras, and plan to survey it in a separate paper.

Here, we give one example of this fascinating phenomena, again using the $P2$ group. Let us consider the PSAs with $\bR^2=(\bL_a\bR)^2=\alpha$ and $(\bL_b\bR)^2=(\bL_a\bL_b\bR)^2=-\alpha$, which can be realized by the canonical model with the flux configuration in Fig.~\ref{Fig-example-model}\textbf{b}. The PSAs lead to $\bR\bL_b\bR^{-1}=-\bL_b^{-1}$ for both $\alpha=\pm 1$. From this relation, one can show that the anti-unitary operator $\bR T$ will shift momentum $\bm{k}$ to $\bm{k}+\bm{G}_b/2$ with $\bm{G}_b$ the reciprocal translation vector corresponding to $\bL_b$ (see Methods).

Now, consider the effect of $\bR T$ on a single energy band with eigenstates $\ket{\psi_{\bm{k}}}$. Recall that spacetime inversion symmetry can quantize the Berry phase, also known as the Zak phase, along any periodic path in the BZ to be either $0$ or $\pi$~\cite{Zak_Phase}. In contrast, here, $\bR T$ with $(\bR T)^2=\alpha$ exerts a stronger constraint on the Zak phase, i.e., it completely determines the Zak phase $\theta_b$ along any $\bm{G}_b$-periodic path as

\begin{equation}\label{Enforced_BP}
\theta_b=i\ln \alpha \mod 2\pi,
\end{equation}

due to the nontrivial action of $\bR T$ discussed above (see Methods). This result means: if $\alpha=-1$, the Zak phase is enforced to be nontrivial. This is confirmed by the concrete model in Fig.~\ref{Fig-example-model}\textbf{b}. This model has four isolated bands, and each band is enforced to have a nontrivial Zak phase $\pi$ along $k_b$. Hence, there must be topological edge states within the first and the third energy gaps, as shown in Fig.~\ref{Fig-example-model}\textbf{d}.

%The result follows from examining the phase factors involved in the action of $\hat{\bR} \hat{T}$ on $\ket{\psi_{\bm{k}}}$. Supposing $\bR T \ket{\psi_{\bm{k}}}= e^{i\phi(\bm{k})} \ket{\psi_{\bm{k}+\bm{G}_{b}/2 }}$, we see that $e^{i \phi(\bm{k} + \bm{G}_{b}/2 ) - i \phi(\bm{k}  ) } =\alpha$ since  $(\bR T)^2=\alpha$. Then, it is straightforward to derive \eqref{Enforced_BP}, and the derivation details can be found in the SM.

(\textit{3}) \textit{Eightfold degenerate nodal point.}
Highly degenerate nodal points protected by crystal symmetries have been a hot topic. Without including the twofold degeneracy of spin-$1/2$, the highest degeneracy protected by wallpaper groups is fourfold~\cite{Zheng_Xin_Eightfold}. Here, we find that PSA can achieve a degeneracy of eightfold, beyond any ordinary representations.

This is exemplified by the PSA of $P3m1$ with $\bL_a\bL_b\bL_a^{-1}\bL_b^{-1}=-1$ and $\bM^2=-1$ (see Table~\ref{coboundary-invariant}). The canonical model is illustrated in Fig.~\ref{Fig-example-model}\textbf{e} and \ref{Fig-example-model}\textbf{f}. Since $\bL_a$ does not commute with $\bL_b$, we choose $\bL_a^2$ and $\bL_b^2$ to generate an invariant subgroup of $P3m1$, and the BZ is specified by $\bL_a^2=-e^{i\bm{k}\cdot \bm{e}_a}$ and $\bL_b^2=-e^{i\bm{k}\cdot \bm{e}_b}$ under the Fourier transform, with $\bm{e}_{a,b}$ being the translation vectors of $\bL_{a,b}^2$. At high-symmetry point $\Gamma$, the little co-group is given by $\mathbb{Z}_2^2 \rtimes D_3 \times \mathcal{Z}_2^T$, where $\mathbb{Z}_2^2$ are generated by $\bL_{a,b}$. This little co-group is projectively represented with factors inherited from that of $P3m1$. We find that it has two $4$D irreducible representations  and one $8$D irreducible representation. The latter gives the eightfold nodal point, which is indeed confirmed via a concrete model as illustrated in Fig.~\ref{Fig-example-model}\textbf{g} and \ref{Fig-example-model}\textbf{h}.
		
\bigskip
	
\noindent\textbf{DISCUSSION}\\
In conclusion, we have established a unified theory for $T$-invariant crystals. Particularly, we classified all PSAs of wallpaper groups, developed a general construction method, presented canonical models to realize each PSA, and revealed remarkable physical consequences. The theory can be directly extended to 3D space groups. 
Our work provides a solid foundation for the study of artificial crystals and opens the door to a wealth of new physics beyond the current paradigm based ordinary symmetry representations.

Notably, although our focus here is on spinless systems (which most artificial crystals belong to), the generalization to spinful systems is straightforward. This is because in the presence of $T$-invariance, it is always sufficient to consider $\mathbb{Z}_2$-valued factor systems, as stressed above. Then, in addition to the phases arising from fluxes, one only needs to take care of reflections and rotations of spin-$1/2$ by $2\pi$, which lead to the phase $-1$. Hence, all the cohomology invariants in classes (\textit{ii}) and (\textit{iii}) are reversed. Formally, we may just replace each $\alpha$ and $\beta$ by $(-1)^{2s}\alpha$ and $(-1)^{2s}\beta$, respectively, with $s=0$ and $1/2$ for spinless and spin-1/2 cases.

Finally, we note that our theory of PSAs is based on two fundamental principles of physics: (a) Physical systems are classified by symmetries (Landau's paradigm); and (b) Symmetries are projectively represented in a physical system (Wigner's principle). Hence, the PSAs derived here are general and classify all $T$-invariant crystal systems, including not only artificial crystals, but also real materials, strongly correlated spin systems, and beyond.

%and all of their lattice realizations have been systematically constructed. Moreover, the novel physics brought by the projective symmetry algebras has been demonstrated from three aspects. Promisingly, there are a wealth of topological novelties waiting to be explored, based on the framework set up here.

%\textcolor{blue}{\textit{Summary and Discussions}}
%Although our main focus here is on artificial crystals with engineerable fluxes, the electronic crystals with spin-orbital coupling can be readily included into our framework. This is because in the presence of $T$-invariance, it is always sufficient to consider $\mathbb{Z}_2$-valued multipliers as previously stressed. In addition to the phases arising from fluxes, rotating a spin-$1/2$ by $2\pi$ or reflecting it leads to the phase $-1$. Hence, all the cohomology invariants $\alpha$ in class \textit{ii}) are inversed. Formally, we just modify each $\alpha$ by $\alpha_s\alpha$ with $\alpha_s=\pm 1$ (see the caption of Tab.xx ) to specify the two types of crystals. %Notably, the two types of crystals are 	algebraically distinguished by $T^2=\pm 1$, which is determined by internal spin degrees of freedom and cannot be switched by external gauge fluxes.

%\newpage
%~\newpage
\bigskip
\noindent \textbf{\large METHODS}\\
\noindent \textbf{Projective representations with time-reversal symmetry}
\begin{figure*}[ht]
	\centering
	\includegraphics[width=\textwidth]{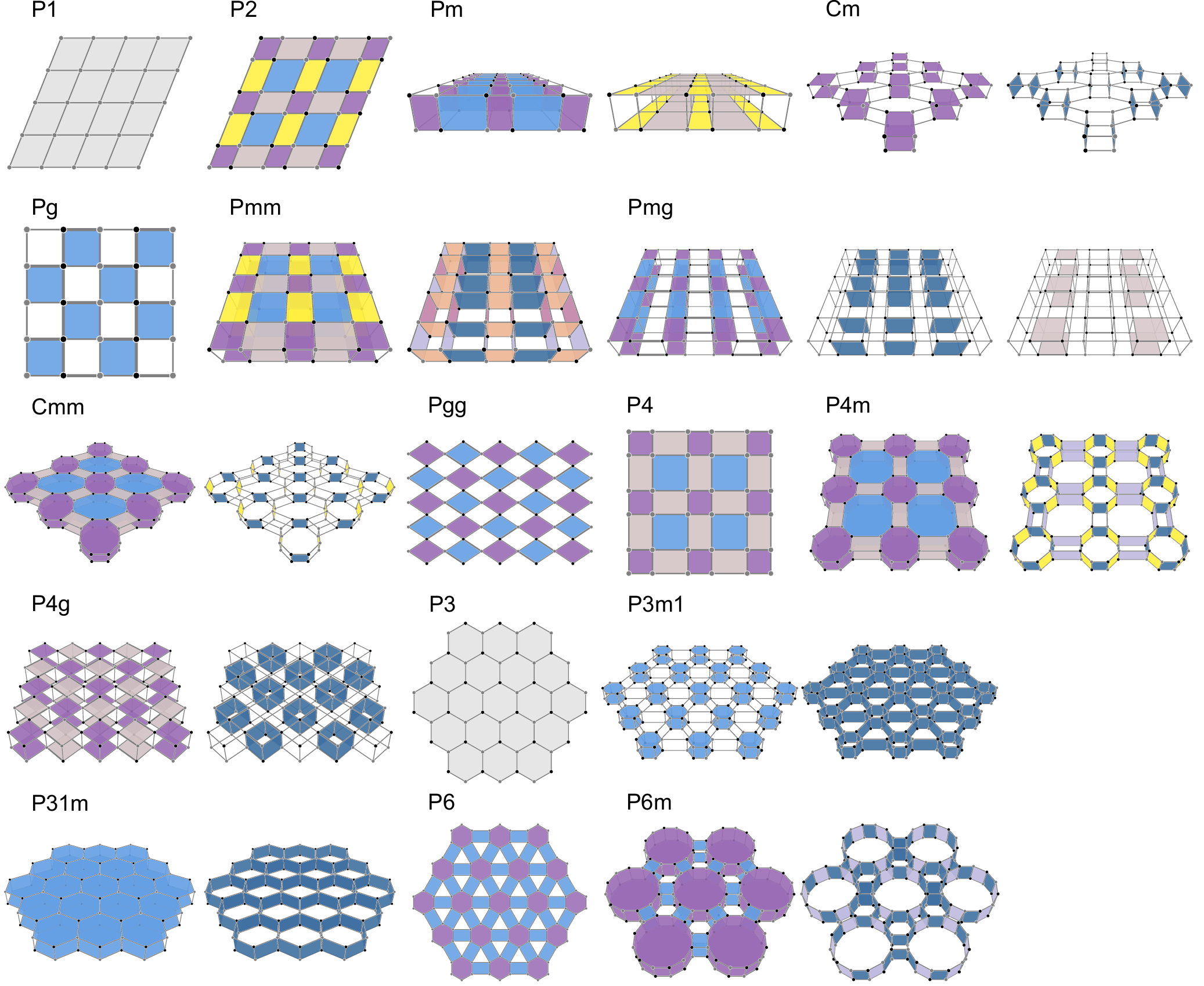}
	\caption{\textbf{Illustration for the canonical models of 17 wallpaper groups.} 
	For each wallpaper group, the cohomology invariants are realized by independent fluxes $\Phi_a\in \{0,\pi\}$ on the lattice, which are distinguished by different colors. }
	\label{fig:Standard_Models}
\end{figure*}	
In the main text, we emphasized that with $T$ symmetry, the phase factors of space group symmetries can be constrained to be valued in $\mathbb{Z}_2$. Here, we present a proof for this proposition.

Let us enlarge the space group $G$ by including $T$ with $T^2=1$. Then, each group element can be written as $gT^a$ with $g\in G$ and $a=0, 1$. Suppose that under the projective representation $\rho$, the phase factor $\lambda$ arises through

\begin{equation}
\rho(g_1T^{a_1})\rho(g_2T^{a_2})=\lambda(g_1T^{a_1},g_2T^{a_2})\rho(g_1g_2T^{a_1+a_2}).
\end{equation}

We shall prove that by appropriately modifying the phase of each operator $\rho(gT^a)$, we can always transform $\lambda(g_1T^{a_1},g_2T^{a_2})$ into the form,

\begin{equation}
\tilde{\lambda}(g_1T^{a_1},g_2T^{a_2})=\nu(g_1,g_2)\omega(T^{a_1},T^{a_2}),
\end{equation}

where $\nu(g_1,g_2)$, $\omega(T^{a_1},T^{a_2})\in \mathbb{Z}_2$.

We start with observing that for all $g\in G$,

\begin{equation}
\rho(g)\rho(T)=\lambda(g,T)\rho(gT)=\frac{\lambda(g,T)}{\lambda(T,g)}\rho(T)\rho(g),
\end{equation}

which motivates us to modify the phase of each $\rho(g)$ as

\begin{equation}
\tilde{\rho}(g):=\sqrt{\frac{\lambda(T,g)}{\lambda(g,T)}}\rho(g),
\end{equation}

Note that $\rho(T)$ is an anti-unitary operator, i.e., $\rho(T)c=c^*\rho(T)$ for $c\in \mathbb{C}$. Hence,

\begin{equation}
\tilde{\rho}(g)\rho(T)=\rho(T)\tilde{\rho}(g).
\end{equation}

We further modify the operators for the other half of group elements as

\begin{equation}
\tilde{\rho}(gT):=\sqrt{\lambda(g,T)\lambda(T,g)}\rho(gT),
\end{equation}

for all $g\in G$. Note that $\tilde{\rho}(T)=\rho(T)$. Then, one observes that

\begin{equation}
\tilde{\rho}(g)\tilde{\rho}(T)=\tilde{\rho}(gT).
\end{equation}

Let $\tilde{\lambda}$ denote the phase factor for $\tilde{\rho}$. Restricting on $G$, $\tilde{\lambda}$ satisfies

\begin{equation}
\tilde{\rho}(g_1)\tilde{\rho}(g_2)=\tilde{\lambda}(g_1,g_2)\tilde{\rho}(g_1g_2)
\end{equation}

for all $g_1,g_2\in G$. The left-hand side commutes with $\tilde{\rho}(T)$, so does the right-hand side. Hence, $\nu:=\tilde{\lambda}|_{G\times G}\in \mathbb{Z}_2=\{\pm 1\}$. On the other hand,  $\tilde{\lambda}(T,T)$ appears in

\begin{equation}
\tilde{\rho}(T)\tilde{\rho}(T)=\tilde{\lambda}(T,T)1.
\end{equation}

Clearly, $\tilde{\lambda}(T,T)$ commutes with $\tilde{\rho}(T)$, and therefore $\omega(T,T):=\tilde{\lambda}(T,T)\in \mathbb{Z}_2$.

Finally, it is straightforward to check that

\begin{equation}
\begin{split}
\tilde{\rho}(g_1T^{a_1})\tilde{\rho}(g_2T^{a_2})&=\tilde{\rho}(g_1)\tilde{\rho}(T^{a_1})\tilde{\rho}(g_2)\tilde{\rho}(T^{a_2})\\ &=\tilde{\rho}(g_1)\tilde{\rho}(g_2)\tilde{\rho}(T^{a_1})\tilde{\rho}(T^{a_2})\\
&=\nu(g_1,g_2)\tilde{\rho}(g_1g_2)\omega(T^{a_1},T^{a_2})\tilde{\rho}(T^{a_1+a_2})\\
&=\nu(g_1,g_2)\omega(T^{a_1},T^{a_2})\tilde{\rho}(g_1g_2T^{a_1+a_2}).
\end{split}
\end{equation}

This concludes the proof of our proposition. In the proof, we have repeatedly used the relations: $\tilde{\rho}(g)\tilde{\rho}(T)=\tilde{\rho}(gT)$ and $\tilde{\rho}(g)\tilde{\rho}(T)=\tilde{\rho}(T)\tilde{\rho}(g)$.
\\

\noindent \textbf{Projective symmetry algebras and gauge fluxes} 
Let us consider a set of lattice sites and hopping amplitudes among them, which give rise to a tight-binding model,

\begin{equation}
\hat{H}=\sum_{ij}H_{ij}a^\dagger_i a_j.
\end{equation}

Here, $a^\dagger_i$ and $a_j$ are the particle creation and annihilation operators at sites $i$ and $j$, respectively. $H_{ij}$ represents the hopping amplitudes $t_{ij}$ from site $j$ to $i$ if $i\ne j$ and the onsite energy $\epsilon_i$ at site $i$ if $i=j$. $H$ is a Hermitian matrix and called the one-particle Hamiltonian of the tight-binding model.

Each hopping amplitude $t_{ij}$ may have a phase $e^{i\phi_{ij}}$ (such that $t_{ij}=|t_{ij}|e^{i\phi_{ij}}$), which is called the gauge connection of the lattice model. Particularly, here we consider the $\mathbb{Z}_2$ gauge connections with $\phi_{ij}\in \{0,\pi\}$. For each closed loop $C$ formed by successive hoppings, one can compute the product $W_{C}$ of the phases of all the hopping amplitudes involved. $W_C$ is called the Wilson loop operator of the loop $C$, and the gauge flux $\Phi_C$ through the loop $C$ is given by $W_C=e^{-i\Phi_C}$. For the $\mathbb{Z}_2$ gauge field, we have $W_C\in\{\pm 1\}$ and $\Phi_C=\{0,\pi\}$.

For each site $i$, we may change the phase of $a^\dagger_i$ for each $i$ by an arbitrary $e^{i\theta^i}$. Particularly, $\theta^i$ is valued in $\{0,\pi\}$ for the $\mathbb{Z}_2$ gauge field considered here. Accordingly, the hopping amplitudes are transformed as $t_{ij}\mapsto e^{i\theta^i}t_{ij}e^{-i\theta^j}$, which is called a gauge transformation. An immediate result is that $W_C=e^{-i\Phi_C}$ is invariant under any gauge transformation. This can be seen from that the ending site of a hopping is the starting site of the next hopping in a loop $C$, and therefore all phase changes involved are cancelled out. To summarize, the gauge fluxes are gauge-invariant quantities, whereas the gauge connections are not.

%The gauge transformation can be described by a diagonal matrix $\mathsf{G}$ with $\mathsf{G}_{ii}=e^{i\chi_i}$. One may immediately observe that the Hamiltonian is transformed as
%\begin{equation}
%H\mapsto \mathsf{G}H\mathsf{G}^\dagger.
%\end{equation}

Only gauge-invariant quantities are physical. In the current case, the gauge flux configuration completely determines the physics of the model. Hence, a spatial transformation $R$ that leaves  the crystal and the gauge flux configuration invariant is regarded as a symmetry of the system. However, $R$ does not necessarily preserve the gauge-connection configuration $A$. After the action of $R$, $A$ is generally changed to another one $A'$. Since the two gauge-connection configurations $A$ and $A'$ describe the same flux configuration, they are related by a gauge transformation $\mathsf{G}_R$.
On the lattice, $R$ is represented by a matrix indexed by lattice sites, which we still denote by $R$. The gauge transformation $\mathsf{G}_R$ is a diagonal matrix with $[\mathsf{G}_R]_{ii}=e^{i\theta_R^i}$, i.e., with the phase assigned to the $i$th site.
Then, the \emph{physical} symmetry operator in this case should be the combination

\begin{equation}\label{physical-operator}
	\mathsf{R}=\mathsf{G}_R R.
\end{equation}

That is, after the spatial transformation $R$, the gauge transformation $\mathsf{G}_R$ is needed to recover the original gauge connection configuration $A$. Notably, it is $\mathsf{R}=\mathsf{G}_R R$ that commutes with the Hamiltonian $H$, i.e.,

\begin{equation}
[\mathsf{R},H]=0.
\end{equation}

%We now enunciate how the physical operator $\mathsf{R}$ acts on $\hat{H}= \sum_{ij} H_{ij} | i\rangle   \langle j|$. $\bR$ transforms $\hat{H}$ to
%\begin{equation}
%\begin{split}
%\hat{H}' &= \sum_{ij} H_{ij} \mathsf{G}_R|  R (i)\rangle \langle  R  (j) |\mathsf{G}_R^{-1} \\
%=&
%\sum_{ij} H_{ij}  \mathsf{G}_R(R(i)) \mathsf{G}_R^*(R(j))  | R (i)\rangle \langle  R  (j) |  \nonumber \\
%=&  \sum_{ij} \mathsf{G}_R(i) \mathsf{G}_R^*(j)   H_{R^{-1}(i)R^{-1}(j)}   | i\rangle  \langle  j |,
%\end{split}
%\end{equation}
%where $G_{R}(i)=e^{i\chi_i}$, namely the phase assigned to site $i$, and $R(i)$ is the site transformed from $i$ by $R$. Thus, if $\hat{H}$ preserves $\bR$, we have
The commutation relation is equivalent to the requirement,

\begin{equation} \label{phase-relation}
	t_{ij} =  \mathsf{G}_R(i)t_{R^{-1}(i)R^{-1}(j)} \mathsf{G}_R^*(j) ,
\end{equation}

where $G_{R}(i)=e^{i\theta^i_R}$, namely the phase assigned to site $i$, and $R(i)$ is the site transformed from $i$ by $R$.

Then, we consider the successive action of two spatial symmetries, $\bR_1 = \bG_{R_1} R_1$ and $\bR_2 = \bG_{R_2} R_2 $. There are two natural operators to implement the action, namely, $\bG_{R_1} R_1\bG_{R_2} R_2$ and $\bG_{R_{12}}R_{12}$ with $R_{12}=R_1R_2$. Their difference is $\Delta_\bG(R_1,R_2)=\bG_{R_1} R_1\bG_{R_2}R_1^{-1}/\bG_{R_1R_2}$. $\Delta_\bG(R_1,R_2)$ is a diagonal matrix with $i$th diagonal entry being $\bG_{R_1}(i)\bG_{R_2}(R_1^{-1}(i))/\bG_{R_1R_2}(i)$, and therefore represents a gauge transformation. It is clear that $\Delta_\bG(R_1,R_2)$ commutes with all possible symmetry-preserving Hamiltonians. Particularly, let us presume the usual case that $H$ is a connected lattice model, i.e., any two sites are connected by hoppings. The presumption sufficiently leads to the fact that $\Delta_\bG(R_1,R_2)$ is proportional to the identity matrix, namely $[\Delta_\bG(R_1,R_2)]_{ij}=\nu(R_1,R_2)\delta_{ij}$ with $\nu(R_1,R_2)\in \mathbb{Z}_2 \subset U(1)$, i.e., the physical symmetry operators satisfy the PSA,

\begin{equation}
	\bR_1 \bR_2  = \nu(R_1,R_2)  \bR_{12}.
\end{equation}

If $\nu$ and $\nu'$ are related by transforming $\bR$ to $\bR' = \chi(R) \bR$ with $\chi(R)\in U(1)$ or $\mathbb{Z}_2$, the two PSAs belong to the same cohomology class. It must be noted that the cohomology class of such a PSA is independent of the choice of gauge connections, and is solely determined by the flux configuration.  \\

\noindent \textbf{Realization of cohomology invariants}
Based on the general discussions in the last section, we now show the flux lattices in Fig.~\ref{fig:Flux_Invariant}\textbf{c}-\textbf{h} can realize the five classes of cohomology invariants, respectively.

	(\textit{i}) Let us start with the cohomology invariant $\sigma = \bL_a \bL_b \bL_a^{-1} \bL_b^{-1}$. Since
	
	\begin{equation}
		\begin{split}
			\bL_a \bL_b \bL_a^{-1} \bL_b^{-1} &= \bG_a  L_a  \bG_b  L_b (\bG_a  L_a )^{-1}  (\bG_b  L_b)^{-1}\\
			 &=  \bG_a  (L_a  \bG_b L_a^{-1})  L_b \bG_a L_b^{-1}   (\bG_b  L_b)^{-1},
		\end{split}
	\end{equation}
	
	the algebraic relation is equivalent to
	
	\begin{equation}
	\bG_a(i)  \bG_b(L_a^{-1}(i))  \bG_a^*(L_b^{-1}(i)) \bG_b^*(i) = \sigma
	\end{equation}
	
    for any lattice site $i$.
	 For the lattice model in Fig.~\ref{fig:Flux_Invariant}\textbf{c}, we have from \eqref{phase-relation} the relations
	 
	\begin{equation}
		t_{23} = t_{14} \bG_{a}(2) \bG_{a}^*(3),\quad t_{43} = t_{12} \bG_{b}(4) \bG_{b}^*(3),
	\end{equation}
	 
	which implies
	 
	\begin{equation}
%	\begin{split}
	e^{i \phi_{23}} =e^{i \phi_{14}} \bG_{a}(2) \bG_{a}^*(3),\ \ \ e^{i \phi_{43}} = e^{i \phi_{12}} \bG_{b}(4) \bG_{b}^*(3).
%	\end{split}	
	\end{equation}
	 
	Here,  $1,2,3,4$ label the four sites in Fig.~\ref{fig:Flux_Invariant}\textbf{c}, which are counted counterclockwise from the lower left corner.
	The flux through the rectangle satisfies
	 
	\begin{equation}
		\begin{split}
			e^{-i\Phi} &= e^{i \phi_{12}} e^{i \phi_{23}} e^{i \phi_{34}} e^{i \phi_{41}}  \\
			&= e^{i \phi_{12}} e^{i \phi_{14}} e^{i \phi_{21}} e^{i \phi_{41}} \bG_{a}(2) \bG_{a}^*(3) \bG_{b}^*(4) \bG_{b}(3)  \\
			&= \bG^*_{a}(3) \bG_{b}^*(L_a^{-1}(3)) \bG_{a}(L_b^{-1}(3)) \bG_{b}(3)  = \sigma^* .
		\end{split}
	\end{equation}
	 
	Thus, in the presence of flux $\Phi$, $L_a$ and $L_b$ satisfy $\bL_a \bL_b \bL_a^{-1} \bL_b^{-1} =e^{i\Phi}$. This argument can be generalized to other lattices. See Supplementary Figures~28 and 37.
	
	(\textit{ii}) 	For a cohomology invariant $ \alpha = \bR^n_{2\pi/n} $, $n$ must be even. Here, we have added the subscript $2\pi/n$ for $\bR$ to specify the rotation angle. When $n$ is even, rotating $n/2$ times is a two-fold rotation $ R^{n/2}_{2\pi/n} = R_{\pi} $. In general, $\bR_{2 \pi/n}^{n/2} = \xi \bR_{\pi}$ with $\xi\in \mathbb{Z}_2$. No matter whether $\xi=\pm 1$, the cohomology invariant can always be expressed as
	 
	\begin{equation}
		\bR_{\pi}^2 =\alpha.
	\end{equation}
	 
	Substituting $\bR_{\pi}=\mathsf{G}_\pi R_\pi$ into the identity above, we see the cohomology invariant can be realized by
	 
	\begin{equation}\label{Rpi-gauge}
		\bG_{\pi}(i) \bG_{\pi} (R_{\pi}(i)) = \alpha
	\end{equation}
	 
	for any site $i$.
	
	Let us label the vertices of the plaquette in Fig.~\ref{fig:Flux_Invariant}\textbf{d} by $i=1,2,3,\cdots 2 l$ with $n=2l$. Then, $R_{\pi}(i) = i + l$. From \eqref{phase-relation}, the hopping amplitudes satisfy
	 
	\begin{equation}
		t_{i,i+1} = t_{i+l,i+l+1} \bG_{\pi}(i) \bG_{\pi}^*(i+1).
	\end{equation}
	 
	Then, the flux is found to be
	 
	\begin{equation}
		\begin{split}
		\quad e^{-i \Phi}   &=  \prod_{i = 1 }^{2l} e^{i \phi_{i,i+1}} = \prod_{i = 1}^{l}  \bG_{r_{\pi}}(i) \bG_{r_{\pi}}^*(i+1) \\
		&=\bG_{r_{\pi}}(1) \bG_{r_{\pi}}^*(l+1)
		\end{split}
	\end{equation}
	 
	From \eqref{Rpi-gauge}, we conclude that $e^{i \Phi} = \alpha$. Note that all phases are restricted in $\mathbb{Z}_2=\{\pm 1\}$.
	
	(\textit{iii}) For $\bM^2=\beta$, the $2$D mirror reflection is interpreted as the twofold rotation through an axis parallel to the $2$D plane. Then, it is clear from (\textit{ii}) that $e^{i \Phi} = \beta$ with $\Phi$ the flux through each vertical plaquettes in Fig.~\ref{fig:Flux_Invariant}\textbf{e}.

	(\textit{iv})
	The cohomology invariant $\eta=\bM_x \bL_y \bM_x^{-1}\bL_y^{-1} $ is translated as
	 
	\begin{equation}
		\begin{split}
			\eta= & \bM_x  \bL_y  \bM_x^{-1}  \bL_y^{-1} =  \bG_{m} M_x  \bG_{y} L_y  (\bG_{m} M_x )^{-1}  (\bG_{y} L_y )^{-1} \\
			=& \bG_{m} ( M_x   \bG_{y} M_x^{-1})  M_x L_y M_x^{-1} L_y^{-1}  (L_y \bG_{m} L_y^{-1} )\bG_{y}^{-1}.
		\end{split}
	\end{equation}
	 
	Hence, we have the identity,
	 
	\begin{equation}
	\bG_{m}(i) \bG_{y}(M_x(i))  \bG_{m}^*(L_y^{-1}(i))   \bG_{y}^* (i) = \eta.
	\end{equation}
	 
	Label the four white sites in Fig.~\ref{fig:Flux_Invariant}\textbf{f} by $1$, $2$, $3$ and $4$, which are counted counterclockwise from the lower left site.  From \eqref{phase-relation}, we have the identities,
	 
	\begin{equation}
		t_{23} = t_{14} \bG_{m}(2) \bG_{m}^*(3),\quad t_{43} = t_{12} \bG_{y}(4) \bG_{y}^*(3),
	\end{equation}
	 
	Then, the flux is computed as
	 
	\begin{equation}
		\begin{split}
				e^{-i \Phi} =&  e^{i \phi_{12}} e^{i \phi_{23}} e^{i \phi_{34}} e^{i \phi_{41}}  \\
			=& e^{i \phi_{12}} e^{i \phi_{14}} e^{i \phi_{21}} e^{i \phi_{41}} \bG_{m}(2) \bG_{m}^*(3) \bG_{y}^*(4) \bG_{y}(3)  \\
			=& \bG^*_{m}(3) \bG_{y}^*(M(3)) \bG_{m}(L_y^{-1}(3)) \bG_{y}(3)  = \eta^* .
		\end{split}
	\end{equation}

	(\textit{v})  The algebraic relation $\tau= \bL_x  \bg_x \bL_x\bg_x^{-1}$ leads to
	 
	\begin{equation}
	\begin{split}
			& \bL_x  \bg_x  \bL_x  \bg_x^{-1} =  \bG_{x} L_x  \bG_{g} g_x  (\bG_{x} L_x )  (\bG_{g} g_x )^{-1}  \\
	=& \bG_{x} ( L_x   \bG_{g} L_x^{-1})  L_x g_x L_x  g_x^{-1}  ((L_x g_x)^{-1} \bG_{x} L_x g_x^{-1} )\bG_{g}^{-1},
	\end{split}
	\end{equation}
	 
	which is equivalent to
	 
	\begin{equation}
	\bG_{x}(\bm{r}) \bG_{g}(L_x(\bm{r}))  \bG_{x}(L_x g^{-1}(\bm{r}))   \bG_{g}^* (\bm{r}) = \tau.
	\end{equation}

	For the lattice in Fig.~\ref{fig:Flux_Invariant}\textbf{h}, the hopping amplitudes satisfy
	 
	\begin{equation}
		t_{23} = t_{14} \bG_{x}(2) \bG_{x}^*(3),\quad  t_{43} = t_{21} \bG_{g_x}(4) \bG_{g_x}^*(3),
	\end{equation}
	 
	referring to \eqref{phase-relation}. Here, $1,2,3,4$ label the four sites  in Fig.~\ref{fig:Flux_Invariant}\textbf{h}, which are counted counterclockwise from the lower left one. Then, the flux through the rectangular plaquette is
 
	\begin{equation}
		\begin{split}
		 e^{-i \Phi}=& e^{i \phi_{12}} e^{i \phi_{23}} e^{i \phi_{34}} e^{i \phi_{41}}  \\
		=& e^{i \phi_{12}} e^{i \phi_{14}} e^{i \phi_{12}} e^{i \phi_{41}} \bG_{x}(2) \bG_{x}^*(3) \bG_{g_x}^*(4) \bG_{g_x}(3)  \\
		=& e^{i 2 \phi_{12}} \bG^*_{x}(3) \bG_{g_x}^*(L_x(3)) \bG_{x}(L_x g_x^{-1}(3)) \bG_{g_x}(3)   \\
		=& \bG^*_{x}(3) \bG_{g_x}^*(L_x(3)) \bG_{x}^*(L_x g_x^{-1}(3)) \bG_{g_x}(3)  = \tau^* .
		\end{split}
	\end{equation}
 
Note again that all phases above are either $+1$ or $-1$.\\

\noindent \textbf{Construction of canonical models} 
In the main text, we have elucidated how to realize cohomology invariants $\sigma$, $\alpha$, $\beta$, $\eta$ and $\tau$ in Table~\ref{coboundary-invariant} by lattice flux configurations. Here, we briefly introduce the general procedure for constructing the canonical models for all $17$ wallpaper groups, which realize all $458$ PSAs. 
 It must be noted that the purpose of these models is to demonstrate the physical realization of all PSAs, so they are made as simple as possible and contain only nearest neighbor hoppings. One can certainly write down more complex models with more complicated lattice structures and hopping processes for a given PSA, just like what one typically does when constructing models based on ordinary representations of space groups.

The illustration for all the canonical models is given in Fig.~\ref{fig:Standard_Models}, and the full details for the model construction can be found in the Supplemental Note 3. In this process, we find it useful to categorize the $17$ wallpaper groups into five classes.

(\textit{a}) Groups $P1$, $P3$, and $Pg$ are quite simple, since each of them has only one cohomology invariant $\sigma$ or $\tau$ (see Table~\ref{coboundary-invariant}). It is straightforward to design lattice models with the flux patterns as introduced in (\textit{i}) or (\textit{iv}).
	
(\textit{b}) For groups $P2$, $Pgg$, $P4$, and $P6$, all cohomology invariants are of type $\alpha$ in (\textit{ii}), i.e.,  each $\alpha_i=\bR^n$ for some $n$-fold rotation $\bR$ through a rotation center in the unit cell. Under lattice translations, each rotation center gives a lattice of rotation centers. Accordingly, each $\alpha_i$ is associated with such a lattice, and different $\alpha_i$'s correspond to different lattices. This has been illustrated with our example $P2$ in the main text. Then, the canonical model is constructed as the dual lattice for the lattice of rotation centers. This means each plaquette in the canonical model hosts a unique rotation center; conversely, each rotation center is the center of a plaquette preserving the rotation symmetry. Then, each $\alpha_i=\pm 1$ is realized by inserting flux $0$ or $\pi$ into the corresponding plaquettes.
	
(\textit{c}) For groups $Cm$, $P3m1$, and $P31m$, each has two cohomology invariants $\sigma$ and $\beta$. We first construct a one-layer lattice model realizing $\sigma$ as described in (\textit{i}). Then, we double the one-layer model into a two-layer model, and introduce the nearest-neighbor interlayer hopping amplitudes to realize $\beta$ as given in (\textit{iii}) or in Fig.~\ref{fig:Flux_Invariant}\textbf{e}.
	
(\textit{d}) Each of groups $Pmm$, $Cmm$, $P4m$, $P4g$, and $P6m$ has the two types of cohomology invariants $\alpha$ and $\beta$ in (\textit{ii}) and (\textit{iii}), respectively. Here, following (\textit{b}), we first construct a one-layer model to accommodate all $\alpha$-invariants. Then, we double the one-layer model into a two-layer model, and appropriately insert fluxes for interlayer plaquettes to realize all $\beta$-invariants. Note that according to Fig.~\ref{fig:Flux_Invariant}\textbf{e}, the vertical mirror planes should cross lattice bonds rather than lattice sites, which can always be satisfied.

(\textit{e}) The remaining two groups are $Pm$ and $Pmg$, both having $\eta$- and $\beta$-invariants. Therefore, we refer to (\textit{iv}) and Fig.~\ref{fig:Flux_Invariant}\textbf{g} for constructing two-layer models for them. Since $Pmg$ also has two $\alpha$-invariants, we first construct the one-layer model according to the $\alpha$-invariants following (\textit{b}), and then double it into a two-layer model to accommodate the $\beta$- and $\eta$-invariants following Fig.~\ref{fig:Flux_Invariant}\textbf{g}.
\\

\noindent \textbf{Shift of high-symmetry points} 
We derive the shift of high-symmetry points in Fig.~\ref{Fig-example-model}\textbf{c}.
From the PSAs for $P2$, it is straightforward to derive that
 
    \begin{equation}\label{example-p2-relation-I}
        \begin{split}
			&\bL_a \bL_b \bL_a^{-1} \bL_b^{-1} = \alpha_1 \alpha_2 \alpha_3 \alpha_4 , \quad \bR^2 = \alpha_1, \\
           & \bR \bL_b \bR^{-1} \bL_b = \alpha_1 \alpha_3, \quad \bR \bL_a \bR^{-1} \bL_a = \alpha_1 \alpha_2  .
        \end{split}
	\end{equation}
 
Alternatively, these relations can be derived from the configuration of the canonical model in Fig.~\ref{Fig-example-model}\textbf{a}. There, the fluxes through the plaquettes colored in red, green, purple, blue correspond to cohomology invarants $\alpha_1,\alpha_2,\alpha_3,\alpha_4$, respectively. If $\alpha_i=1$ $(\alpha_i=-1)$, the corresponding flux is $0$ $(\pi)$.

When $\alpha_1 = \alpha_2 = \alpha$ and $\alpha_3 = \alpha_4 = - \alpha$, plaquettes in each row have the same flux, and the flux values alternate across the rows (see Fig.~\ref{Fig-example-model}\textbf{b}). Accordingly, the third equation above gives $\bR \bL_b \bR^{-1} = -\bL_b^{-1}$. In momentum space, $\bL_b$ is diagonalized as $e^{i\bm{k}\cdot b}$. Then, $\bR e^{i\bm{k}\cdot b} \bR^{-1} = e^{-i(\bm{k}-\bm{G}_b/2)\cdot b}$, with $\bm{G}_b$ the reciprocal lattice vector for $b$. Hence, we see that $\bm{k}$ is transformed to $-\bm{k}+\bm{G}_b/2$ under $\bR$, i.e.,
 
\begin{equation}
	\bR:~\bm{k} \mapsto -\bm{k}+\bm{G}_b/2.
\end{equation}
 
Thus, the $\bR$-invariant momenta are shifted to $\pm \bm{G}_b/4$ and $\bm{G}_a/2\pm\bm{G}_b/4$, as shown in Fig.~\ref{Fig-example-model}\textbf{c}.
	\\

\noindent \textbf{Enforced topology by projective symmetry algebras} 
Here, we provide the details for the nontrivial Zak phase enforced by $\bR T$ symmetry discussed in the main text. The $\bR T$ symmetry puts the following constraint for a Hamiltonian:
 
\begin{equation}
U_{\bR T}\mathcal{H}^*(\bm{k}+\bm{G}_b/2)U_{\bR T}^\dagger=\mathcal{H}(\bm{k})
\end{equation}
 
where $U_{\bR T}$ is a unitary operator determined by $\bR T$. Suppose that $\mathcal{H}(\bm{k})$ has a single band $|\psi_{\bm{k}}\rangle$ over the $\bm{G}_b$ period from $\bm{k}=0$ to $\bm{G}_b$. The action of  $\bR T$ on $\ket{\psi_{\bm{k}}}$ will give a band eigenstate at $\bm{k}+\bm{G}_b/2$, generally differing from $\ket{\psi_{\bm{k}+\bm{G}_b/2}}$ by a $\bm k$ dependent phase, i.e.,
 
\begin{equation}
U_{\bR T}\ket{\psi_{\bm{k}}}^*= e^{i\phi(\bm{k})} \ket{\psi_{\bm{k}+\bm{G}_{b}/2 }}.
\end{equation}
 
Accordingly, $\bR T$ relates the Berry connection $\mathcal{A}_b(\bm{k})=\langle \psi_{\bm k}|i\partial_{k_b}|\psi_{\bm k}\rangle$ at $\bm{k}$ and $\bm{k}+\bm{G}_b/2$ as
 
\begin{equation}
\mathcal{A}_b(\bm{k})+\mathcal{A}_b(\bm{k}+\bm{G}_b/2)=\partial_{k_b}\phi(\bm{k}).
\end{equation}
 
Because of this, the Zak phase $\theta_b=\oint dk_b~\mathcal{A}_b(\bm{k})$ over any $\bm{G}_b$-periodic path can be expressed as
 
\begin{equation}
\theta_b=\phi(\bm{G}_b/2)-\phi(0).
\end{equation}
 
Moreover, the PSA relation $(\bR T)^2=\alpha$ requires that $U_{\bR T}U^*_{\bR T}=\alpha$, which in turn leads to
 
\begin{equation}
e^{i \phi(\bm{k} + \bm{G}_{b}/2 ) - i \phi(\bm{k})} =\alpha.
\end{equation}
 
Thus, we arrive at $\theta_b=i\ln \alpha \mod 2\pi$ as claimed in the main text. For $\alpha=-1$, this Zak phase is guaranteed to be nontrivial.
\\

\noindent \textbf{Eightfold degenerate nodal point} 
For the group $P3m1$, when  $\sigma = \alpha = -1$, the little co-algebra at the $\Gamma$ point has two irreducible $4$D representations and one irreducible $8$D representation. The $8$D irreducible representation can be expressed as
 
\begin{equation}
    \begin{split}
       & \bL_a = i \sigma_1 \otimes \sigma_0 \otimes \sigma_0,\quad \bL_b = i\sigma_3 \otimes \sigma_0 \otimes \sigma_0,\\
      &   \bR = U_R \otimes D_R \otimes \sigma_0,\quad \bM = U_M \otimes D_M \otimes \sigma_3,\\
      &    T = \sigma_2 \otimes \sigma_2 \otimes \sigma_1 K.
    \end{split}
\end{equation}
Here, $K$ denotes the complex conjugation, and 	
\begin{equation}
    \begin{split}
       &U_R = \exp(i\bm{n}_1\cdot\bm{\sigma}2\pi/3),\quad  D_R =\exp(-i\sigma_2 2\pi/3), \\
        &	  U_M = \exp (i\bm{n}_2\cdot \bm{\sigma}\pi/2),\quad D_M = \sigma_3,
    \end{split}
\end{equation}
 
with $\bm{n}_2=(1,-1,-1)/\sqrt{3}$ and $\bm{n}_2=(0,1,-1)/\sqrt{2}$.

Following (\textit{c}), we can construct the canonical model for $P3m1$. First, we construct the one-layer lattice that realizes the $\sigma$-invariant for translations as illustrated in Fig.~\ref{Fig-example-model}\textbf{d}. Then, we double the one-layer model into the bilayer model as illustrated in Fig.~\ref{Fig-example-model}\textbf{e}. In order to realize all cohomology invariants, we add flux at all regular hexagons and rectangles. The expression for this lattice model is given in the Supplementary Note 4.

\bigskip
\noindent \textbf{\large DATA AVAILABILITY}\\
The data generated and analyzed during this study are available from the corresponding author upon request.

\bigskip
\noindent \textbf{\large CODE AVAILABILITY}\\
All code used to generate the plotted band structures is available from the corresponding author upon request.

\bigskip
\def\bibsection{\ }
\noindent \textbf{REFERENCES}
\bibliographystyle{naturemag}

%\bibliography{WP_Ref}

\bigskip

\noindent \textbf{\large ACKNOWLEDGEMENTS}\\
This work is supported by  National Natural Science Foundation of China (Grants No.~12161160315 and No.~12174181), Basic Research Program of Jiangsu Province (Grant No.~BK20211506), and  Singapore MOE AcRF Tier 2 (MOE2019-T2-1-001).

\bigskip
\noindent \textbf{\large AUTHOR CONTRIBUTIONS}\\
Z.C. and Y.Z. conceived the idea. S.Y. and Y.Z. supervised the project. Z.C., Z.Z. and Y.Z. did the theoretical analysis.
Z.C., Z.Z., S.Y. and Y.Z. wrote the manuscript.

\bigskip
\noindent \textbf{\large COMPETING INTERESTS}\\
The authors declare no competing interests.

\widetext
\clearpage



\section*{Supplementary material}

# Supplementary information for
# "Classification of time-reversal-invariant crystals with gauge structures"

Z. Y. Chen,[1] Zheng Zhang,[1] Shengyuan A. Yang,[2] and Y. X. Zhao[1, 3, *]

[1]*National Laboratory of Solid State Microstructures and Department of Physics, Nanjing University, Nanjing 210093, China*
[2]*Research Laboratory for Quantum Materials, Singapore University of Technology and Design, Singapore 487372, Singapore*
[3]*Collaborative Innovation Center of Advanced Microstructures, Nanjing University, Nanjing 210093, China*

## Contents

## Supplementary Note 1.  Background

In this section, we introduce the basics of projective symmetry algebras for a given symmetry group $G$. First, we introduce the multipliers or factor systems of projective representations, which are classified by the second cohomology group $H^2(G, U(1))$. Then, we discuss the consequences of including time-reversal symmetry into the symmetry group. Finally, we establish the correspondence between projective symmetry algebras and cohomology invariants.

### a.  Introduction to projective representation

A projective representation of a group $G$ with coefficients in an Abelian group $\mathcal{A}$ is a map from $G$ to linear transfomations on a vector space $V$:

$$\rho : G \to \mathrm{GL}(V), \tag{1}$$

which satisfies a modified multiplication rule

$$\rho(g_1)\rho(g_2) = \nu(g_1, g_2)\rho(g_1 g_2), \quad \forall g_1, g_2 \in G. \tag{2}$$

Here, $\nu(g_1, g_2)$ is a function from $G \times G$ to $\mathcal{A}$, which is called a factor system or multiplier. The associativity requires that

$$\nu(g_1, g_2)\nu(g_1 g_2, g_3) = \nu(g_2, g_3)\nu(g_1, g_2 g_3). \tag{3}$$

A function from $G \times G$ to $\mathcal{A}$ satisfies Eq. (3) is called a 2-cocyle. We denote the set of all possible 2-cocyles as $Z^2(G, \mathcal{A})$, which is an abelian group under the multiplication of functions.

$\rho(g)$ can be redefined by a phase factor $\chi(g) \in \mathcal{A}$: $\rho(g) \to \rho'(g) = \chi(g)\rho(g)$, with which the factor system $\nu$ is transformed as

$$\nu(g_1, g_2) \to \nu'(g_1, g_2) = \nu(g_1, g_2)\frac{\chi(g_1)\chi(g_2)}{\chi(g_1 g_2)}. \tag{4}$$

————————
*zhaoyx@nju.edu.cn

We regard $\nu$ and $\nu'$ as equivalent factor systems. They differs each other by a trivial 2-cocyle $\frac{\chi(g_1)\chi(g_2)}{\chi(g_1 g_2)}$. All trivial 2-cocyles form an abelian group $B^2(G, \mathcal{A}) \subset Z^2(G, \mathcal{A})$. Thus, the non-equivalent classes of factor systems are given by the quotient group

$$H^2(G, \mathcal{A}) = Z^2(G, \mathcal{A})/B^2(G, \mathcal{A}), \tag{5}$$

which is also called a second cohomology group.

A projective representation of $G$ with multiplier $\nu \in \mathcal{A}$ corresponds to a group extension of $G$ by $\mathcal{A}$:

$$1 \xrightarrow{i} \mathcal{A} \to \tilde{G} \xrightarrow{p} G \to 1. \tag{6}$$

Here, $i$ is an injective homomorphism, i.e., $\mathcal{A}$ is a subgroup of $\tilde{G}$, and $p$ is a surjective homomorphism. We now consider a lift from $G$ to $\tilde{G}$, i.e., for each $g \in G$ we assign $\mathsf{g} \in \tilde{G}$ with $p(\mathsf{g}) = g$. Then, $g_1 g_2 = g_3$ is lifted to $\mathsf{g}_1 \mathsf{g}_2 = \tilde{\nu}(g_1, g_2) \mathsf{g}_3$ with $\tilde{\nu}(g_1, g_2) \in \mathcal{A}$. It can be shown that $\tilde{\nu}$ is a 2-cocycle, and $\nu$ and $\tilde{\nu}$ are in the same cohomology class.

### b. Projective symmetry algebras with time reversal

Time reversal $T$ generates a twofold cyclic group, which we denote as $\mathcal{Z}_2^T$. In our work, we consider symmetry groups in the form of $G \times \mathcal{Z}_2^T$, where $G$ is a spatial symmetry group.

The equivalence classes of multipliers form an abelian group, termed as the twisted second-cohomology group

$$H^{c,2}(G \times \mathcal{Z}_2^T, U(1)), \tag{7}$$

where the superscript $c$ indicates the complex conjugation of $T$ on $U(1)$. Below, we show that each multiplier $\lambda$ of $G \times \mathcal{Z}_2^T$ is similar to the following decomposition

$$\lambda(g_1 T^{a_1}, g_2 T^{a_2}) = \nu(g_1, g_2)\omega(T^{a_1}, T^{a_2}), \quad \nu, \omega \in \mathbb{Z}_2 = \{\pm 1\}, \tag{8}$$

where $g_1, g_2 \in G$ and $a_1, a_2 \in \{0, 1\}$. This decomposition implies that

$$H^{c,2}(G \times \mathcal{Z}_2^T, U(1)) = H^2(G, \mathbb{Z}_2) \times H^2(\mathcal{Z}_2^T, \mathbb{Z}_2). \tag{9}$$

Before proving this result, let us first give some remarks on its physical meanings.

(i) With time-reversal invariance, the multipliers can be restricted to be valued in $\mathbb{Z}_2$, rather than in $U(1)$. Conversely, it is sufficient to consider the second group cohomology with $\mathbb{Z}_2$ as coefficient. Since the time-reversal operator acts trivially on $\mathbb{Z}_2 \subset U(1)$, the twisted cohomology is "untwisted". This is particularly interesting in quantum physics. For instance, recall that according to Wigner each multiplier of the inhomogeneous Lorentz group (the connected component containing the identity) is similar to one valued in $\mathbb{Z}_2$. Then, all multipliers of the inhomogeneous Lorentz groups can preserve time-reversal symmetry. This seems an accidental nice property, given that time-reversal is not included in the inhomogeneous Lorentz group. Wigner's result can be traced back to the fundamental group of the inhomogeneous Lorentz group, which is $\mathbb{Z}_2$. Analogous analysis can be given for $SO(3)$.

(ii) $H^2(\mathcal{Z}_2^T, \mathbb{Z}_2) \cong \mathbb{Z}_2$ is specified by $\mathcal{T}^2 = (-1)^j$. Here, $j\hbar/2$ is the spin number of the particles under consideration, and $\mathcal{T}$ denotes the anti-unitary operator representing $T$.

(iii) The classification of projective representations of $G \times \mathcal{Z}_2^T$, namely $H^{c,2}(G \times \mathcal{Z}_2^T, U(1))$, has been reduced to calculating $H^2(G, \mathbb{Z}_2)$ with coefficient $\mathbb{Z}_2$.

(iv) That the spin is integral or half integral specifies a $\mathbb{Z}_2$-multiplier of $G \times \mathcal{Z}_2^T$, which we denote by $\lambda_s$ with $s = 0, 1/2$ for integral and half integral spins, respectively. Restricted on $G$, let us denote $\nu_s = \lambda_s|_{G \times G}$.

(v) If $G$ consists of spatial transformations over a lattice, the $\mathbb{Z}_2$ gauge flux configuration on the lattice can endow a $\mathbb{Z}_2$-multiplier $\nu_f$ for $G$, and can preserve time-reversal symmetry at the same time.

(vi) Still consider $G$ consisting of spatial transformations over a lattice. Then, the internal spin degrees of freedom and the external gauge fluxes together lead to the $\mathbb{Z}_2$-multiplier for $G$:

$$\nu(g_1, g_2) = \nu_s(g_1, g_2)\nu_f(g_1, g_2). \tag{10}$$

(vii) It is possible that some $\mathbb{Z}_2$ gauge flux configurations $F$ can realize $\nu_{s=1/2}$. Then, such as flux configuration can exchange the multiplier classes of integral and half integral spins.

We now proceed to prove every $U(1)$-multiplier $\sigma$ is similar to the decomposition into two $\mathbb{Z}_2$-multipliers: $\lambda(g_1 T^{a_1}, g_2 T^{a_2}) = \nu(g_1, g_2)\omega(T^{a_1}, T^{a_2})$. We observe that

$$\rho(g)\rho(T) = \lambda(g, T)\rho(gT) = \frac{\lambda(g, T)}{\lambda(T, g)}\rho(T)\rho(g), \tag{11}$$

which motivates us to perform the transformation,

$$\tilde{\rho}(g) := \sqrt{\frac{\lambda(T, g)}{\lambda(g, T)}}\rho(g), \tag{12}$$

for all $g \in G$. Then, since $\rho(T)c = c^*\rho(T)$ for all $c \in \mathbb{C}$,

$$\tilde{\rho}(g)\rho(T) = \rho(T)\tilde{\rho}(g). \tag{13}$$

We further transform the operators for the other half of group elements as

$$\tilde{\rho}(gT) := \sqrt{\lambda(g, T)\lambda(T, g)}\rho(gT), \tag{14}$$

for all $g \in G$. Note that $\tilde{\rho}(T) = \rho(T)$. Hence,

$$\tilde{\rho}(g)\tilde{\rho}(T) = \tilde{\rho}(gT). \tag{15}$$

We further restrict the transformed multiplier $\tilde{\lambda}$ to $G$. Considering

$$\tilde{\rho}(g_1)\tilde{\rho}(g_2) = \tilde{\lambda}(g_1, g_2)\tilde{\rho}(g_1 g_2) \tag{16}$$

we see the left side commutes with $\tilde{\rho}(T)$, so does the right side. Hence, $\nu = \tilde{\lambda}|_{G \times G} \in \mathbb{Z}_2 = \{\pm 1\}$. Restricting $\tilde{\lambda}$ on $\mathcal{Z}_2^T$, there is only one variable $\tilde{\lambda}(T, T)$, which appears in

$$\tilde{\rho}(T)\tilde{\rho}(T) = \tilde{\lambda}(T, T)\mathbb{1}. \tag{17}$$

Clearly, $\tilde{\lambda}(T, T)$ commutes with $\tilde{\rho}(T)$, and therefore $\omega = \tilde{\lambda}|_{\mathcal{Z}_2^T \times \mathcal{Z}_2^T} \in \mathbb{Z}_2$.

Finally, it is straightforward to check

$$\begin{aligned}
\tilde{\rho}(g_1 T^{a_1})\tilde{\rho}(g_2 T^{a_2}) &= \tilde{\rho}(g_1)\tilde{\rho}(T^{a_1})\tilde{\rho}(g_2)\tilde{\rho}(T^{a_2}) \\
&= \tilde{\rho}(g_1)\tilde{\rho}(g_2)\tilde{\rho}(T^{a_1})\tilde{\rho}(T^{a_2}) \\
&= \nu(g_1, g_2)\tilde{\rho}(g_1 g_2)\omega(T^{a_1}, T^{a_2})\tilde{\rho}(T^{a_1 + a_2}) \\
&= \nu(g_1, g_2)\omega(T^{a_1}, T^{a_2})\tilde{\rho}(g_1 g_2 T^{a_1 + a_2}).
\end{aligned} \tag{18}$$

which verifies the decomposition form. Note that we have repeatedly used the facts: $\tilde{\rho}(g)\tilde{\rho}(T) = \tilde{\rho}(gT)$ and $\tilde{\rho}(g)\tilde{\rho}(T) = \tilde{\rho}(T)\tilde{\rho}(g)$.

### c. Projective symmetry algebras and cohomology invariants

Let us now discuss how to present all cohomology classes of multipliers for a symmetry group $G$ in terms of cohomology invariants constructed as algebraic relations of symmetry operators.

Mathematically, a wallpaper group can be presented by generators and relations (the algebra of generators)

$$G = \langle \mathcal{S} | \mathcal{R} \rangle, \tag{19}$$

where $\mathcal{S}$ is the set of generators $\{s_1, s_2, \cdots, s_{n_s}\} \subset G$, and $\mathcal{R}$ is the set of relations among generators $\{r_1, r_2, \cdots, r_{n_r}\}$, which satisfy

$$r_1(\boldsymbol{s}) = 1, r_2(\boldsymbol{s}) = 1, \cdots, r_{n_r}(\boldsymbol{s}) = 1. \tag{20}$$

For any element $g \in G$, we can specify a "word" of generators to form it,

$$w(g) = s_{i_1} s_{i_2} \cdots . \tag{21}$$

Two "words" can be equal under the relations, for example,

$$
\begin{aligned}
w_i &= s_{i_1} s_{i_2} \cdots s_{i_{m-1}} \underbrace{s_{i_m} \cdots s_{i_{m+k-1}}}_{r_j = 1} s_{i_{m+k}} \cdots \\
&= s_{i_1} s_{i_2} \cdots s_{i_{m-1}} s_{i_{m+k}} \cdots = w_f.
\end{aligned} \tag{22}
$$

The projective representation of a wallpaper group can be described in a more compact way by modifying relations of the presentation:

$$r_1(\boldsymbol{s}) = \alpha_1, r_2(\boldsymbol{s}) = \alpha_2, \cdots, r_{n_r}(\boldsymbol{s}) = \alpha_{n_r}, \tag{23}$$

where $\alpha_1, \alpha_2, \cdots, \alpha_{n_r} \in \mathcal{A}$. We denote the modified presentation as

$$\tilde{G} = \langle \mathcal{S} | \mathcal{R}, \mathcal{F} \rangle, \tag{24}$$

where $\mathcal{F} : \mathcal{R} \to \mathcal{A}$ is a function describing the modification, $\mathcal{F}(r_i) = \alpha_i$. We call the modified presentation $\tilde{G}$ a *projective symmetry algebra*, or shortly, *projective algebra*. It can be considered as a modified group algebra over $\mathcal{A}$.

Now, two "words" can be equal up to a phase under the new relations, for example,

$$
\begin{aligned}
w_i &= s_{i_1} s_{i_2} \cdots s_{i_{m-1}} \underbrace{s_{i_m} \cdots s_{i_{m+k-1}}}_{r_j = \alpha_j} s_{i_{m+k}} \cdots \\
&= \alpha_j s_{i_1} s_{i_2} \cdots s_{i_{m-1}} s_{i_{m+k}} \cdots = \alpha_j w_f.
\end{aligned} \tag{25}
$$

Now, we show that all the cohomology classes of factor systems can be described in the form of Eq. (24).

Suppose we have a group $G$, which has a presentation $G = \langle \mathcal{S} | \mathcal{R} \rangle$. Given a projective representation of $G$, $\rho : G \to \mathrm{GL}(V)$, the representation of generators $\rho(s), s \in \mathcal{S}$ will satisfy modified relations

$$r_1(\rho(\boldsymbol{s})) = \alpha_1, r_2(\rho(\boldsymbol{s})) = \alpha_2, \cdots, r_{n_r}(\rho(\boldsymbol{s})) = \alpha_{n_r}. \tag{26}$$

where $\alpha_i$ is determined by the factor system. So the representation of generators form a modified presentation $\tilde{G} = \langle \rho(\mathcal{S}) | \mathcal{R}, \mathcal{F} \rangle$, where $\rho(\mathcal{S}) = \{\rho(s) | s \in \mathcal{S}\}$. In general, $\rho(\mathcal{S})$ does not generate the original projective representation, since for each group element $w(g) = s_{i_1} s_{i_2} \cdots$, the representation $\rho(g)$ in general has the form

$$\rho(g) = \chi_{\mathcal{S}}(g) \rho(s_{i_1}) \rho(s_{i_2}) \cdots, \tag{27}$$

where $\chi_{\mathcal{S}}(g)$ is a phase due to the factor system. However, we can redefining $\rho'(g) = \chi_{\mathcal{S}}(g)^{-1} \rho(g)$, which is exactly the projective representation generated by $\rho(\mathcal{S})$. Moreover, its factor system is in the same cohomology class with $\rho$. Thus, every cohomology classes of factor systems of $G$ can be described by a projective algebra $\tilde{G}$. So, to discuss cohomology classes of factor systems, we can concentrate on projective algebras, which are much more simpler than factor systems.

Similar to the factor system $\nu$, we cannot choose the factor function $\mathcal{F} \in \mathcal{A}^{n_r}$ arbitrarily because the modified relations may not be consistent with the associativity of $\tilde{G}$. If we start with a "word" $w_i$, associate generators in two different paths $P_1, P_2$ and result in the same final "word" $w_f$, the associativity requires

$$w_i = \left( \prod_{j \in P_1} \mathcal{F}(r_j) \right) w_f = \left( \prod_{j \in P_2} \mathcal{F}(r_j) \right) w_f. \tag{28}$$

This is a general requirement of associativity for projective algebra, and we will see that the cocyle equation Eq. (3) is a special form of it in the following. The set of factor functions $\mathcal{F}$ satisfy the associative condition Eq. (28) form a group $\mathrm{AMap}(\mathcal{R}, \mathcal{A}) \subset \mathcal{A}^{n_r}$. Furthermore, we can also redefine each generator $s_i \to s_i' = \chi_i s_i$, and values of factor function will be transformed as

$$\mathcal{F}(r_i(\boldsymbol{s})) \to \mathcal{F}'(r_i(\boldsymbol{s})) = \mathcal{F}(r_i(\boldsymbol{s})) r_i^{-1}(\boldsymbol{\chi}). \tag{29}$$

We call this transformation *coboundary transformation*. The sets of factor functions given by $r_i(\boldsymbol{\chi})$ are considered as trivial factor functions, whose set we denote as $\mathrm{TMap}(\mathcal{R}, \mathcal{A}) = \{\mathcal{F} : \mathcal{R} \to \mathcal{A} | \mathcal{F}(r_i(\boldsymbol{s})) = r_i(\boldsymbol{\chi}), \chi \in \{\mathcal{S} \to \mathcal{A}\}\} \subset$

AMap$(\mathcal{R}, \mathcal{A})$. The quotient group of AMap$(\mathcal{R}, \mathcal{A})$ by TMap$(\mathcal{R}, \mathcal{A})$ will give the same result as the second group cohomology

$$\text{AMap}(\mathcal{R}, \mathcal{A}) / \text{TMap}(\mathcal{R}, \mathcal{A}) = H^2(G, \mathcal{A}). \tag{30}$$

Here is an example. Every group has a trivial presentation $G = \langle \mathcal{S} | \mathcal{R} \rangle$, where $\mathcal{S}$ is the group $G$ itself and $\mathcal{R} = \{r_{ij} = g_i g_j g_{ij}^{-1} = 1 | i, j \in G\}$ is the set of multiplication relations of $G$. When we consider the projective algebra over $\mathcal{A}$, The factor function is just the factor system $\mathcal{F}(r_{ij}) = \nu(g_i, g_j) = \alpha_{i,j} \in \mathcal{A}$. For a word of length three $g_1 g_2 g_3$, we have two paths to associate it into $g_{123}$

$$
\begin{aligned}
& \overbrace{g_1 g_2 g_3} \\
= \quad & \underbrace{g_1 g_2}_{\alpha_{1,2} g_{12}} \; g_3 = \alpha_{1,2} \underbrace{g_{12} g_3}_{\alpha_{12,3} g_{123}} = \alpha_{1,2}\alpha_{12,3} \; g_{123} \\
= \quad & g_1 \underbrace{g_2 g_3}_{\alpha_{2,3} g_{23}} = \alpha_{2,3} \underbrace{g_1 g_{23}}_{\alpha_{1,23} g_{123}} = \alpha_{2,3}\alpha_{1,23} \; g_{123}.
\end{aligned}
$$

We see the cocyle equation of factor system Eq. (3) is derived as expectated. In this example, AMap$(\mathcal{R}, \mathcal{A}) = Z^2(G, \mathcal{A})$, and trivial maps take the form as $r_{ij}(\chi) = (\chi_i \chi_j)/\chi_{ij}$, thus TMap$(\mathcal{R}, \mathcal{A}) = B^2(G, \mathcal{A})$.

For another example, we take a look at the group $G = \mathbb{Z}_2^2$, the presentation is given by

$$\mathbb{Z}_2^2 = \langle e_1, e_2 | e_1^2, e_2^2, e_1 e_2 e_1^{-1} e_2^{-1} \rangle. \tag{31}$$

When we consider the projective algebra over $U(1)$, the modified relations are

$$e_1^2 = \alpha_1, e_2^2 = \alpha_2, e_1 e_2 e_1^{-1} e_2^{-1} = \alpha_3, \quad \alpha_1, \alpha_2, \alpha_3 \in U(1). \tag{32}$$

For $w = e_1 e_2 e_1^{-1} e_2^{-1} e_2 e_1 e_2 e_1^{-1} e_2^{-1} e_2$, we have two paths to associate the generators:

$$
\begin{aligned}
& w = e_1 e_2 e_1^{-1} \underbrace{e_2^{-1} e_2}_{1} e_1 e_2 e_1^{-1} \underbrace{e_2^{-1} e_2}_{1} = \\
& e_1 e_2 \underbrace{e_1^{-1} e_1}_{1} e_2 e_1^{-1} = e_1 \underbrace{e_2 e_2}_{\alpha_2} e_1^{-1} = \alpha_2 \underbrace{e_1 e_1^{-1}}_{1} = \alpha_2,
\end{aligned}
\tag{33}
$$

and

$$
\begin{aligned}
& w = \underbrace{e_1 e_2 e_1^{-1} e_2^{-1}}_{\alpha_3} e_2 \underbrace{e_1 e_2 e_1^{-1} e_2^{-1}}_{\alpha_3} e_2 = \\
& \alpha_3^2 \underbrace{e_2 e_2}_{\alpha_2} = \alpha_2 \alpha_3^2.
\end{aligned}
\tag{34}
$$

Thus, the associativity requires $\alpha_2 \alpha_3^2 = \alpha_2$, i.e., $\alpha_3$ can only takes value in $\{\pm 1\} \subset U(1)$. Since there are no other restriction on $\alpha_1$ and $\alpha_2$, AMap$(\mathcal{R}, U(1)) = U(1) \otimes U(1) \otimes \mathbb{Z}_2$. By redefining $e_1 \to e_1' = e_1 \chi_1$, $e_2 \to e_2' = e_2 \chi_2$, factors $\alpha_1, \alpha_2, \alpha_3$ will be transformed as $\alpha_1 \to \alpha_1' = \alpha_1 \chi_1^2$, $\alpha_2 \to \alpha_2' = \alpha_2 \chi_2^2$. Thus $U(1) \otimes U(1) \subset$ AMap$(\mathcal{R}, U(1))$ can be trivialized. Finally, the corresponding second group cohomology is $H^2(\mathbb{Z}_2^2, U(1)) = \mathbb{Z}_2$.

For a general group presentation $G = \langle \mathcal{S} | \mathcal{R} \rangle$, we do not find general form of self-consistent equations required by Eq. (28). However, for a wallpaper group $G$, there exist a standard presentation

$$w(g) = L_a^i L_b^j R^k M^l, \forall g \in G, \tag{35}$$

where $L_a, L_b$ are generators of translations, and $R, M$ are generators of rotation and reflection. For the production of two group elements $g_1 g_2 = g_3$, we can always use relations to turn $w(g_1) w(g_2)$ into the standard presentation with an additional factor $\nu(g_1, g_2)$,

$$w(g_1) w(g_2) \quad = L_a^{i_1} L_b^{j_1} R^{k_1} M^{l_1} L_a^{i_2} L_b^{j_2} R^{k_2} M^{l_2} = \nu(g_1, g_2) L_a^{i_3} L_b^{j_3} R^{k_3} M^{l_3} = \nu(g_1, g_2) w(g_3). \tag{36}$$

With the factor system $\nu(g_1, g_2)$ obtained from the presentation, all the self-consistent equations can be found from cocyle equations of $\nu$.

Solving cocyle equations is a tedious work, fortunately, the second group cohomology $H^2(G, U(1))$ and $H^2(G, \mathbb{Z}_2)$ are already known for all wallpaper groups. Thus, in this work, we use some self-consistent equations to reduce factors $\alpha_i = \mathcal{F}(r_i)$ and check the consistency with group cohomology.

In general, values of $\alpha_i = \mathcal{F}(r_i)$ is not invariant under coboundary transformation $\mathcal{F}(r_i) \to \mathcal{F}(r_i) r_i(\chi), \forall \chi \in \{\mathcal{S} \to \mathcal{A}\}$. We can recombine original relations into new one $r_i' = r_{i_1} r_{i_2} \cdots$, whose value is $\mathcal{F}(r_i') := \mathcal{F}(r_{i_1}) \mathcal{F}(r_{i_1}) \cdots$. If the new value is invariant under coboundary transformation $\mathcal{F}(r_i') = \mathcal{F}(r_i') r_i'(\chi), \forall \chi \in \{\mathcal{S} \to \mathcal{A}\}$, we call it a *cohomology invariant*. Different classes of projective algebras of a wallpaper group can be labelled by a complete set of cohomology invariants, and thus all the factor systems of the wallpaper group can be obtained by enumerating all the possible values of cohomology invariants in this complete set. In the following, we will use projective algebras and cohomology invariants to describe factor systems of wallpaper groups.

## Supplementary Note 2.  Projective symmetry algebras of wallpaper groups and their cohomology invariants

Based on the basics of projective symmetry algebras introduced in the previous section, we proceed to construct all the $(U(1)$ and $\mathbb{Z}_2)$ projective algebras and the corresponding cohomology invariants for the 17 wallpaper groups.

Let us start with summarizing our notations. Wallpaper groups contain the following primary symmetry operations: translation, rotation, reflection and glide-reflection. In general, $L, R, M$ and $g$ are used to represent them respectively. The modified symmetry operations are written in the bold font like $\mathsf{L}, \mathsf{R}, \mathsf{M}, \mathsf{g}$. We will use diagrams to visualize the wallpaper groups. In these diagrams, the rotation centers, reflection axes and glide-reflection axes are represented by the shapes in Fig. 1.

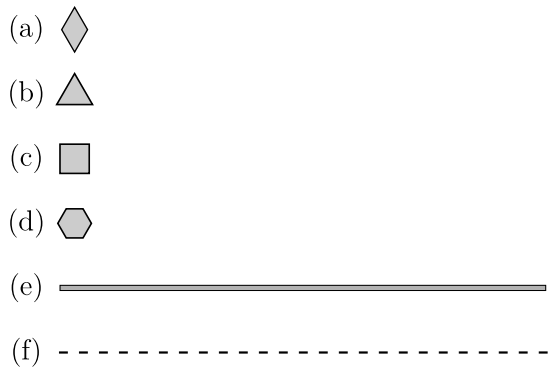

Supplementary Fig.1: Notations for cell structure. (a) A center of rotation of order two ($\pi$). (b) A center of rotation of order three ($2\pi/3$). (c) A center of rotation of order four ($\pi/2$). (d) A center of six-fold rotation ($\pi/3$). (d) An axis of reflection. (d) An axis of glide-reflection.

### a.   $C_n$ and $D_n$

Before introducing projective algebras of wallpaper groups, we first introduce projective algebras of point group $C_n$ and $D_n$, which can help the reader understand our method.

#### i.   $C_n$

The group $C_n$ is generated by rotation operation $R$ with relation $R^n = 1$. The relation acquires a factor $\alpha$ after extension,

$$\mathsf{R}^n = \alpha. \tag{37}$$

For $U(1)$ extension, $\alpha \in U(1)$. However, this factor can be cancelled by redefining the generator $\mathsf{R} \to \mathsf{R}'$:

$$\mathsf{R}' = \alpha^{-1/n} \mathsf{R}, \quad (\mathsf{R}')^n = \alpha(\alpha^{-1/n})^n = 1. \tag{38}$$

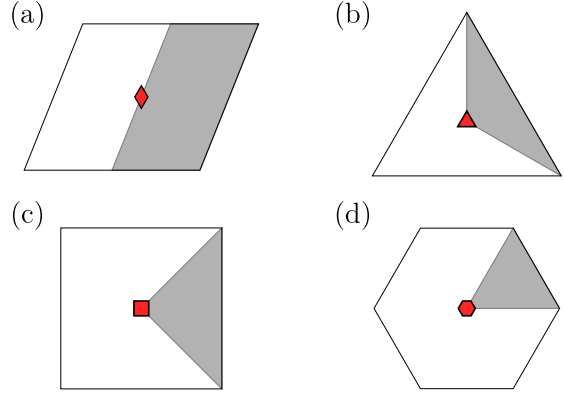

Supplementary Fig.2: Illustration of point groups $C_n$. (a) A parallelogram under point group $C_2$, the shaded region is the fundamental domain. (b) A triangle under point group $C_3$, the shaded region is the fundamental domain. (c) A square under point group $C_4$, the shaded region is the fundamental domain. (d) A hexagon under point group $C_6$, the shaded region is the fundamental domain.

So $C_n$ only has the trivial $U(1)$ projective algebra, which is in agreement with the result of the group cohomology

$$H^2(C_n, U(1)) = 1. \tag{39}$$

For $\mathbb{Z}_2$ extension, $\alpha \in \{\pm 1\}$. When $n$ is odd, we can redefine $\mathsf{R} \to \mathsf{R}' = \alpha\mathsf{R}$, thus $(\mathsf{R}')^n = \alpha\alpha^n = \alpha^{n+1} = 1$. So $C_n$ only has the trivial $\mathbb{Z}_2$ projective algebra when $n$ is odd. However, when $n$ is even, factor $\alpha$ cannot be cancelled by redefining the generator $\mathsf{R}$. So there are two nonequivalent projective algebras of $C_n$ when $n$ is even, where $\alpha = -1$ corresponds to the nontrivial projective representation and $\alpha = 1$ corresponds to the trivial one. One can check $\alpha$ is a cohomology invariant. These results are consistent with the results of the group cohomology

$$H^2(C_n, \mathbb{Z}_2) = \begin{cases} 1 & n \text{ odd} \\ \mathbb{Z}_2 & n \text{ even} \end{cases}. \tag{40}$$

### ii. $D_n$

Now we proceed to analyze group $D_n$. The group $D_n$ is generated by rotation operation $R$ and mirror reflection operation $M$ with relations $R^n = 1, M^2 = 1, MRM^{-1} = R^{-1}$. After extension, these relations become

$$\mathsf{R}^n = \alpha_r, \tag{41a}$$
$$\mathsf{M}^2 = \alpha_m, \tag{41b}$$
$$\mathsf{MRM}^{-1} = \alpha_{rm}\mathsf{R}^{-1}. \tag{41c}$$

To see the restriction on factors by the self-consistency condition, we do conjugate transformation for Eq. (41a), and we have

$$\begin{aligned} \alpha_r &= \mathsf{MR}^n\mathsf{M}^{-1} = (\mathsf{MRM}^{-1})^n \\ &= \alpha_{rm}^n \mathsf{R}^{-n} = \alpha_{rm}^n \alpha_r^{-1}, \end{aligned} \tag{42}$$

where the associativity Eq. (28) is implied in the derivation. $\alpha_r$ and $\alpha_{rm}$ satisfy

$$\alpha_{rm}^n \alpha_r^{-2} = 1. \tag{43}$$

For $U(1)$ extension there are two solution of $\alpha_r$ in terms of $\alpha_{rm}$,

$$\alpha_r = \pm\alpha_{rm}^{n/2} \equiv \alpha\alpha_{rm}^{n/2}. \tag{44}$$

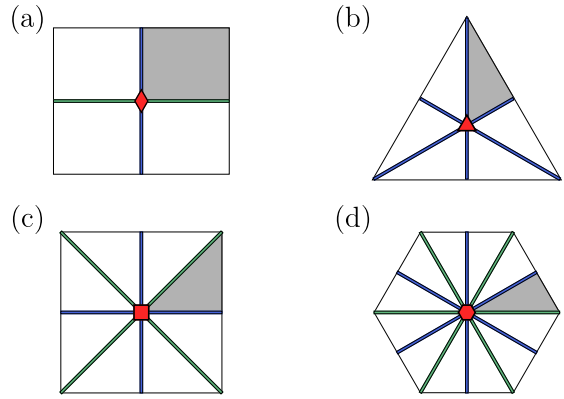

Supplementary Fig.3: Illustration of point groups $D_n$.(a) A rectangle under point group $D_2$, the shaded region is the fundamental domain. There are two conjugacy classes of reflection which are colored differently. (b) A triangle under point group $D_3$, the grey shaded region is the fundamental domain. There is only one conjugacy class of reflection. (c) A square under point group $D_4$, the shaded region is the fundamental domain. There are two conjugacy classes of reflection which are colored differently. (d) A hexagon under point group $D_6$, the shaded region is the fundamental domain. There are two conjugacy classes of reflection which are colored differently.

When $n$ is odd, we can redefine $R \to R' = \pm\alpha_{rm}^{-1/2}R$, which cancels factor $\alpha_r$ and $\alpha_{rm}$ simultaneously. Factor $\alpha_m$ can also be cancelled by redefining $M \to M' = \alpha_m^{-1/2}M$. Thus we do not have any cohomology invariant in this case. When $n$ is even, however, we can only reduce the relation to $R'^n = \pm 1$ by redefining $R \to R' = \pm\alpha_{rm}^{-1/2}R$. So there is one independent cohomology invariant $\alpha = \alpha_r\alpha_{rm}^{-n/2} = \pm 1$ in this case. The cohomology invariant $\alpha \in \{\pm 1\}$ is equal to the commutator of two pependicular reflection when $n$ is even

$$\alpha = \left[ M, R^{n/2}M \right]. \tag{45}$$

These results are consistent with that of the group cohomology

$$H^2(D_n, U(1)) = \begin{cases} 1 & n \text{ odd} \\ \mathbb{Z}_2 & n \text{ even} \end{cases}. \tag{46}$$

For $\mathbb{Z}_2$ extension, $\alpha_{rm}, \alpha_m, \alpha_r \in \mathbb{Z}_2$. When $n$ is odd, Eq. (43) has two solutions $\alpha_{rm} = 1, \alpha_r = \pm 1$. However, the nontrivial one $\alpha_{rm} = 1, \alpha_r = -1$ can be cancelled by redefining $R \to R' = \alpha_r R$. The factor $\alpha_m = \pm 1$ cannot be reduced anymore, which is the only cohomology invariant in this case. When $n$ is even, Eq. (43) is satisfied automatically, and we cannot reduce any of the three factors by redefining the generators, so we have three independent cohomology invariants $\alpha_r, \alpha_m, \alpha_{rm}$ in this case. The results are consistent with that of the group cohomology

$$H^2(D_n, \mathbb{Z}_2) = \begin{cases} \mathbb{Z}_2 & n \text{ odd} \\ \mathbb{Z}_2^3 & n \text{ even} \end{cases}. \tag{47}$$

The $\mathbb{Z}_2$ extensions of $D_n$ can also be analyzed by another way based on conjugacy classes. For a group $G$, the conjugacy class of group element $r$ is defined as $C_s(r) \equiv \{grg^{-1}|g \in G\}$. Here is a general conclusion: Suppose $r$ is an element of order $n$, i.e., $r^n = 1$. After extension, we have $r^n = \alpha$. If $r'$ is in the conjugate class $C_s(r)$, then it also satisfies $r'^n = \alpha$ after extension. This conclusion is easy to show. Suppose $r' = grg^{-1}$, then after extension, we have

$$(r')^n = (grg^{-1})^n = gr^ng^{-1} = g\alpha g^{-1} = \alpha. \tag{48}$$

In wallpaper groups, rotation $R$ and reflection $M$ are both order $n$ elements, so this conclusion can be applied to the conjugacy classes $C_s(R)$ and $C_s(M)$. This may help us to reduce the factors.

For group $D_n$, to apply the above conclusion, we rewrite Eq. (41) as

$$R^n = \alpha_r \equiv \alpha_1, \tag{49a}$$
$$M^2 = \alpha_m \equiv \beta_1, \tag{49b}$$
$$(RM)^2 = \alpha_{cm}\alpha_m \equiv \beta_2. \tag{49c}$$

When $n$ is odd, we can set $n = 2k - 1$ for some integer $k$. Now, $M$ and $RM$ are two reflections and they belong to the same conjugacy class, which can be seen by

$$R^{1-k}MR^{-(1-k)} = R^{1-k}MR^{k-1}M^{-1}M$$
$$= R^{2-2k}M = R^{1-2k}RM = RM.$$

According to Eq. (48), reflections in the same conjugacy class have the same factor, so $\beta_1 = \beta_2$. On the other hand, as we show before, the factor $\alpha_1$ of rotation can be cancelled when $n$ is odd. Thus, we only have one independent cohomology invariant to label the projective representations of $D_n$ in this case. When $n$ is even, $M$ and $RM$ belong to different conjugacy classes, so $\beta_1$ and $\beta_2$ are independent. Futhermore, $\alpha_1$ cannot be cancelled. Thus we have three independent cohomology invariants in this case.

### b. P1

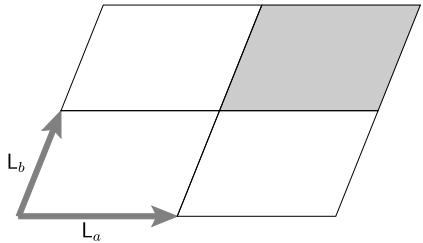

Supplementary Fig.4: Cell structure of $P1$. The shaded region is the fundamental domain.

The group $P1$ has two independent generators of translation $L_a$, $L_b$. The two generators commute with each other, $[L_a, L_b] = L_a L_b L_a^{-1} L_b^{-1} = 1$. So the presentation of $P1$ is given by

$$P1 = \langle L_a, L_b | [L_a, L_b] \rangle. \tag{50}$$

The projective algebra can be obtained by adding an additional factor to the relation,

$$\mathsf{L}_a \mathsf{L}_b \mathsf{L}_a^{-1} \mathsf{L}_b^{-1} = \sigma. \tag{51}$$

For $U(1)$ extension, $\sigma$ takes values in $U(1)$, and for $\mathbb{Z}_2$ extension, $\sigma$ takes values in $\mathbb{Z}_2$. $\sigma$ is the cohomology invariant. These results are consistent with that of the group cohomology

$$H^2(P1, U(1)) = U(1), \tag{52}$$
$$H^2(P1, \mathbb{Z}_2) = \mathbb{Z}_2. \tag{53}$$

### c. P2

The group $P2$ is obtained by adding two-fold rotation to $P1$. The generator of rotation $R$ reverses the directions of translation $L_a$, $L_b$, so the presentation of $P2$ is given by

$$P2 = \quad \langle L_a, L_b, R | [L_a, L_b], RL_a R^{-1} = L_a^{-1},$$
$$RL_b R^{-1} = L_b^{-1}, R^2 \rangle. \tag{54}$$

In group $P2$, there are four different conjugacy classes of rotation $C_s(R)$, $C_s(L_a R)$, $C_s(L_b R)$, $C_s(L_a L_b R)$, as shown in Fig. 5 with four different colors. The relations of presentation can also be expressed in terms of the squares of the four rotation centers:

$$P2 = \langle L_a, L_b, R | R^2, (L_a R)^2, (L_b R)^2, (L_a L_b R)^2 \rangle. \tag{55}$$

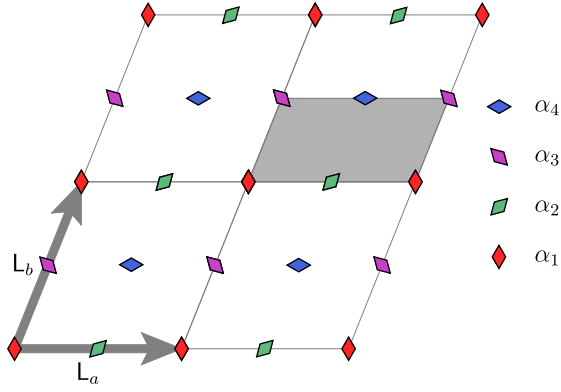

Supplementary Fig.5: Cell structure of $P2$. The shaded region is the fundamental domain. There are four different classes of rotations in group $P2$, their corresponding rotation centers are drawn in different colors. Rotation centers in the same class have the same factor.

When we consider projective algebra, relations of presentation (54) become

$$\mathsf{L}_a\mathsf{L}_b\mathsf{L}_a^{-1}\mathsf{L}_b^{-1} = \sigma, \tag{56a}$$

$$\mathsf{R}\mathsf{L}_a\mathsf{R}^{-1}\mathsf{L}_a = \eta_a, \tag{56b}$$

$$\mathsf{R}\mathsf{L}_b\mathsf{R}^{-1}\mathsf{L}_b = \eta_b, \tag{56c}$$

$$\mathsf{R}^2 = \alpha. \tag{56d}$$

For $\mathbb{Z}_2$ extension, four factors $\sigma, \eta_x, \eta_y, \alpha \in \mathbb{Z}_2$ serve as independent cohomology invariants of projective representations.

For $U(1)$ extension, factors $\eta_a, \eta_b, \alpha$ can be cancelled by redefining $\mathsf{L}_a \to \mathsf{L}_a' = \eta_a^{-1/2}\mathsf{L}_a, \mathsf{L}_b \to \mathsf{L}_b' = \eta_b^{-1/2}\mathsf{L}_b, \mathsf{R} \to \mathsf{R}' = \alpha^{-1/2}\mathsf{R}$. Thus we only have one independent cohomology invariant $\sigma \in U(1)$.

These results are consistent with that of the group cohomology

$$H^2(P2, \mathbb{Z}_2) = \mathbb{Z}_2^4, \tag{57a}$$

$$H^2(P2, U(1)) = U(1). \tag{57b}$$

For $\mathbb{Z}_2$ extension, we can also obtain the projective algebra by modifying the relations in Eq. (55) as

$$\mathsf{R}^2 = \alpha \equiv \alpha_1, \tag{58a}$$

$$(\mathsf{L}_a\mathsf{R})^2 = \eta_a\alpha \equiv \alpha_2, \tag{58b}$$

$$(\mathsf{L}_b\mathsf{R})^2 = \eta_b\alpha \equiv \alpha_3, \tag{58c}$$

$$(\mathsf{L}_a\mathsf{L}_b\mathsf{R})^2 = \sigma\eta_a\eta_b\alpha \equiv \alpha_4. \tag{58d}$$

These relations mean that each conjugacy class of rotation has an independent cohomology invariant, as shown in Fig. 5.

### d. Pm

For group Pm, two translations $L_x, L_y$ are in perpendicular directions. The reflection $M_x$ reverses $L_x$ to $L_x^{-1}$ but leaves $L_y$ invariant. The presentation is given by

$$Pm = \quad \langle L_x, L_y, M_x | [L_x, L_y], M_xL_xM_x^{-1} = L_x^{-1},$$
$$M_xL_yM_x^{-1} = L_y, M_x^2\rangle. \tag{59}$$

There are two conjugacy classes of reflection $C_s(M_x), C_s(L_xM_x)$. They are parallel to each other, with a distance of half of the lattice constant in the $x$ direction, as shown in Fig. 6. The presentation can also be written in terms of

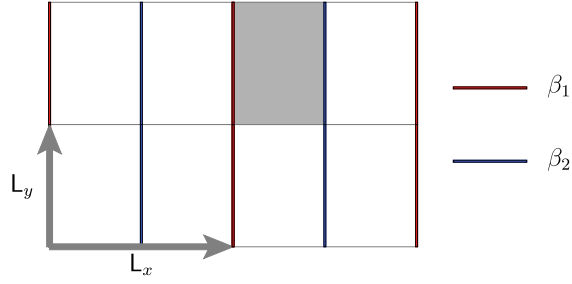

Supplementary Fig.6: Cell structure of $Pm$. There are two conjugacy classes of reflection in $Pm$. The gray shaded region is the fundamental domain of the group.

squares of the two reflections $M_x, L_x M_x$ and their commutators with $L_y$,

$$
\begin{aligned}
Pm = \quad & \langle L_x, L_y, M_x | [M_x, L_y], [L_x M_x, L_y], \\
& (L_x M_x)^2, M_x^2 \rangle.
\end{aligned}
\tag{60}
$$

The projective relations of the presentation Eq. (59) are

$$
\mathsf{L}_x \mathsf{L}_y \mathsf{L}_x^{-1} \mathsf{L}_y^{-1} = \sigma,
\tag{61a}
$$

$$
\mathsf{M}_x \mathsf{L}_y \mathsf{M}_x^{-1} \mathsf{L}_y^{-1} = \eta_y,
\tag{61b}
$$

$$
\mathsf{M}_x \mathsf{L}_x \mathsf{M}_x^{-1} \mathsf{L}_x = \eta_x,
\tag{61c}
$$

$$
\mathsf{M}_x^2 = \alpha.
\tag{61d}
$$

The self-consistency condition requires $\sigma, \eta_y \in \{\pm 1\}$, this can be derived by conjugate relations Eq.(61)(a)(d).

$$
\begin{aligned}
\sigma \quad & = \mathsf{M}_x \sigma \mathsf{M}_x^{-1} = \mathsf{M}_x \mathsf{L}_x \mathsf{L}_y \mathsf{L}_x^{-1} \mathsf{L}_y^{-1} \mathsf{M}_x^{-1} \\
& = \mathsf{L}_x^{-1} \mathsf{L}_y \mathsf{L}_x \mathsf{L}_y^{-1} = (\mathsf{L}_y \mathsf{L}_x^{-1} \mathsf{L}_y^{-1} \mathsf{L}_x)^{-1} = \sigma^{-1},
\end{aligned}
\tag{62}
$$

$$
\alpha \quad = \mathsf{L}_y \alpha \mathsf{L}_y^{-1} = \mathsf{L}_y \mathsf{M}_x^2 \mathsf{L}_y^{-1} = \eta_y^{-2} \alpha^2.
\tag{63}
$$

Factors $\eta_x, \alpha$ are trivial in $U(1)$ projective representations because they can be cancelled by redefining $\mathsf{L}_x \to \mathsf{L}_x' = \eta_x^{-1/2} \mathsf{L}_x, \mathsf{M}_x \to \mathsf{M}_x' = \alpha^{-1/2} \mathsf{M}_x$. Thus, we have two independent cohomology invariants $\sigma, \eta_y \in \{\pm 1\}$.

For $\mathbb{Z}_2$ extension, $\sigma, \eta_x, \eta_y, \alpha \in \{\pm 1\}$ are four nontrivial independent cohomology invariants.

Above results are consistent with that of the group cohomology

$$
H^2(Pm, \mathbb{Z}_2) = \mathbb{Z}_2^4,
\tag{64a}
$$

$$
H^2(Pm, U(1)) = \mathbb{Z}_2^2.
\tag{64b}
$$

For $\mathbb{Z}_2$ extension, we can also obtain the projective algebra by modifying the relations in Eq. (60) as

$$
(\mathsf{M}_x)^2 = \alpha \equiv \beta_1,
\tag{65a}
$$

$$
(\mathsf{L}_x \mathsf{M}_x)^2 = \eta_x \alpha \equiv \beta_2,
\tag{65b}
$$

$$
[\mathsf{M}_x, \mathsf{L}_y] = \eta_y \equiv \eta_1,
\tag{65c}
$$

$$
[\mathsf{L}_x \mathsf{M}_x, \mathsf{L}_y] = \sigma \eta_y \equiv \eta_2.
\tag{65d}
$$

The first two relations means that the two classes of reflections have two independent cohomology invariants, and the the last two relations means that they also have two cohomology invariants which are their commutators with $L_y$.

### e. $Pg$

The group $Pg$ contains glide-reflection operations. Here we suppose the reflection operation reverses the $x$ direction, so the glide-reflection operator is $g_x = L_{\frac{y}{2}} M_x$, where $L_{\frac{y}{2}}$ is a translation in the $y$ direction with half lattice constant.

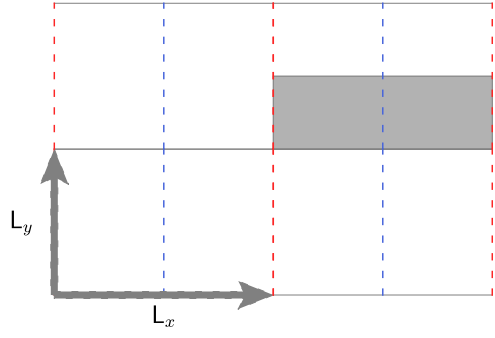

Supplementary Fig.7: Cell structure of $Pg$. The gray shaded region is the fundamental domain of the group.

Since $g_x^2 = L_y$ , we can take $L_x$ and $g_x$ as generators of group $Pg$. Because the glide-reflection $g_x$ reverses the direction of $L_x$, the presentation is given by

$$Pg = \langle g_x, L_y | g_x L_x g_x^{-1} = L_x^{-1} \rangle. \tag{66}$$

The projective relations of presentation Eq. (66) is

$$\mathsf{g}_x \mathsf{L}_x \mathsf{g}_x^{-1} = \tau \mathsf{L}_x^{-1}. \tag{67}$$

For $\mathbb{Z}_2$ extension, $\tau \in \{\pm 1\}$ is the only one independent cohomology invariant. For $U(1)$ extension, $\tau$ is trivial because we can cancel it by redefining $\mathsf{L}_x \to \mathsf{L}_x' = \tau^{-1/2} \mathsf{L}_x$. These are consistent with the results of the group cohomology

$$H^2(Pg, \mathbb{Z}_2) = \mathbb{Z}_2, \tag{68a}$$
$$H^2(Pg, U(1)) = 1. \tag{68b}$$

### f. Cm

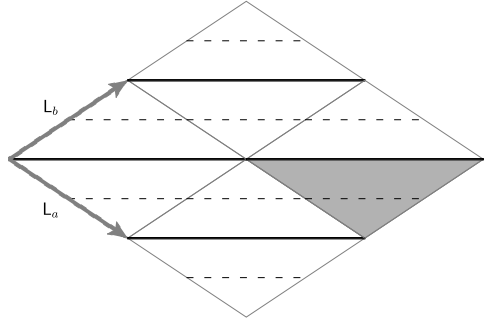

Supplementary Fig.8: Cell structure of $Cm$. The gray shaded region is the fundamental domain of the group.

For group $Cm$ the reflection $M$ interchanges the two translations $L_a, L_b$. The presentation is given by

$$Cm = \langle L_a, L_b, M | [L_a, L_b], M L_a M^{-1} = L_b, M^2 \rangle. \tag{69}$$

When we consider projective algebra, relations of presentation become

$$\mathsf{L}_a \mathsf{L}_b \mathsf{L}_a^{-1} \mathsf{L}_b^{-1} = \sigma, \tag{70a}$$
$$\mathsf{M} \mathsf{L}_a \mathsf{M}^{-1} = \mathsf{L}_b, \tag{70b}$$
$$\mathsf{M}^2 = \beta. \tag{70c}$$

Notice that (70b) does not contribute a factor because the factor $\eta = \mathsf{M}\mathsf{L}_a\mathsf{M}^{-1}\mathsf{L}_b^{-1}$ can be trivialized by redefining $\mathsf{L}_b \to \mathsf{L}_b' = \eta^{-1}\mathsf{L}_b$.

For $\mathbb{Z}_2$ extension, we have two independent cohomology invariants $\sigma, \alpha \in \mathbb{Z}_2$.

For $U(1)$ extension, $\sigma \in \mathbb{Z}_2$ because

$$
\begin{aligned}
\sigma &= \mathsf{M}\sigma\mathsf{M}^{-1} = \mathsf{M}\mathsf{L}_a\mathsf{L}_b^{-1}\mathsf{L}_b^{-1}\mathsf{M}^{-1} \\
&= \mathsf{L}_b\mathsf{L}_a\mathsf{L}_b^{-1}\mathsf{L}_a^{-1} = (\mathsf{L}_a\mathsf{L}_b\mathsf{L}_a^{-1}\mathsf{L}_b^{-1})^{-1} = \sigma^{-1}.
\end{aligned}
\tag{71}
$$

The factor $\alpha$ is trivial in $U(1)$ extension because we can cancel it by redefining $\mathsf{M} \to \mathsf{M}' = \alpha^{-1/2}\mathsf{M}$. Thus we have only one cohomology invariant $\sigma \in \mathbb{Z}_2$.

These results are consistent with that of the group cohomology

$$
H^2(Cm, \mathbb{Z}_2) = \mathbb{Z}_2^2,
\tag{72a}
$$

$$
H^2(Cm, U(1)) = \mathbb{Z}_2.
\tag{72b}
$$

### g. Pmm

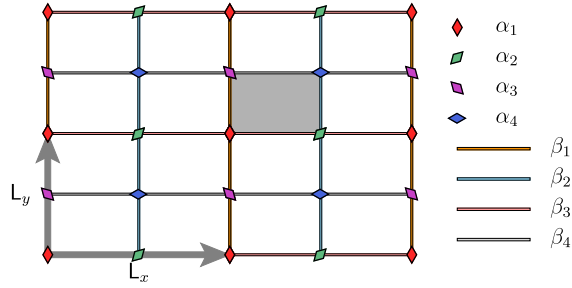

Supplementary Fig.9: Cell structure of $Pmm$. There are four different conjugacy classes of rotations and reflections respectively. The gray shaded region is the fundamental domain of the group.

The group $Pmm$ contains two perpendicular reflection $M_x, M_y$. The combination $M_xM_y$ is a two-fold rotation $R$. The two reflections commute with each other. $M_x$ reverses $L_x$ and preserves $L_y$, while $M_y$ preserves $L_x$ and reverses $L_y$. The presentation of $Pmm$ is

$$
\begin{aligned}
Pmm = \ &\langle L_x, L_y, M_x, M_y | [L_x, L_y], M_xL_xM_x^{-1} = L_x^{-1}, \\
&M_xL_yM_x^{-1} = L_y, M_yL_yM_y^{-1} = L_y^{-1}, \\
&M_yL_xM_y^{-1} = L_x, M_x^2, M_y^2, [M_x, M_y]\rangle.
\end{aligned}
\tag{73}
$$

In group $Pmm$, there are four different conjugacy classes of rotation centers $C_s(R), C_s(L_xR), C_s(L_yR), C_s(L_xL_yR)$, and four different conjugacy classes of reflection axes $C_s(M_y), C_s(L_yM_y), C_s(M_x), C_s(L_xM_x)$, as shown in Fig. 9.

The relations of presentation can also be expressed in terms of the squares of the four rotations and the four reflections:

$$
\begin{aligned}
Pmm = \ &\langle L_x, L_y, M_x, M_y | (M_xM_y)^2, (L_xM_xM_y)^2, \\
&(L_yM_xM_y)^2, (L_xL_yM_xM_y)^2, M_y^2, \\
&(L_yM_y)^2, M_x^2, (L_xM_x)^2\rangle.
\end{aligned}
\tag{74}
$$

When we consider projective algebra, relations of presentation become

$$\mathsf{L}_x\mathsf{L}_y\mathsf{L}_x^{-1}\mathsf{L}_y^{-1} = \sigma, \tag{75a}$$

$$\mathsf{M}_x\mathsf{L}_x\mathsf{M}_x^{-1} = \eta_{m_xx}\mathsf{L}_x^{-1}, \tag{75b}$$

$$\mathsf{M}_x\mathsf{L}_y\mathsf{M}_x^{-1}\mathsf{L}_y^{-1} = \eta_{m_xy}, \tag{75c}$$

$$\mathsf{M}_y\mathsf{L}_y\mathsf{M}_y^{-1} = \eta_{m_yy}\mathsf{L}_y^{-1}, \tag{75d}$$

$$\mathsf{M}_y\mathsf{L}_x\mathsf{M}_y^{-1}\mathsf{L}_x^{-1} = \eta_{m_yx}, \tag{75e}$$

$$\mathsf{M}_x^2 = \alpha_x, \tag{75f}$$

$$\mathsf{M}_y^2 = \alpha_y, \tag{75g}$$

$$\mathsf{M}_x\mathsf{M}_y\mathsf{M}_x^{-1}\mathsf{M}_y^{-1} = \alpha_{xy}. \tag{75h}$$

For $\mathbb{Z}_2$ extension, $\sigma, \eta_{m_xx}, \eta_{m_xy}, \eta_{m_yy}, \eta_{m_yx}, \alpha_x, \alpha_y, \alpha_{xy} \in \mathbb{Z}_2$, and we have eight independent cohomology invariants. For $U(1)$ extension $\sigma, \eta_{m_xy}, \eta_{m_yx}, \alpha_{xy} \in \mathbb{Z}_2$ due to the self-consistency condition in the presence of reflection symmetries, and the proof is similar to Eq. (62) in the case of group $Pm$. Factors $\eta_{m_xx}, \eta_{m_yy}, \alpha_x, \alpha_y$ are trivial because we can cancel them by redefining $\mathsf{L}_x \to \mathsf{L}_x' = \eta_{m_xx}^{-1/2}\mathsf{L}_x, \mathsf{L}_y \to \mathsf{L}_y' = \eta_{m_yy}^{-1/2}\mathsf{L}_y, \mathsf{M}_x \to \mathsf{M}_x' = \alpha_x^{-1/2}\mathsf{M}_x, \mathsf{M}_y \to \mathsf{M}_y' = \alpha_y^{-1/2}\mathsf{M}_y$. Thus, we only have four independent $\mathbb{Z}_2$ cohomology invariants.

Above results are consistent with that of the group cohomology

$$H^2(Pmm, \mathbb{Z}_2) = \mathbb{Z}_2^8, \tag{76a}$$

$$H^2(Pmm, U(1)) = \mathbb{Z}_2^4. \tag{76b}$$

For $\mathbb{Z}_2$ extension, we can also obtain the projective algebra by modifying the relations in Eq. (74) as

$$\mathsf{R}^2 = (\mathsf{M}_x\mathsf{M}_y)^2 = \alpha_r \equiv \alpha_1, \tag{77a}$$

$$(\mathsf{L}_x\mathsf{R})^2 = \eta_{rx}\alpha_r \equiv \alpha_2, \tag{77b}$$

$$(\mathsf{L}_y\mathsf{R})^2 = \eta_{ry}\alpha_r \equiv \alpha_3, \tag{77c}$$

$$(\mathsf{L}_x\mathsf{L}_y\mathsf{R})^2 = \sigma\eta_{rx}\eta_{ry}\alpha_r \equiv \alpha_4, \tag{77d}$$

$$\mathsf{M}_x^2 = \alpha_x \equiv \beta_1, \tag{77e}$$

$$(\mathsf{L}_x\mathsf{M}_x)^2 = \eta_{m_xx}\alpha_x \equiv \beta_2, \tag{77f}$$

$$\mathsf{M}_y^2 = \alpha_y \equiv \beta_3, \tag{77g}$$

$$(\mathsf{L}_y\mathsf{M}_y)^2 = \eta_{m_yy}\alpha_y \equiv \beta_4. \tag{77h}$$

where

$$\alpha_r \equiv \alpha_x\alpha_y\alpha_{xy}, \tag{78a}$$

$$\eta_{rx} \equiv \eta_{m_xx}\eta_{m_yx}, \tag{78b}$$

$$\eta_{ry} \equiv \eta_{m_xy}\eta_{m_yy}. \tag{78c}$$

Eq. (77) means that each conjugacy class of rotations and reflections has an independent cohomology invariant, as Fig. 9 shows.

### h. Pmg

If we replace reflection $M_y$ by glide-reflection $g_y$ in group $Pmm$, we get group $Pmg$. Since $M_x$ reverses $g_y$ and preserves $L_y$. The presentation of $Pmg$ is given by

$$Pmg = \langle g_y, L_y, M_x | g_yL_yg_y^{-1} = L_y^{-1}, M_xL_yM_x^{-1} = L_y, \\ M_xg_yM_x^{-1} = g_y^{-1}, M_x^2 \rangle. \tag{79}$$

There are two conjugacy classes of rotation $C_s(M_xg_y)$, $C_s(L_yM_xg_y)$ and one conjugacy class of reflection $C_s(M_x)$ in $Pmg$. The presentation can also be given in terms of the commutator $[M_x, L_y]$ and the square of each rotation and

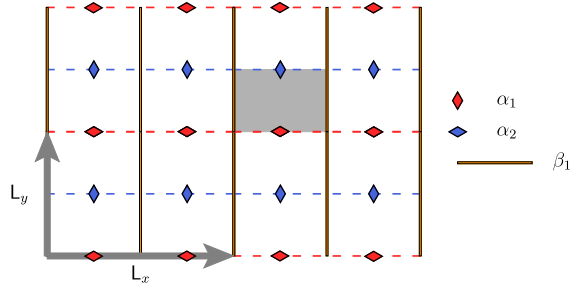

Supplementary Fig.10: Cell structure of $Pmg$. There are two conjugacy classes of rotations and one conjugacy class of reflections. Centers of rotations all lie on glide-reflection axes. The gray shaded region is the fundamental domain of the group.

reflection,

$$
\begin{aligned}
Pmg \quad &= \langle g_y, L_y, M_x | [M_x, L_y], (L_y M_x g_y)^2, \\
&\quad (M_x g_y)^2, M_x^2 \rangle.
\end{aligned}
\tag{80}
$$

Projective relations of Eq. (79) are

$$
g_y L_y g_y^{-1} = \tau_{g_y y} L_y^{-1},
\tag{81a}
$$

$$
M_x L_y M_x^{-1} L_y^{-1} = \eta_{m_x y},
\tag{81b}
$$

$$
M_x g_y M_x^{-1} g_y = \alpha_{m_x g_y},
\tag{81c}
$$

$$
M_x^2 = \alpha_{m_x}.
\tag{81d}
$$

For $\mathbb{Z}_2$ extension, we have four independent cohomology invariants $\tau_{g_y y}, \eta_{m_x y}, \alpha_{m_x g_y}, \alpha_{m_x} \in \mathbb{Z}_2$.

For $U(1)$ extension, we have only one cohomology invariant $\eta_{m_x y}, \in \mathbb{Z}_2$ (the $\mathbb{Z}_2$ value is due to the self-consistency condition). Factors $\tau_{g_y y}, \alpha_{m_x g_y}, \alpha_{m_x}$ can be cancelled by redefining $L_y \to L_y' = \tau_{g_y y}^{-1/2} L_y, g_y \to g_y' = \alpha_{m_x g_y}^{-1/2} g_y, M_x \to M_x' = \alpha_{m_x}^{-1/2} M_x$.

Above results are consistent with that of the group cohomology

$$
H^2(Pmg, \mathbb{Z}_2) = \mathbb{Z}_2^4,
\tag{82a}
$$

$$
H^2(Pmg, U(1)) = \mathbb{Z}_2.
\tag{82b}
$$

For $\mathbb{Z}_2$ extension, we can also obtain the projective algebra by modifying the relations in Eq. (80) as

$$
(M_x g_y)^2 = \alpha_r \equiv \alpha_1,
\tag{83a}
$$

$$
(L_y M_x g_y)^2 = \eta_{ry} \alpha_r \equiv \alpha_2,
\tag{83b}
$$

$$
M_x^2 = \alpha_{m_x} \equiv \beta,
\tag{83c}
$$

$$
M_x L_y M_x^{-1} L_y^{-1} = \eta_{m_x y} \equiv \eta.
\tag{83d}
$$

where

$$
\alpha_r \equiv \alpha_{m_x g_y} \alpha_{m_x},
\tag{84a}
$$

$$
\eta_{ry} \equiv \eta_{m_x y} \tau_{g_y y}.
\tag{84b}
$$

Eq. (83) means that each conjugacy class of rotations and reflections has an independent cohomology invariants, and the commutator of the reflection and $L_y$ is also a cohomology invariant.

### i. Pgg

The group $Pgg$ has two perpendicular glide-reflections $g_x, g_y$. Since $g_x^2 = L_y, g_y^2 = L_x$, we can take $g_x, g_y$ as generators. If we choose the intersection point of the two glide axes as origin, then $g_x g_y$ and $g_x g_y^{-1}$ act on coordinates

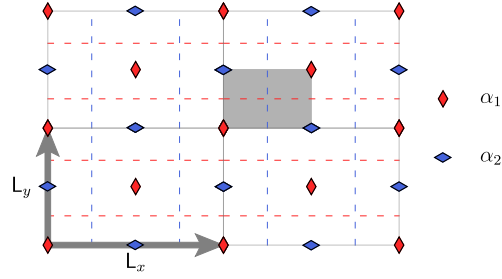

Supplementary Fig.11: Cell structure of $Pgg$. The two glide reflections are in perpendicular directions. There are two conjugacy classes of rotation. The rotation centers do not lie on glide-reflection axes. The gray shaded region is the fundamental domain.

as $g_x g_y(x,y) = g_y(x+1/2, -y) = (-x-1/2, -y+1/2)$, $g_x g_y^{-1}(x,y) = g_x(x-1/2, -y) = (-x+1/2, -y+1/2)$. So they are $R_\pi$ rotation operations around $(-1/4, -1/4)$ and $(1/4, 1/4)$ respectively, as shown in Fig. 11. The presentation of $Pgg$ can be given by

$$Pgg = \langle g_x, g_y | (g_x g_y)^2, (g_x g_y^{-1})^2 \rangle. \tag{85}$$

When we consider projective algebra, relations of the presentation become

$$(\mathsf{g}_x \mathsf{g}_y)^2 = \alpha \equiv \alpha_1, \tag{86a}$$

$$(\mathsf{g}_x \mathsf{g}_y^{-1})^2 = \tau \alpha \equiv \alpha_2. \tag{86b}$$

.

For $\mathbb{Z}_2$ extension, there are two cohomology invariants $\alpha, \tau \in \mathbb{Z}_2$, while for $U(1)$ extension these factors are trivial because we can cancel them by redefining $\mathsf{g}_x \to \mathsf{g}_x' = \alpha^{-1/2} \tau^{-1/4} \mathsf{g}_x, \mathsf{g}_y \to \mathsf{g}_y' = \tau^{1/4} \mathsf{g}_y$. These are in agreement with the results of group cohomology

$$H^2(Pgg, \mathbb{Z}_2) = \mathbb{Z}_2^2, \tag{87a}$$

$$H^2(Pgg, U(1)) = 1. \tag{87b}$$

We can also get cohomology invariants between $\mathsf{g}_y, \mathsf{g}_x$ and translation operator $\mathsf{L}_x = \mathsf{g}_y^2, \mathsf{L}_y = \mathsf{g}_x^2$,

$$\mathsf{L}_x \mathsf{g}_x \mathsf{L}_x \mathsf{g}_x^{-1} = \mathsf{g}_y^2 \mathsf{g}_x \mathsf{g}_y^2 \mathsf{g}_x^{-1}$$
$$= \mathsf{g}_y(\mathsf{g}_y \mathsf{g}_x \mathsf{g}_y) \mathsf{g}_y \mathsf{g}_x^{-1} = \alpha \mathsf{g}_y \mathsf{g}_x^{-1} \mathsf{g}_y \mathsf{g}_x^{-1} = \tau, \tag{88a}$$

$$\mathsf{L}_y \mathsf{g}_y \mathsf{L}_y \mathsf{g}_y^{-1} = \mathsf{g}_x^2 \mathsf{g}_y \mathsf{g}_x^2 \mathsf{g}_y^{-1}$$
$$= \mathsf{g}_x(\mathsf{g}_x \mathsf{g}_y \mathsf{g}_x) \mathsf{g}_x \mathsf{g}_y^{-1} = \alpha \mathsf{g}_x \mathsf{g}_y^{-1} \mathsf{g}_x \mathsf{g}_y^{-1} = \tau. \tag{88b}$$

### j. Cmm

If we add another reflection perpendicular to the reflection of group $Cm$, we get group $Cmm$. The two perpendicular reflections in $Cmm$ commute with each other. One reflection transforms $L_a$ to $L_b$ while the other transforms $L_a$ to $L_b^{-1}$. Thus, the presentation is given by

$$Cmm = \langle L_a, L_b, M_x, M_y | [L_a, L_b], M_x L_a M_x^{-1} = L_b^{-1},$$
$$M_y L_a M_y^{-1} = L_b, [M_x, M_y], M_x^2, M_y^2 \rangle. \tag{89}$$

There are three conjugacy classes of rotations $C_s(M_x M_y), C_s(L_a M_x M_y), C_s(L_a M_x L_a M_y)$ and two conjugacy classes of reflections $C_s(M_x), C_s(M_y)$. We can also present the group by

$$Cmm = \langle L_a, M_x, M_y | (M_x M_y)^2, (L_a M_x M_y)^2,$$
$$(L_a M_x L_a^{-1} M_y)^2, M_x^2, M_y^2 \rangle. \tag{90}$$

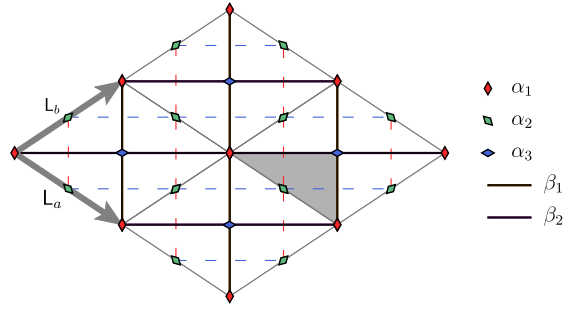

Supplementary Fig.12: Cell structure of $Cmm$. There are two (conjugacy class of) reflections in perpendicular directions. One class of rotation centers are not on reflection axes, while the other two classes of rotation centers are on reflection axes. The gray shaded region is the fundamental domain of the group.

When we consider projective algebra, relations of the presentation Eq. (89) can be modified as

$$\mathsf{L}_a\mathsf{L}_b\mathsf{L}_a^{-1}\mathsf{L}_b^{-1} = \sigma, \tag{91a}$$

$$\mathsf{M}_y\mathsf{L}_a\mathsf{M}_y^{-1} = \eta_1\mathsf{L}_b, \tag{91b}$$

$$\mathsf{M}_x\mathsf{L}_a\mathsf{M}_x^{-1} = \eta_2\mathsf{L}_b^{-1}, \tag{91c}$$

$$\mathsf{M}_y\mathsf{M}_x\mathsf{M}_y^{-1}\mathsf{M}_x^{-1} = \alpha_{xy}, \tag{91d}$$

$$\mathsf{M}_x^2 = \alpha_x, \tag{91e}$$

$$\mathsf{M}_y^2 = \alpha_y. \tag{91f}$$

For $\mathbb{Z}_2$ extension, factor $\eta_1$ or $\eta_2$ can be cancelled by redefining generators, but they cannot be cancelled simultaneously. The factor $\eta = \eta_1\eta_2$ is a cohomology invariant. We have five independent cohomology invariants $\sigma, \eta, \alpha_{xy}, \alpha_x, \alpha_y \in \mathbb{Z}_2$.

For $U(1)$ extension, we have two independent cohomology invariants $\sigma, \alpha_{xy} \in \mathbb{Z}_2$ (the $\mathbb{Z}_2$ value is due to the self-consistency condition). Factors $\eta_1, \eta_2, \alpha_x, \alpha_y$ are trivial because we can cancel them by redefining $\mathsf{L}_a \to \mathsf{L}_a' = (\eta_1\eta_2)^{-1/2}\mathsf{L}_a, \mathsf{L}_b \to \mathsf{L}_b' = (\eta_1\eta_2^{-1})^{1/2}\mathsf{L}_b, \mathsf{M}_x \to \mathsf{M}_x' = \alpha_x^{-1/2}\mathsf{M}_x, \mathsf{M}_y \to \mathsf{M}_y' = \alpha_y^{-1/2}\mathsf{M}_y$. These are consistent with the results of the group cohomology

$$H^2(Cmm, \mathbb{Z}_2) = \mathbb{Z}_2^5, \tag{92a}$$

$$H^2(Cmm, U(1)) = \mathbb{Z}_2^2. \tag{92b}$$

For $\mathbb{Z}_2$ extension, we can also obtain the projective algebra by modifying the relations in Eq. (90) as

$$\mathsf{R}^2 = (\mathsf{M}_x\mathsf{M}_y)^2 = \alpha_r \equiv \alpha_1, \tag{93a}$$

$$(\mathsf{L}_a\mathsf{R})^2 = (\mathsf{L}_a\mathsf{M}_x\mathsf{M}_y)^2 = \eta\alpha_r \equiv \alpha_2, \tag{93b}$$

$$(\mathsf{L}_a\mathsf{L}_b\mathsf{R})^2 = (\mathsf{L}_a\mathsf{M}_x\mathsf{L}_a^{-1}\mathsf{M}_y)^2 = \sigma\alpha_r \equiv \alpha_3, \tag{93c}$$

$$\mathsf{M}_x^2 = \alpha_x \equiv \beta_1, \tag{93d}$$

$$\mathsf{M}_y^2 = \alpha_y \equiv \beta_2, \tag{93e}$$

where

$$\alpha_r \equiv \alpha_x\alpha_y\alpha_{xy}. \tag{94a}$$

These relations mean that each conjugacy class of rotations and reflections has an independent cohomology invariant, as shown in Fig. 12.

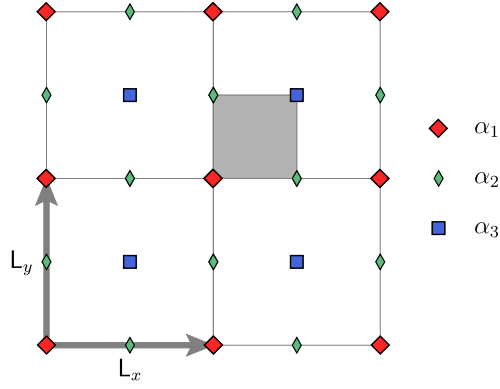

Supplementary Fig.13: Cell structure of $P4$. There are three conjugacy classes of rotation centers. The gray shaded region is the fundamental domain of the group.

### k. P4

The group $P4$ contains a four-fold rotation, which rotates translations $L_x, L_y$ as $RL_xR^{-1} = L_y, RL_yR^{-1} = L_x^{-1}$. The presentation is given by

$$
\begin{aligned}
P4 = \quad &\langle L_x, L_y, R|[L_x, L_y], RL_xR^{-1} = L_y, \\
&RL_yR^{-1} = L_x^{-1}, R^4 \rangle.
\end{aligned}
\tag{95}
$$

There are two conjugacy classes of four-fold rotation $C_s(R), C_s(L_xR)$ and one conjugacy class of two-fold rotation $C_s(L_xR^2)$, as shown in Fig. 13. The presentation can also be given by

$$
P4 = \langle L_x, R|(L_xR)^4, (L_xR^2)^2, R^4 \rangle.
\tag{96}
$$

When we consider projective algebra, the relations of presentation (95) become

$$
\mathsf{L}_x\mathsf{L}_y\mathsf{L}_x^{-1}\mathsf{L}_y^{-1} = \sigma,
\tag{97a}
$$

$$
\mathsf{R}\mathsf{L}_x\mathsf{R}^{-1} = \eta_1\mathsf{L}_y,
\tag{97b}
$$

$$
\mathsf{R}\mathsf{L}_y\mathsf{R}^{-1} = \eta_2\mathsf{L}_x^{-1},
\tag{97c}
$$

$$
\mathsf{R}^4 = \alpha.
\tag{97d}
$$

For $U(1)$ extension, we have only one independent cohomology invariant $\sigma \in U(1)$. $\eta_1, \eta_2, \alpha$ are trivial because we can cancel them by redefining $\mathsf{L}_x \to \mathsf{L}_x' = (\eta_1\eta_2)^{-1/2}\mathsf{L}_x, \mathsf{L}_y \to \mathsf{L}_y' = (\eta_1\eta_2^{-1})^{1/2}\mathsf{L}_y, \mathsf{R} \to \mathsf{R}' = \alpha^{-1/4}\mathsf{R}$.

For $\mathbb{Z}_2$ extension, although $\eta_1$ or $\eta_2$ can be cancelled by redefinition, $\eta = \eta_1\eta_2$ is a cohomology invariant. We have three independent cohomology invariants $\sigma, \eta, \alpha \in \mathbb{Z}_2$.

These results are consistent with that of the group cohomology

$$
H^2(P4, U(1)) = U(1),
\tag{98a}
$$

$$
H^2(P4, \mathbb{Z}_2) = \mathbb{Z}_2^3.
\tag{98b}
$$

For $\mathbb{Z}_2$ extension, we can also obtain the projective algebra by modifying the relations in Eq. (96) as

$$
\mathsf{R}^4 = \alpha \equiv \alpha_1,
\tag{99a}
$$

$$
(\mathsf{L}_x\mathsf{R}^2)^2 = \eta\alpha \equiv \alpha_2,
\tag{99b}
$$

$$
(\mathsf{L}_x\mathsf{R})^4 = \sigma\alpha \equiv \alpha_3.
\tag{99c}
$$

These relations mean that each conjugacy class of rotation has an independent cohomology invariant, as shown in Fig. 13.

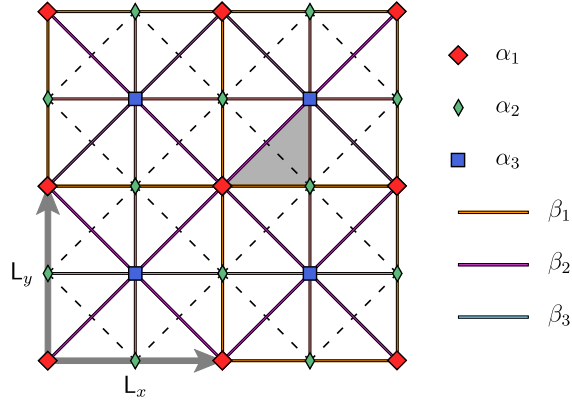

Supplementary Fig.14: Cell structure of $P4m$. There are three conjugacy classes of rotation and three conjugacy classes of reflection. The gray shaded region is the fundamental domain of the group.

### 1. P4m

If we add reflection symmetry to the group $P4$, we obtain group $P4m$. Here we choose the reflection $M$ whose axis is perpendicular to $\bm{e}_x$ as a generator. $M$ reverses translation $L_x$ and rotation $C$. Thus, the presentation is given by

$$
\begin{aligned}
P4m = \quad &\langle L_x, L_y, R, M | [L_x, L_y], RL_xR^{-1} = L_y, \\
&RL_yR^{-1} = L_x^{-1}, ML_xM^{-1} = L_x^{-1}, \\
&MCM^{-1} = R^{-1}, R^4, M^{-1} \rangle.
\end{aligned} \tag{100}
$$

There are two conjugacy classes of four-fold rotation $C_s(R), C_S(L_xR)$, one conjugacy class of two-fold rotation $C_S(L_xR^2)$ and three conjugacy classes of reflection $C_S(M), C_S(RM), C_S(L_xM)$, as shown in Fig. 14. The group can also be presented in terms of rotations and reflections as

$$
\begin{aligned}
P4m = \quad &\langle L_x, R, M | (L_xC)^4, (L_xR^2)^2, R^4, \\
&(L_xM)^2, (RM)^2, M^2 \rangle.
\end{aligned} \tag{101}
$$

When we consider projective algebra, we can modify the relations of generators as

$$
\begin{aligned}
\mathsf{L}_x\mathsf{L}_y\mathsf{L}_x^{-1}\mathsf{L}_y^{-1} &= \sigma, & \text{(102a)} \\
\mathsf{R}\mathsf{L}_x\mathsf{R}^{-1} &= \eta_1\mathsf{L}_y, & \text{(102b)} \\
\mathsf{R}\mathsf{L}_y\mathsf{R}^{-1} &= \eta_2\mathsf{L}_x^{-1}, & \text{(102c)} \\
\mathsf{M}\mathsf{L}_x\mathsf{M}^{-1} &= \eta_m\mathsf{L}_x^{-1}, & \text{(102d)} \\
\mathsf{R}^4 &= \alpha_r, & \text{(102e)} \\
\mathsf{M}^2 &= \alpha_m, & \text{(102f)} \\
\mathsf{M}\mathsf{R}\mathsf{M}^{-1} &= \alpha_{rm}\mathsf{R}^{-1}. & \text{(102g)}
\end{aligned}
$$

Like other examples of group which contains reflection, the self-consistency condition requires $\sigma \in \mathbb{Z}_2$. Furthermore, $\mathsf{R}, \mathsf{M}$ generate a projective algebra of $D_4$. As we analyzed before, factors $\alpha_{rm}, \alpha_r$ must satisfy the relation Eq.(43) $\alpha_{rm}^4\alpha_r^{-2} = 1$, and $\alpha_r$ has two solutions $\alpha_r = \pm\alpha_{rm}^2 \equiv \alpha\alpha_{rm}^2$.

To see the requirement of self-consistency condition on other factors, we conjugate Eq.(102)(b)(c) by $\mathsf{M}$. Using Eq.(102)(d)(g), we have

$$
\begin{aligned}
\eta_m\mathsf{R}^{-1}\mathsf{L}_x^{-1}\mathsf{R} &= \eta_1\mathsf{M}\mathsf{L}_y\mathsf{M}^{-1}, \\
\mathsf{R}^{-1}\mathsf{M}\mathsf{L}_y\mathsf{M}^{-1}\mathsf{R} &= \eta_2\eta_m^{-1}\mathsf{L}_x.
\end{aligned}
$$

Cancel $\mathsf{M}\mathsf{L}_y\mathsf{M}^{-1}$ in the two equations above, we get

$$
\eta_2\eta_m^{-1}\mathsf{L}_x = \eta_m\eta_1^{-1}\mathsf{R}^{-1}(\mathsf{R}^{-1}\mathsf{L}_x^{-1}\mathsf{R})\mathsf{R}.
$$

Because $R^{-1}L_x^{-1}R = \eta_2^{-1}L_y$, $R^{-1}L_yR = \eta_1^{-1}L_x$, the above equation becomes

$$\eta_2\eta_m^{-1}L_x = \eta_m\eta_1^{-2}\eta_2^{-1}L_x.$$

Thus, $\eta_1, \eta_2, \eta_m$ must satisfy

$$(\eta_m^{-1}\eta_1\eta_2)^2 = 1. \tag{103}$$

It has two solutions for $\eta_m$,

$$\eta_m = \pm\eta_1\eta_2 \equiv \eta(\eta_1\eta_2) \tag{104}$$

For $U(1)$ extension, factors $\eta_1, \eta_2, \alpha_m, \alpha_{rm}$ can be trivialized by redefining $L_x \to L_x' = (\eta_1\eta_2)^{-1/2}L_x, L_y \to L_y' = (\eta_1\eta_2^{-1})^{1/2}L_y, R \to R' = \alpha_{rm}^{-1/2}R, M \to M' = \alpha_m^{-1/2}M$. We have three independent cohomology invariants $\sigma, \eta = \eta_1^{-1}\eta_2^{-1}\eta_m, \alpha = \alpha_r\alpha_{rm}^{-2} \in \mathbb{Z}_2$.

For $\mathbb{Z}_2$ extension, we have six independent cohomology invariants $\sigma, \eta_r = \eta_1\eta_2, \alpha_r, \eta_m, \alpha_m, \alpha_{rm} \in \mathbb{Z}_2$.

These results are in agreement with that of the group cohomology

$$H^2(p4m, U(1)) = \mathbb{Z}_2^3, \tag{105a}$$

$$H^2(p4m, \mathbb{Z}_2) = \mathbb{Z}_2^6. \tag{105b}$$

For $\mathbb{Z}_2$ extension, we can also obtain the projective algebra by modifying the relations in Eq. (101) as

$$R^4 = \alpha_r \equiv \alpha_1, \tag{106a}$$

$$(L_xR^2)^2 = \eta_r\alpha_r \equiv \alpha_2, \tag{106b}$$

$$(L_xR)^4 = \sigma\alpha_r \equiv \alpha_3, \tag{106c}$$

$$M^2 = \alpha_m \equiv \beta_1, \tag{106d}$$

$$(RM)^2 = \alpha_{rm}\alpha_m \equiv \beta_2, \tag{106e}$$

$$(L_xM)^2 = \eta_m\alpha_m \equiv \beta_3. \tag{106f}$$

These relations mean that each conjugacy class of rotations and reflections has an independent cohomology invariant, as shown in Fig. 14.

### m. P4g

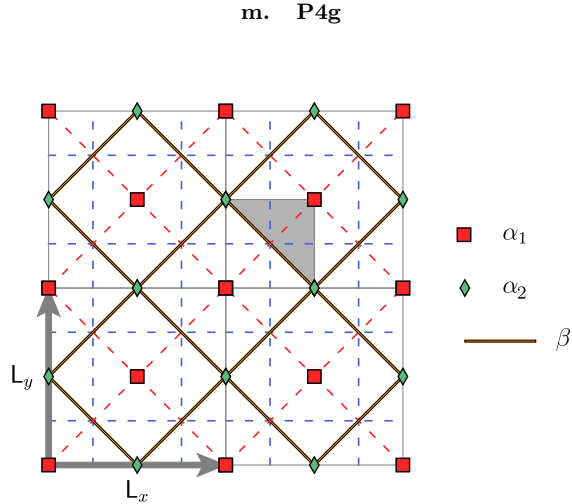

Supplementary Fig.15: Cell structure of $P4g$. There are two conjugacy classes of rotation and one conjugacy class of reflection. The gray shaded region is the fundamental domain of the group.

For group $P4g$, we can take $g_y, R$ as generators. If we choose origin as the rotation center of $R$, then $R, g_y$ act on coordinates as $R(x, y) = (-y, x), g_y(x, y) = (x + 1/2, -y + 1/2)$. $L_xR^2 = g_y^2R^2$ is a two-fold rotation and $(g_y^2R^2)^2 = 1$.

Observe that $g_y R(x, y) = (-y + 1/2, -x + 1/2)$, so $Rg_y$ is a reflection whose reflection axis is in the direction of $\boldsymbol{e}_y - \boldsymbol{e}_x$, thus we have $(g_y R)^2 = 1$.

There is one class of four-fold rotation centers $C_s(C)$, one class of two-fold rotation centers $C_s(R^2 g_y^2)$ and one class of reflections $C_s(g_y C)$, as shown in Fig. 15. The presentation is given by

$$P4g = \langle g_y, R | (g_y R)^2, (g_y^2 R^2)^2, R^4 \rangle. \tag{107}$$

When we consider projective algebra, the relations of presentation Eq. (107) become

$$\mathsf{R}^4 = \alpha_1, \tag{108a}$$

$$(\mathsf{g}_y^2 \mathsf{R}^2)^2 = \alpha_2, \tag{108b}$$

$$(\mathsf{g}_y \mathsf{R})^2 = \beta_1. \tag{108c}$$

To get the restriction of self-consistency condition, we first conjugate $\mathsf{g}_y^2 \mathsf{R}^2$ by $\mathsf{g}_y \mathsf{R}$

$$(\mathsf{g}_y \mathsf{R})(\mathsf{g}_y^2 \mathsf{R}^2)(\mathsf{g}_y \mathsf{R})^{-1} = \beta_1 (\mathsf{g}_y \mathsf{R})^{-1} (\mathsf{g}_y^2 \mathsf{R}^2)(\mathsf{g}_y \mathsf{R})^{-1}$$
$$= \beta_1 (\mathsf{R}^{-1} \mathsf{g}_y^{-1})(\mathsf{g}_y \mathsf{g}_y \mathsf{R} \mathsf{R})(\mathsf{R}^{-1} \mathsf{g}_y^{-1}) = \alpha_3 \mathsf{R}^{-1} (\mathsf{g}_y \mathsf{R}) \mathsf{g}_y^{-1}$$
$$= \beta_1^2 \mathsf{R}^{-1} (\mathsf{g}_y \mathsf{R})^{-1} \mathsf{g}_y^{-1} = \beta_1^2 \mathsf{R}^{-2} \mathsf{g}_y^{-2} = \beta_1^2 (\mathsf{g}_y^2 \mathsf{R}^2)^{-1}.$$

Then we conjugate Eq.(108)(b) by $\mathsf{g}_y \mathsf{R}$,

$$\alpha_2 = (\mathsf{g}_y \mathsf{R})(\mathsf{g}_y^2 \mathsf{R}^2)^2 (\mathsf{g}_y \mathsf{R})^{-1} = \beta_1^4 (\mathsf{g}_y^2 \mathsf{R}^2)^{-2} = \beta_1^4 \alpha_2^{-1}. \tag{109}$$

Thus $\alpha_2, \beta_1$ satisfy

$$\beta_1^4 \alpha_2^{-2} = 1. \tag{110}$$

It has two solutions

$$\alpha_2 = \pm \beta_1 \equiv \alpha \beta_1^2. \tag{111}$$

For $U(1)$ extension, factors $\alpha_1, \beta_1$ are trivial because we can cancel them by redefining $\mathsf{R} \to \mathsf{R}' = \alpha_1^{-1/4} \mathsf{R}$, $\mathsf{g}_y \to \mathsf{g}_y' = \beta_1^{-1/2} \mathsf{g}_y$. There is only one independent cohomology invariant $\alpha = \alpha_2 \beta_1^{-2} \in \mathbb{Z}_2$.

For $\mathbb{Z}_2$ extension, we have three independent cohomology invariants $\alpha_1, \alpha_2, \beta_1 \in \mathbb{Z}_2$, corresponding to conjugacy classes of rotations and reflections respectively, as shown in Fig. 14.

These results are in agreement with that of the group cohomology

$$H^2(P4g, U(1)) = \mathbb{Z}_2, \tag{112a}$$

$$H^2(P4g, \mathbb{Z}_2) = \mathbb{Z}_2^3. \tag{112b}$$

### n. P3

In the group $P3$, the angle between two translation $L_a$ and $L_b$ is $\pi/3$. The rotation operator $R$ rotates translations as $R L_a R^{-1} = L_a^{-1} L_b$, $R L_b R^{-1} = L_a^{-1}$. The presentation is given by

$$P3 = \langle L_a, L_b, R | [L_a, L_b], R L_a R^{-1} = L_a^{-1} L_b, \\ R L_b R^{-1} = L_a^{-1}, R^3 \rangle. \tag{113}$$

When we consider projective algebra, the relations in Eq. (113) are modified as

$$\mathsf{L}_a \mathsf{L}_b \mathsf{L}_a^{-1} \mathsf{L}_b^{-1} = \sigma,$$

$$\mathsf{R} \mathsf{L}_a \mathsf{R}^{-1} = \eta_1 \mathsf{L}_a^{-1} \mathsf{L}_b,$$

$$\mathsf{R} \mathsf{L}_b \mathsf{R}^{-1} = \eta_2 \mathsf{L}_a^{-1},$$

$$\mathsf{R}^3 = \alpha.$$

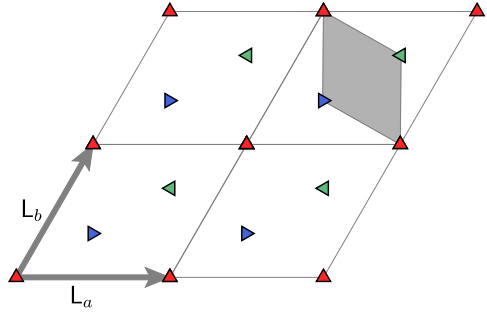

Supplementary Fig.16: Cell structure for $P3$. There are three different conjugacy classes of three-fold rotation centers. The gray shaded region is the fundamental domain of the group.

For $\mathbb{Z}_2$ extension, $\eta_1, \eta_2$ can be trivialized by redefining $\mathsf{L}_a \to \mathsf{L}_a' = (\eta_1\eta_2)\mathsf{L}_a$, $\mathsf{L}_b \to \mathsf{L}_b' = \eta_1\mathsf{L}_b$. For $U(1)$ extension, $\eta_1, \eta_2$ can also be trivialized by redefining $\mathsf{L}_a \to \mathsf{L}_a' = (\eta_1\eta_2)^{-1/3}\mathsf{L}_a$, $\mathsf{L}_b \to \mathsf{L}_b' = (\eta_1^{-1}\eta_2^2)^{-1/3}\mathsf{L}_b$. Because when $n$ is odd, factor system of $C_n$ is always trivial for both $U(1)$ and $\mathbb{Z}_2$ extension, so $\alpha$ can also be trivialized. We only have one independent cohomology invariant $\sigma$ for the translation subgroups. Finally, the projective algebra is

$$\mathsf{L}_a\mathsf{L}_b\mathsf{L}_a^{-1}\mathsf{L}_b^{-1} = \sigma, \tag{114a}$$

$$\mathsf{R}\mathsf{L}_a\mathsf{R}^{-1} = \mathsf{L}_a^{-1}\mathsf{L}_b, \tag{114b}$$

$$\mathsf{R}\mathsf{L}_b\mathsf{R}^{-1} = \mathsf{L}_a^{-1}, \tag{114c}$$

$$\mathsf{R}^3 = 1, \tag{114d}$$

where $\sigma \in \mathbb{Z}_2$ for $\mathbb{Z}_2$ extension and $\sigma \in U(1)$ for $U(1)$ extension. These results are in agreement with the group cohomology

$$H^2(P3, \mathbb{Z}_2) = \mathbb{Z}_2, \tag{115a}$$

$$H^2(P3, U(1)) = U(1). \tag{115b}$$

### o. $P3m1$

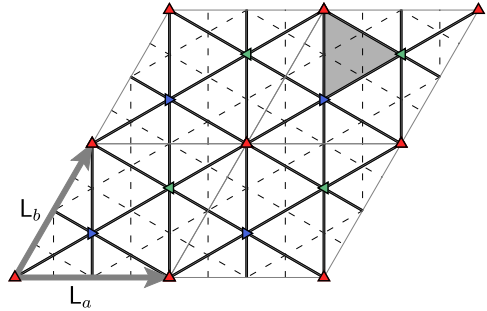

Supplementary Fig.17: Cell structure of $P3m1$. The reflection axes pass through three different rotation centers. The gray shaded region is the fundamental domain of the group.

As shown in Fig. 17, the group $P3m1$ contains reflections whose axes pass through three different rotation centers of $P3$. Here we choose the reflection $M$ whose axis is perpendicular to $\boldsymbol{e}_a$ as one generator. $M$ reverses translation $L_a$ and rotation $R$. The presentation is given by

$$\begin{aligned} P3m1 = \quad & \langle L_a, L_b, R, M | [L_a, L_b], RL_aR^{-1} = L_a^{-1}L_b, \\ & RL_bR^{-1} = L_a^{-1}, ML_aM^{-1} = L_a^{-1}, \\ & MRM^{-1} = R^{-1}, M^2, R^3 \rangle. \end{aligned} \tag{116}$$

When we consider projective algebra, we need to modify the relations in Eq. (116).

For both $\mathbb{Z}_2$ and $U(1)$ extension, we can first trivialize the factor system between rotation and translations as we proved in group $P3$ and trivialize the factor system of point group $D_3$.

We proceed to look at the requirement of self-consistency condition on the factor system between $M$ and $L_a$. Suppose

$$\mathsf{M}\mathsf{L}_a\mathsf{M}^{-1} = \eta\mathsf{L}_a^{-1}.$$

Then we take relations between rotation and translations

$$\mathsf{R}\mathsf{L}_a\mathsf{R}^{-1} = \mathsf{L}_a^{-1}\mathsf{L}_b,$$
$$\mathsf{R}\mathsf{L}_b\mathsf{R}^{-1} = \mathsf{L}_a^{-1},$$

conjugate them by $\mathsf{M}$, and use relations $\mathsf{M}\mathsf{L}_a\mathsf{M}^{-1} = \eta\mathsf{L}_a^{-1}$ and $\mathsf{M}\mathsf{R}\mathsf{M}^{-1} = \mathsf{R}^{-1}$, we get

$$\eta\mathsf{R}^{-1}\mathsf{L}_a^{-1}\mathsf{R} = \eta^{-1}\mathsf{L}_a\mathsf{M}\mathsf{L}_b\mathsf{M}^{-1},$$
$$\mathsf{R}^{-1}\mathsf{M}\mathsf{L}_b\mathsf{M}^{-1}\mathsf{R} = \eta^{-1}\mathsf{L}_a.$$

By cancelling $\mathsf{M}\mathsf{L}_b\mathsf{M}^{-1}$ from above equations, we get

$$\eta\mathsf{R}^{-1}\mathsf{L}_a^{-1}\mathsf{R} = \eta^{-2}\mathsf{L}_a(\mathsf{R}\mathsf{L}_a\mathsf{R}^{-1}).$$

With $\mathsf{R}\mathsf{L}_a\mathsf{R}^{-1} = \mathsf{L}_a^{-1}\mathsf{L}_b$ and $\mathsf{R}^{-1}\mathsf{L}_a^{-1}\mathsf{R} = \mathsf{L}_b$, we have

$$\eta\mathsf{L}_b = \eta^{-2}\mathsf{L}_b.$$

So factor $\eta$ satisfies $\eta^3 = 1$. For $\mathbb{Z}_2$ extension, $\eta = 1$, while for $U(1)$ extension, $\eta = e^{i\theta}, \theta = 0, 2\pi/3, 4\pi/3$. But $\eta$ can be trivialized by redefining $\mathsf{L}_a \to \mathsf{L}_a' = e^{i\xi}\mathsf{L}_a, \mathsf{L}_b \to \mathsf{L}_b' = e^{-i\xi}\mathsf{L}_a, \xi = 0, 4\pi/3, 2\pi/3$ respectively, and the redefining does not influence the original trivialization of factor system between translations and rotation. Hence, there are no nontrivial factor between $\mathsf{M}$ and $\mathsf{L}_a$ for both $\mathbb{Z}_2$ and $U(1)$ extension.

Finally, projective relations of presentation are

$$\mathsf{L}_a\mathsf{L}_b\mathsf{L}_a^{-1}\mathsf{L}_b^{-1} = \sigma, \tag{117a}$$
$$\mathsf{R}\mathsf{L}_a\mathsf{R}^{-1} = \mathsf{L}_a^{-1}\mathsf{L}_b, \tag{117b}$$
$$\mathsf{R}\mathsf{L}_b\mathsf{R}^{-1} = \mathsf{L}_a^{-1}, \tag{117c}$$
$$\mathsf{R}^3 = 1, \tag{117d}$$
$$\mathsf{M}\mathsf{R}\mathsf{M}^{-1} = \mathsf{R}^{-1}, \tag{117e}$$
$$\mathsf{M}\mathsf{L}_a\mathsf{M}^{-1} = \mathsf{L}_a^{-1}, \tag{117f}$$
$$\mathsf{M}^2 = \beta. \tag{117g}$$

For $\mathbb{Z}_2$ extension we have two independent cohomology invariants $\sigma, \beta \in \mathbb{Z}_2$, while for $U(1)$ extension we have only one independent cohomology invariantr $\sigma \in \mathbb{Z}_2$.

Above results are in agreement with the group cohomology

$$H^2(P3m1, \mathbb{Z}_2) = \mathbb{Z}_2^2, \tag{118a}$$
$$H^2(P3m1, U(1)) = \mathbb{Z}_2. \tag{118b}$$

### p. P31m

As shown in Fig. 18, the group $P31m$ contains reflections whose axes pass through only one class of rotation centers. Here we choose reflection $M$ whose axis is parallel to $\boldsymbol{e}_a$ as one generator. $M$ preserves translation $L_a$ and reverses rotation $R$. Thus, The presentation is given by

$$
\begin{aligned}
P31m = \quad & \langle L_a, L_b, R, M | [L_a, L_b], RL_aR^{-1} = L_a^{-1}L_b, \\
& RL_bR^{-1} = L_a^{-1}, ML_aM^{-1} = L_a \\
& MRM = R^{-1}, M^2, R^3 \rangle.
\end{aligned}
\tag{119}
$$

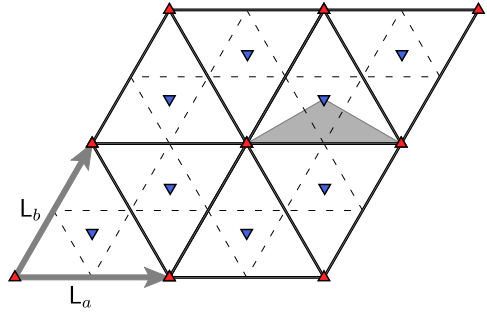

Supplementary Fig.18: Cell structure of $P31m$. Reflection axes pass through only one class of rotation centers. The gray shaded region is the fundamental domain of the group.

When we consider projective algebra, we need to modify the relations in Eq. (119).

For both $\mathbb{Z}_2$ and $U(1)$ extension, we can first trivialize the factor system between rotation and translations as we proved in group $P3$ and trivialize the factor system of point group $D_3$.

We proceed to look at the requirement of self-consistency condition on the factor system between $M$ and $L_a$. Suppose

$$\mathsf{M}\mathsf{L}_a\mathsf{M}^{-1} = \eta\mathsf{L}_a.$$

Then we take relations between rotation and translations

$$\mathsf{R}\mathsf{L}_a\mathsf{R}^{-1} = \mathsf{L}_a^{-1}\mathsf{L}_b,$$
$$\mathsf{R}\mathsf{L}_b\mathsf{R}^{-1} = \mathsf{L}_a^{-1},$$

conjugate them by $\mathsf{M}$, and use relations $\mathsf{M}\mathsf{L}_a\mathsf{M}^{-1} = \eta\mathsf{L}_a$ and $\mathsf{M}\mathsf{R}\mathsf{M}^{-1} = \mathsf{R}^{-1}$, we get

$$\eta\mathsf{R}^{-1}\mathsf{L}_a\mathsf{R} = \eta^{-1}\mathsf{L}_a^{-1}\mathsf{M}\mathsf{L}_b\mathsf{M}^{-1},$$
$$\mathsf{R}^{-1}\mathsf{M}\mathsf{L}_b\mathsf{M}^{-1}\mathsf{R} = \eta^{-1}\mathsf{L}_a^{-1}.$$

By cancelling $\mathsf{M}\mathsf{L}_b\mathsf{M}^{-1}$ from above equations, we get

$$\eta\mathsf{R}^{-1}\mathsf{L}_a\mathsf{R} = \eta^{-2}\mathsf{L}_a^{-1}(\mathsf{R}\mathsf{L}_a^{-1}\mathsf{R}^{-1}).$$

With $\mathsf{R}\mathsf{L}_a^{-1}\mathsf{R}^{-1} = \mathsf{L}_b^{-1}\mathsf{L}_a$ and $\mathsf{R}^{-1}\mathsf{L}_a\mathsf{R} = \mathsf{L}_b^{-1}$, we have

$$\eta\mathsf{L}_b^{-1} = \eta^{-2}\mathsf{L}_a^{-1}\mathsf{L}_b^{-1}\mathsf{L}_a = \eta^{-2}\sigma\mathsf{L}_b^{-1}. \tag{120}$$

Thus we have $\eta^3 = \sigma$. Futhermore, reflection symmetry requires $\eta, \sigma \in \mathbb{Z}_2$, so the only possible solution is $\eta = \sigma$ for both $\mathbb{Z}_2$ and $U(1)$ extension.

Finally, the projective relations of presentation are

$$\mathsf{L}_a\mathsf{L}_b\mathsf{L}_a^{-1}\mathsf{L}_b^{-1} = \sigma, \tag{121a}$$
$$\mathsf{R}\mathsf{L}_a\mathsf{R}^{-1} = \mathsf{L}_a^{-1}\mathsf{L}_b, \tag{121b}$$
$$\mathsf{R}\mathsf{L}_b\mathsf{R}^{-1} = \mathsf{L}_a^{-1}, \tag{121c}$$
$$\mathsf{R}^3 = 1, \tag{121d}$$
$$\mathsf{M}\mathsf{R}\mathsf{M}^{-1} = \mathsf{R}^{-1}, \tag{121e}$$
$$\mathsf{M}\mathsf{L}_a\mathsf{M}^{-1} = \sigma\mathsf{L}_a, \tag{121f}$$
$$\mathsf{M}^2 = \beta. \tag{121g}$$

For $\mathbb{Z}_2$ extension we have two independent cohomology invariants $\sigma, \beta$ $\sigma \in \mathbb{Z}_2$.

These results are in agreement with that of the group cohomology

$$H^2(P31m, \mathbb{Z}_2) = \mathbb{Z}_2^2, \tag{122a}$$
$$H^2(P31m, U(1)) = \mathbb{Z}_2. \tag{122b}$$

## q. P6

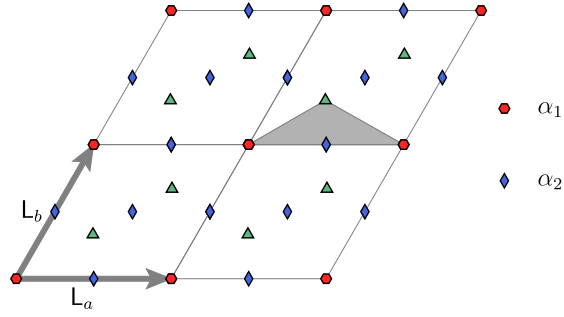

Supplementary Fig.19: Cell structure of $P6$. There is one conjugacy class of two-fold, three-fold and six-fold rotation centers respectively. The gray shaded region is the fundamental domain of the group.

In the group $P6$, the angle between two translation $L_a$ and $L_b$ is $\pi/3$, the rotation operator $R$ rotates translations as $RL_aR^{-1} = L_b, RL_bR^{-1} = L_a^{-1}L_b$. The presentation is given by

$$\begin{aligned}
P6 = \quad &\langle L_a, L_b, R | [L_a, L_b], RL_aR^{-1} = L_b, \\
&RL_bR^{-1} = L_a^{-1}L_b, R^6 \rangle.
\end{aligned} \tag{123}$$

Since $L_b$ can be expressed in terms of $L_a$ and $R$, we can choose $L_a$ and $R$ as independent generators and the presentation can also be given by

$$P6 = \langle L_a, R | (L_aR^2)^3, (L_aR^3)^2, R^6 \rangle. \tag{124}$$

When we consider projective algebra, we need to modify the relations in Eq. (123). For the relations between rotation and translations, we have

$$\begin{aligned}
RL_aR^{-1} &= \eta_1 L_b, \\
RL_bR^{-1} &= \eta_2 L_a^{-1}L_b.
\end{aligned}$$

The factors $\eta_1, \eta_2$ can be trivialized by $L_a \to L_a' = \eta_2^{-1}L_a$, $L_b \to L_b' = (\eta_1^{-1}\eta_2)L_b$. Thus projective relations of presentation are

$$L_aL_bL_a^{-1}L_b^{-1} = \sigma, \tag{125a}$$
$$RL_aR^{-1} = L_b, \tag{125b}$$
$$RL_bR^{-1} = L_a^{-1}L_b, \tag{125c}$$
$$R^6 = \alpha. \tag{125d}$$

For $\mathbb{Z}_2$ extension, we have two independent cohomology invariants $\sigma, \alpha \in \mathbb{Z}_2$, while for $U(1)$ extension we have only one cohomology invariant $\sigma \in U(1)$ since $\alpha$ can be trivialized.

These results are consistent with that of the group cohomology

$$H^2(P6, \mathbb{Z}_2) = U(1), \tag{126a}$$
$$H^2(P6, U(1)) = \mathbb{Z}_2^2. \tag{126b}$$

For $\mathbb{Z}_2$ extension, we can also obtain the projective algebra by modifying the relations in Eq. (124) as

$$(L_aR^2)^3 = 1, \tag{127a}$$
$$R^6 = \alpha \equiv \alpha_1, \tag{127b}$$
$$(L_aR^3)^2 = \sigma\alpha_r \equiv \alpha_2. \tag{127c}$$

These relations mean that the conjugacy class of six-fold rotation centers $C_s(R)$ and two-fold rotation centers $C_s(L_aR^3)$ has an independent cohomology invariant respectively, as shown in Fig. 19.

## r. P6m

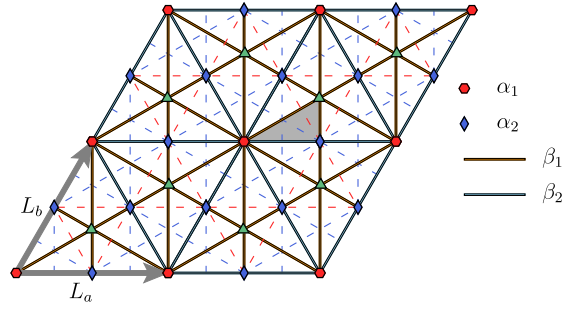

Supplementary Fig.20: Cell structure of $P6m$. There are two conjugacy classes of rotation and two conjugacy classes of reflection. The gray shaded region is the fundamental domain of the group.

The group $P6m$ is obtained by adding reflection symmetry to group $P6$. Here we choose the reflection $M$ whose axis is perpendicular to $\boldsymbol{e}_a$ as a generator. $M$ reverses translation $L_a$ and rotations $R$. The presentation is given by

$$
\begin{aligned}
P6m = \langle L_a, L_b, R, M | [L_a, L_b], R L_a R^{-1} = L_b, \\
R L_b R^{-1} = L_a^{-1} L_b, M L_a M^{-1} = L_a^{-1}, \\
M R M^{-1} = R^{-1}, M^2, R^6 \rangle.
\end{aligned}
\tag{128}
$$

The group can also be presented in terms of rotations and reflections as

$$
\begin{aligned}
P6m = \langle L_a, R, M | (L_a R^2)^3, (L_a R^3)^2, R^6, M^2, \\
(R M)^2, (L_a M)^2 \rangle.
\end{aligned}
\tag{129}
$$

When we consider projective algebra, we need to modify the relations in Eq. (128).

For both $\mathbb{Z}_2$ and $U(1)$ extension, we can first trivialize the factor system between rotation and translations as we proved in group $P6$, and label the factor system of subgroup $D_6$ by cohomology invariants $\alpha_r, \alpha_m, \alpha_{rm}$.

We proceed to look at the requirement of self-consistency condition on the factor system between $M$ and $L_a$. Suppose

$$
\mathsf{M} \mathsf{L}_a \mathsf{M}^{-1} = \eta \mathsf{L}_a^{-1}.
$$

Then we take relations between rotation and translations

$$
\begin{aligned}
\mathsf{R} \mathsf{L}_a \mathsf{R}^{-1} &= \mathsf{L}_b, \\
\mathsf{R} \mathsf{L}_b \mathsf{R}^{-1} &= \mathsf{L}_a^{-1} \mathsf{L}_b,
\end{aligned}
$$

conjugate them by $\mathsf{M}$, and use relations $\mathsf{M} \mathsf{L}_a \mathsf{M}^{-1} = \eta \mathsf{L}_a^{-1}$ and $\mathsf{M} \mathsf{R} \mathsf{M}^{-1} = \mathsf{R}^{-1}$, we get

$$
\begin{aligned}
\eta \mathsf{R}^{-1} \mathsf{L}_a^{-1} \mathsf{R} &= \mathsf{M}_x \mathsf{L}_b \mathsf{M}_x^{-1}, \\
\mathsf{R}^{-1} \mathsf{M}_x \mathsf{L}_b \mathsf{M}_x^{-1} \mathsf{R} &= \eta^{-1} \mathsf{L}_a \mathsf{M}_x \mathsf{L}_b \mathsf{M}_x^{-1}.
\end{aligned}
$$

By cancelling $\mathsf{M} \mathsf{L}_b \mathsf{M}^{-1}$ from above equations, we get

$$
\mathsf{R}^{-1}(\mathsf{R}^{-1} \mathsf{L}_a^{-1} \mathsf{R}) \mathsf{R} = \eta^{-1} \mathsf{L}_a (\mathsf{R}^{-1} \mathsf{L}_a^{-1} \mathsf{R}).
$$

Conjugate above equation with $\mathsf{R}^2$, we have

$$
\mathsf{L}_a^{-1} = \eta^{-1}(\mathsf{R}^2 \mathsf{L}_a \mathsf{R}^{-2})(\mathsf{R} \mathsf{L}_a^{-1} \mathsf{R}^{-1}).
$$

Use $\mathsf{R}^2 \mathsf{L}_a \mathsf{R}^{-2} = \mathsf{L}_a^{-1} \mathsf{L}_b$ and $\mathsf{R} \mathsf{L}_a^{-1} \mathsf{R}^{-1} = \mathsf{L}_b^{-1} \mathsf{L}_b$, we obtain

$$
\mathsf{L}_a^{-1} = \eta^{-1} \mathsf{L}_a^{-1}.
$$

So $\eta = 1$, i,e., $\eta$ is trivial.

Thus, the projective relations of presentation of $P6m$ are

$$L_a L_b L_a^{-1} L_b^{-1} = \sigma, \tag{130a}$$

$$RL_b R^{-1} = L_a^{-1} L_b, \tag{130b}$$

$$RL_a R^{-1} = L_b, \tag{130c}$$

$$ML_a M^{-1} = L_a^{-1}, \tag{130d}$$

$$R^6 = \alpha_r, \tag{130e}$$

$$MRM^{-1} = \alpha_{rm} R^{-1}, \tag{130f}$$

$$M^2 = \alpha_m. \tag{130g}$$

For $\mathbb{Z}_2$ extension, we have four independent cohomology invariants $\sigma, \alpha_r, \alpha_{rm}, \alpha_m \in \mathbb{Z}_2$, while for $U(1)$ extension we have two independent cohomology invariants $\sigma, \alpha = \alpha_r \alpha_{rm}^{-3} \in \mathbb{Z}_2$, since $\alpha_{rm}, \alpha_m$ are trivial as we analyzed in group $D_n$.

These results are in agreement with that of the group cohomology

$$H^2(P6m, \mathbb{Z}_2) = \mathbb{Z}_2^4, \tag{131a}$$

$$H^2(P6m, U(1)) = \mathbb{Z}_2^2. \tag{131b}$$

For $\mathbb{Z}_2$ extension, we can also obtain the projective algebra by modifying the relations in Eq. (129) as

$$(L_a R^2)^3 = 1, \tag{132a}$$

$$R^6 = \alpha_r \equiv \alpha_1, \tag{132b}$$

$$(L_a R^3)^2 = \sigma \alpha_r \equiv \alpha_2, \tag{132c}$$

$$M^2 = \alpha_m \equiv \beta_1, \tag{132d}$$

$$(L_a M)^2 = \alpha_m = \beta_1, \tag{132e}$$

$$(RM)^2 = \alpha_{rm} \alpha_m \equiv \beta_2. \tag{132f}$$

The meaning of these relations are illustrated in Fig. 20.

### s. Table of relations and cohomology invariants

| $G$ | $H^2(G, \mathbb{Z}_2)$ | Generators | Cohomology invariants | $N_G$ |
|---|---|---|---|---|
| $P1$ | $\mathbb{Z}_2$ | $\mathsf{L}_a, \mathsf{L}_b$ | $\sigma = \mathsf{L}_x \mathsf{L}_y \mathsf{L}_x^{-1} \mathsf{L}_y^{-1}$. | 2 |
| $P2$ | $\mathbb{Z}_2^4$ | $\mathsf{L}_a, \mathsf{L}_b, \mathsf{R}$ | $\alpha_1 = \mathsf{R}^2, \quad \alpha_2 = (\mathsf{L}_a \mathsf{R})^2, \quad \alpha_3 = (\mathsf{L}_b \mathsf{R})^2, \quad \alpha_4 = (\mathsf{L}_a \mathsf{L}_b \mathsf{R})^2$. | 5 |
| $Pm$ | $\mathbb{Z}_2^4$ | $\mathsf{L}_x, \mathsf{L}_y, \mathsf{M}_x$ | $\beta_1 = \mathsf{M}_x^2, \quad \beta_2 = (\mathsf{L}_x \mathsf{M}_x)^2, \quad \eta_1 = \mathsf{M}_x \mathsf{L}_y \mathsf{M}_x^{-1} \mathsf{L}_y^{-1}, \quad \eta_2 = (\mathsf{L}_x \mathsf{M}_x) \mathsf{L}_y (\mathsf{L}_x \mathsf{M}_x)^{-1} \mathsf{L}_y^{-1}$. | 10 |
| $Pg$ | $\mathbb{Z}_2$ | $\mathsf{L}_x, \mathsf{g}_x$ | $\tau = \mathsf{g}_x \mathsf{L}_x \mathsf{g}_x^{-1} \mathsf{L}_x$. | 2 |
| $Cm$ | $\mathbb{Z}_2^2$ | $\mathsf{L}_a, \mathsf{L}_b, \mathsf{M}$ | $\sigma = \mathsf{L}_a \mathsf{L}_b \mathsf{L}_a^{-1} \mathsf{L}_b^{-1}, \quad \beta = \mathsf{M}^2, \quad 1 = \mathsf{M} \mathsf{L}_a \mathsf{M}^{-1} \mathsf{L}_b^{-1}$. | 4 |
| $Pmm$ | $\mathbb{Z}_2^8$ | $\mathsf{L}_x, \mathsf{L}_y, \mathsf{M}_x, \mathsf{M}_y$ | $\alpha_1 = (\mathsf{M}_x \mathsf{M}_y)^2, \quad \alpha_2 = (\mathsf{L}_x \mathsf{M}_x \mathsf{M}_y)^2, \quad \alpha_3 = (\mathsf{L}_y \mathsf{M}_x \mathsf{M}_y)^2, \quad \alpha_4 = (\mathsf{L}_x \mathsf{L}_y \mathsf{M}_x \mathsf{M}_y)^2,$ $\beta_1 = \mathsf{M}_x^2, \quad \beta_2 = (\mathsf{L}_x \mathsf{M}_x)^2, \quad \beta_3 = \mathsf{M}_y^2, \quad \beta_4 = (\mathsf{L}_y \mathsf{M}_y)^2$. | 51 |
| $Pmg$ | $\mathbb{Z}_2^4$ | $\mathsf{g}_y, \mathsf{L}_y, \mathsf{M}_x$ | $\alpha_1 = (\mathsf{M}_x \mathsf{g}_y)^2, \quad \alpha_2 = (\mathsf{L}_y \mathsf{M}_x \mathsf{g}_y)^2, \quad \beta_1 = \mathsf{M}_x^2, \quad \eta = \mathsf{M}_x \mathsf{L}_y \mathsf{M}_x^{-1} \mathsf{L}_y^{-1}$. | 12 |
| $Pgg$ | $\mathbb{Z}_2^2$ | $\mathsf{g}_x, \mathsf{g}_y$ | $\alpha_1 = (\mathsf{g}_x \mathsf{g}_y)^2, \quad \alpha_2 = (\mathsf{g}_x \mathsf{g}_y^{-1})^2$. | 3 |
| $Cmm$ | $\mathbb{Z}_2^5$ | $\mathsf{L}_a, \mathsf{M}_x, \mathsf{M}_y$ | $\alpha_1 = (\mathsf{M}_x \mathsf{M}_y)^2, \quad \alpha_2 = (\mathsf{L}_a \mathsf{M}_x \mathsf{M}_y)^2, \quad \alpha_3 = (\mathsf{L}_a \mathsf{M}_x \mathsf{L}_a^{-1} \mathsf{M}_y)^2, \quad \beta_1 = \mathsf{M}_x^2, \quad \beta_2 = \mathsf{M}_y^2$. | 18 |
| $P4$ | $\mathbb{Z}_2^3$ | $\mathsf{L}_x, \mathsf{R}$ | $\alpha_1 = \mathsf{R}^4, \quad \alpha_2 = (\mathsf{L}_x \mathsf{R}^2)^2, \quad \alpha_3 = (\mathsf{L}_x \mathsf{R})^4$. | 6 |
| $P4m$ | $\mathbb{Z}_2^6$ | $\mathsf{L}_x, \mathsf{R}, \mathsf{M}$ | $\alpha_1 = \mathsf{R}^4, \quad \alpha_2 = (\mathsf{L}_x \mathsf{R}^2)^2, \quad \alpha_3 = (\mathsf{L}_x \mathsf{R})^4, \quad \beta_1 = \mathsf{M}^2, \quad \beta_2 = (\mathsf{R}\mathsf{M})^2, \quad \beta_3 = (\mathsf{L}_x \mathsf{M})^2$. | 40 |
| $P4g$ | $\mathbb{Z}_2^3$ | $\mathsf{g}_y, \mathsf{R}$ | $\alpha_1 = \mathsf{R}^4, \quad \alpha_2 = (\mathsf{g}_y^2 \mathsf{R}^2)^2, \quad \beta_1 = (\mathsf{g}_y \mathsf{R})^2$. | 6 |
| $P3$ | $\mathbb{Z}_2$ | $\mathsf{L}_a, \mathsf{L}_b, \mathsf{R}$ | $\sigma = \mathsf{L}_a \mathsf{L}_b \mathsf{L}_a^{-1} \mathsf{L}_b^{-1}, \quad 1 = \mathsf{R}\mathsf{L}_a \mathsf{R}^{-1} \mathsf{L}_b^{-1} \mathsf{L}_a, \quad 1 = \mathsf{R}\mathsf{L}_b \mathsf{R}^{-1} \mathsf{L}_a, \quad 1 = \mathsf{R}^3$. | 2 |
| $P3m1$ | $\mathbb{Z}_2^2$ | $\mathsf{L}_a, \mathsf{L}_b, \mathsf{R}, \mathsf{M}$ | $\sigma = \mathsf{L}_a \mathsf{L}_b \mathsf{L}_a^{-1} \mathsf{L}_b^{-1}, \quad \beta = \mathsf{M}^2 = (\mathsf{R}\mathsf{M})^2 = (\mathsf{L}_a \mathsf{M})^2, \quad 1 = \mathsf{R}\mathsf{L}_a \mathsf{R}^{-1} \mathsf{L}_b^{-1} \mathsf{L}_a,$ $1 = \mathsf{R}\mathsf{L}_b \mathsf{R}^{-1} \mathsf{L}_a, \quad 1 = \mathsf{R}^3$. | 4 |
| $P31m$ | $\mathbb{Z}_2^2$ | $\mathsf{L}_a, \mathsf{L}_b, \mathsf{R}, \mathsf{M}$ | $\sigma = \mathsf{L}_a \mathsf{L}_b \mathsf{L}_a^{-1} \mathsf{L}_b^{-1} = \mathsf{L}_a \mathsf{M} \mathsf{L}_a^{-1} \mathsf{M}^{-1}, \quad \beta = \mathsf{M}^2 = (\mathsf{R}\mathsf{M})^2, \quad 1 = \mathsf{R}\mathsf{L}_a \mathsf{R}^{-1} \mathsf{L}_b^{-1} \mathsf{L}_a,$ $1 = \mathsf{R}\mathsf{L}_b \mathsf{R}^{-1} \mathsf{L}_a, \quad 1 = \mathsf{R}^3$. | 4 |
| $P6$ | $\mathbb{Z}_2^2$ | $\mathsf{L}_a, \mathsf{R}$ | $\alpha_1 = \mathsf{R}^6, \quad \alpha_2 = (\mathsf{L}_a \mathsf{R}^3)^2, \quad 1 = (\mathsf{L}_a \mathsf{R}^2)^3$. | 4 |
| $P6m$ | $\mathbb{Z}_2^4$ | $\mathsf{L}_a, \mathsf{R}, \mathsf{M}$ | $\alpha_1 = \mathsf{R}^6, \quad \alpha_2 = (\mathsf{L}_a \mathsf{R}^3)^2, \quad \beta_1 = \mathsf{M}^2 = (\mathsf{L}_a \mathsf{M})^2, \quad \beta_2 = (\mathsf{R}\mathsf{M})^2, \quad 1 = (\mathsf{L}_a \mathsf{R}^2)^3$. | 16 |

Supplementary Table.I: $\mathbb{Z}_2$ cohomology invariants for all wallpaper groups.

## Supplementary Note 3. The construction of the canonical models from the cohomology invariants

A complete set of the cohomology invariants for all 17 wallpaper groups have been constructed in the previous section. In this section, we further present general procedures to translate these cohomology invariants into lattice models with appropriate flux configurations. Following the general procedure for model construction, we can in the end arrive at 17 canonical lattice models, i.e., for each wallpaper group we construct a canonical model that can realize all possible values of cohomology invariants and therefore all cohomology classes of multipliers.

### a. Flux interpretation of cohomology invariants

In this subsection, we provide technical details for how each type of cohomology invariant can be interpreted as certain flux configurations on the appropriate lattice structure.

#### i. Crystal Symmetries with Gauge Fields

First of all, we present a formalism of crystal symmetries on a lattice with gauge fields.

In gauge field theory, gauge flux configurations are gauge invariant. A given gauge flux configuration can be described by many equivalent gauge connection configurations, which differ from each other by gauge transformations. In lattice systems, a gauge transformation $\mathsf{G}$ simply multiplies a phase on each lattice site, and therefore can be regarded as a diagonal matrix indexed by the lattice sites.

Now, suppose we have a lattice with the gauge field. A spatial symmetry transformation $R$ that preserves the lattice and flux configuration in general changes the connection configuration. The transformed connection configuration must be related to the original one by a gauge transformation $\mathsf{G}_R$, since they correspond to the same flux configuration.

Thus, the symmetry operator should be modified as

$$\mathsf{R} = \mathsf{G}_R R, \tag{133}$$

which is a combination of the pure spatial operation $R$ and the gauge transformation $\mathsf{G}_R$.

We now consider how the physical operator $\mathsf{R}$ acts on a tight-binding model, $H = \sum_{ij} t_{ij} a_i^\dagger a_j$. The action of $\mathsf{R}$ on $H$ is given by

$$
\begin{aligned}
H' &= \sum_{ij} t_{ij} \mathsf{G}_R(R(i)) \mathsf{G}_R^*(R(j)) a_{R(i)}^\dagger a_{R(j)} \\
&= \sum_{ij} \mathsf{G}_R(i) \mathsf{G}_R^*(j) t_{R^{-1}(i)R^{-1}(j)} a_i^\dagger a_j,
\end{aligned}
\tag{134}
$$

where $G_R(i)$ is the phase of gauge tansformation on site $i$, and $R(i)$ is the site transformed from $i$ by $R$. Thus, the invariance under $\mathsf{R}$ means

$$t_{ij} = \mathsf{G}_R(i) \mathsf{G}_R^*(j) t_{R^{-1}(i)R^{-1}(j)}. \tag{135}$$

### ii. The cohomology invariant of the translation subgroups

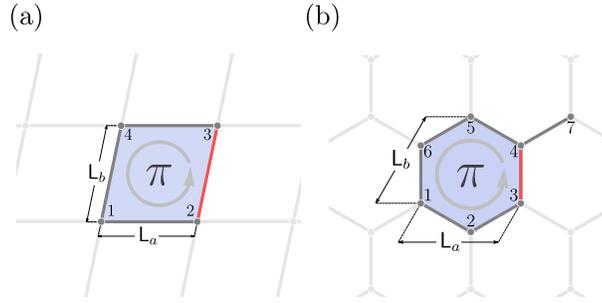

Supplementary Fig.21: The Flux determines the cohomology invariant of translation subgroup. (a) A rectangle plaquette spanned by $L_x$ and $L_y$ in two directions. (b) A hexagon plaquette spanned by $L_a$ and $L_b$ in two directions.

The factor system of the translation subgroup is determined by the cohomology invariant $\sigma = \mathsf{L}_a \mathsf{L}_b \mathsf{L}_a^{-1} \mathsf{L}_b^{-1}$, from which we can derive the relation between gauge transformations:

$$
\begin{aligned}
\mathsf{L}_a \mathsf{L}_b \mathsf{L}_a^{-1} \mathsf{L}_b^{-1} &= \mathsf{G}_a L_a \mathsf{G}_b L_b (\mathsf{G}_a L_a)^{-1} (\mathsf{G}_b L_b)^{-1} \\
&= \mathsf{G}_a(\boldsymbol{r}) (L_a \mathsf{G}_b(\boldsymbol{r}) L_a^{-1}) L_a L_b L_a^{-1} L_b^{-1} (L_b \mathsf{G}_a^*(\boldsymbol{r}) L_b^{-1}) \mathsf{G}_b^*(\boldsymbol{r}) \\
&= \mathsf{G}_a(\boldsymbol{r}) \mathsf{G}_b(L_a^{-1}(\boldsymbol{r})) \mathsf{G}_a^*(L_b^{-1}(\boldsymbol{r})) \mathsf{G}_b^*(\boldsymbol{r}) = \sigma,
\end{aligned}
\tag{136}
$$

where the conjugacy relation between a spatial operator and gauge transformation $L_a G(r) L_a^{-1} = G(L_a^{-1}(r))$ can be derived from its action on a spatial state $|r_0\rangle$,

$$
\begin{aligned}
(L_a G(r) L_a^{-1}) |r_0\rangle &= L_a G(r) |L_a^{-1} r_0\rangle \\
&= L_a G(L_a^{-1} r_0) |L_a^{-1} r_0\rangle = G(L_a^{-1} r_0) |r_0\rangle.
\end{aligned}
\tag{137}
$$

Consider a rectangle plaquette spanned by $L_x, L_y$, as shown in Fig. 21(a). In the presence of gauge field, the modified translation symmetry requires the hoppings satisfy

$$t_{23} = t_{14} \mathsf{G}_a(2) \mathsf{G}_a^*(3), \quad t_{43} = t_{12} \mathsf{G}_b(4) \mathsf{G}_b^*(3).$$

Thus the phases of hoppings $e^{\phi_{ij}} = t_{ij}/|t_{ij}|$ satisfy

$$e^{i\phi_{23}} = e^{i\phi_{14}} \mathsf{G}_a(2) \mathsf{G}_a^*(3), \quad e^{i\phi_{43}} = e^{i\phi_{12}} \mathsf{G}_b(4) \mathsf{G}_b^*(3),$$

The flux surrounding the rectangle satisfies

$$
\begin{aligned}
e^{-i\Phi} &= e^{i\phi_{1\to2\to3\to4\to1}} = e^{i\phi_{12}}e^{i\phi_{23}}e^{i\phi_{34}}e^{i\phi_{41}} \\
&= e^{i\phi_{12}}e^{i\phi_{14}}e^{i\phi_{21}}e^{i\phi_{41}}\mathsf{G}_a(2)\mathsf{G}_b^*(3)\mathsf{G}_b^*(4)\mathsf{G}_b(3) \\
&= \mathsf{G}_a^*(3)\mathsf{G}_b^*(L_a^{-1}(3))\mathsf{G}_a(L_b^{-1}(3))\mathsf{G}_b(3) = \sigma^*.
\end{aligned}
$$

(138)

where the minus sign in exponent $e^{-i\Phi}$ comes from the convention of Peierls substitution under gauge fields $t_{ij} \to t_{ij}e^{-i\int_i^j \mathbf{A}d\mathbf{l}}$.

So when the flux in the plaquette form by $L_a$ and $L_b$ is $\Phi$, the relation between $L_a$ and $L_b$ is modified to $\mathsf{L}_a\mathsf{L}_b\mathsf{L}_a^{-1}\mathsf{L}_b^{-1} = e^{i\Phi}$. This can be understood by the the Ahanorov-Bohm effect, that circling a region with flux $\Phi$ causes an additional phase $e^{i\Phi}$. For $\mathbb{Z}_2$ case, the nontivial $\sigma = -1$ corresponds to $\Phi = \pi$, as shown in Fig. 21.(a), where we choose a gauge that gray bonds have positive hopping amplitude and the red bonds have negative hopping amplitude.

The same argument also works for hexagon lattice as in Fig. 21.(b). In the presence of gauge field, the phases of hoppings satisfy

$$
\begin{aligned}
e^{i\phi_{3\to4\to7}} &= e^{i\phi_{1\to6\to5}}\mathsf{G}_x(3)\mathsf{G}_x^*(7), \\
e^{i\phi_{5\to4\to7}} &= e^{i\phi_{1\to2\to3}}\mathsf{G}_b(5)\mathsf{G}_b^*(7).
\end{aligned}
$$

Thus, the flux surrounding the hexagon satisfies

$$
\begin{aligned}
e^{-i\Phi} &= e^{i\phi_{1\to2\to3\to4\to5\to6\to1}} = \\
&= e^{i\phi_{1\to2\to3}}e^{i\phi_{3\to4\to7}}e^{i\phi_{7\to4\to5}}e^{i\phi_{5\to6\to1}} \\
&= \mathsf{G}_a(3)\mathsf{G}_a^*(7)\mathsf{G}_b^*(5)\mathsf{G}_b(7) \\
&= \mathsf{G}_a^*(7)\mathsf{G}_b^*(L_a^{-1}(7))\mathsf{G}_a(L_b^{-1}(7))\mathsf{G}_b(7) = \sigma^*.
\end{aligned}
$$

(139)

We obtain $e^{i\Phi} = \sigma$ again. This argument can be generalized to arbitrary lattice, where $\Phi$ is the flux in the foundamental domain of the translation subgroup. For some examples, see Fig. 28(a), Fig. 37(a), and Fig. 38(a).

### iii. Cohomology invariants of point groups

For the factor system of point groups, we consider cohomology invariants $\alpha_r = \mathsf{R}^n$ and $\alpha_m = M^2$. As we have shown in Sec. Supplementary Note 2 a, $\alpha_r$ and $\alpha_m$ are trivial for $U(1)$ extension, so here we only discuss $\mathbb{Z}_2$ extension.

To have a nontrivial cohomology invariant $\alpha_r$, $n$ must be even. When $n$ is even, rotating $n/2$ times is a two-fold rotation $R^{n/2} = R_\pi$. After extension, this relation in general gains an additional factor $\xi$, i.e., $\mathsf{R}_{2\pi/n}^{n/2} = \xi\mathsf{R}_\pi$, but it has no influence on the cohomology invariant of $\mathsf{R}_\pi^2$:

$$
\mathsf{R}^n = (\mathsf{R}^{n/2})^2 = (\xi)^2(\mathsf{R}_\pi)^2 = \mathsf{R}_\pi^2 = \alpha_r.
$$

(140)

The equation satisfied by gauge transformation of $R_\pi$ can be obtained by

$$
\begin{aligned}
\mathsf{R}_\pi^2 &= \mathsf{G}_{r_\pi}R_\pi\mathsf{G}_{r_\pi}R_\pi = \mathsf{G}_{r_\pi}R_\pi\mathsf{G}_{r_\pi}R_\pi^{-1}R_\pi^2 \\
&= \mathsf{G}_{r_\pi}(\mathbf{r})\mathsf{G}_{r_\pi}(R_\pi(\mathbf{r})) = \alpha_r.
\end{aligned}
$$

(141)

Consider a circuit invariant under $C_n$ ($n$ is even), and the circuit contains no fixed point and fixed link under $C_n$. We label orbits in the circuit with $i = 1, 2, 3, \cdots 2l$, then $R_\pi(i) = i + l$. The hopping amplitudes satisfy

$$
t_{i,i+1} = t_{i+l,i+l+1}\mathsf{G}_{r_\pi}(i)\mathsf{G}_{r_\pi}^*(i+1).
$$

Thus, the flux surrounding the circuit satisfies

$$
\begin{aligned}
e^{-i\Phi} &= \prod_{i=1,\bmod\ 2l}^{2l} e^{i\phi_{i,i+1}} \\
&= \prod_{i=1,\bmod\ 2l}^{l}\mathsf{G}_{r_\pi}(i)\mathsf{G}_{r_\pi}^*(i+1) \\
&= \mathsf{G}_{r_\pi}(1)\mathsf{G}_{r_\pi}^*(l+1) = \alpha_r.
\end{aligned}
$$

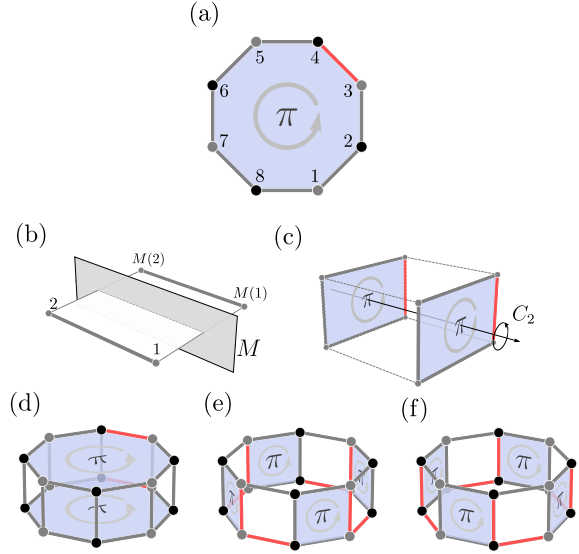

Supplementary Fig.22: Relation of fluxes and cohomology invariants of point group $C_n$ and $D_n$ . (a) The flux determines projective $C_4$ symmetry. When there is $\pi$-flux in the loop, $\mathsf{R}^4 = -1$. (b) Forbidden hoppings for $\mathsf{M}^2 = -1$. (c) Replace reflection with $C_2$ rotation around a horizental axis. (d)-(f) Flux determines projective $D_4$ symmetry. (d) Flux corresponding to the cohomology invariant of rotation. (e)(f) Flux corresponding to cohomology invariants of the two conjugacy classes of reflections ($C_2$ rotations around horizen axes).

We have the relation $e^{i\Phi} = \alpha_r$.

Now we consider $\mathsf{M}^2 = \alpha_m$, the equation satisfied by the gauge transformation of $M$ can be obtained by

$$\mathsf{M}^2 = \mathsf{G}_m M \mathsf{G}_m M = \mathsf{G}_m M \mathsf{G}_m M^{-1} M^2$$
$$= \mathsf{G}_m(\boldsymbol{r}) \mathsf{G}_m(M(\boldsymbol{r})) = \alpha_m. \tag{142}$$

However, things become more subtle now. Consider a hopping from one point $i$ to its reflection image $M(i)$, then reflection symmetry requires the hopping satisfies

$$t_{M(i),i} = t_{i,M(i)} \mathsf{G}(M(i)) \mathsf{G}(i)^* = \alpha_m t_{i,M(i)}. \tag{143}$$

For the nontrivial cohomology invariant $\alpha_m = -1$, we have $t_{i,M(i)} = -t_{M(i),i}$. If we require hoppings to be real, this condition cannot be satisfied, so this kind of hoppings (as in Fig. 22.(b)) are forbidden for $\mathsf{M}^2 = -1$ case. With this reason, the case $\mathsf{M}^2 = -1$ fails to be realized by nearest real hopping models.

In two dimensions, we can replace the reflection with a $C_2$ rotation around a horizental axis, as Fig. 22.(c) shows. Then the rotaion $\mathsf{R}_\pi^2 = -1$ can be realized by nearest hopping models.

For example, to realize all the possible projective algebras of $D_4$, we need to realize the cohomology invariants of rotations and the two conjugacy classes of reflections according to Eq. (49). We can take a bilayer version of model $C_4$ as illustrated in Fig. 22.(d)-(f).

### iv. Cohomology invariants between translation and reflection

Now we proceed to look at the factor system between translation and reflection. The relation $\mathsf{M}_x \mathsf{L}_y \mathsf{M}_x^{-1} = \eta \mathsf{L}_y$ implies the gauge transforms satisfy

$$\mathsf{M}_x \mathsf{L}_y \mathsf{M}_x^{-1} \mathsf{L}_y^{-1} = \mathsf{G}_m M_x \mathsf{G}_y L_y (\mathsf{G}_m M_x)^{-1} (\mathsf{G}_y L_y)^{-1}$$
$$= \mathsf{G}_m(M_x \mathsf{G}_y M_x^{-1}) M_x L_y M_x^{-1} L_y^{-1} (L_y \mathsf{G}_m L_y^{-1}) \mathsf{G}_y^{-1}$$
$$= \mathsf{G}_m(\boldsymbol{r}) \mathsf{G}_y(M_x(\boldsymbol{r})) \mathsf{G}_m^*(L_y^{-1}(\boldsymbol{r})) \mathsf{G}_y^*(\boldsymbol{r}) = \eta. \tag{144}$$

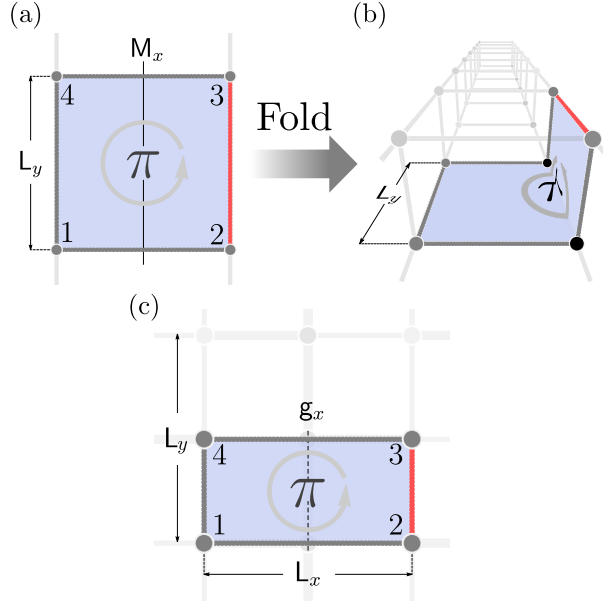

Supplementary Fig.23: The relation of fluxes and the cohomology invariants between translation and (glide-)reflection. (a) A rectangle plaquette spanned by $M_x$ and $L_y$ in two directions. (b) The rectangle is folded when we replace the reflection with a rotaion around a horizental axis. (c) A rectangle plaquette spanned by $L_x$ and $g_x$.

Consider a rectangle plaquette spanned by $M_x$ and $L_y$ as illustrated in Fig. 23(a). $M_x$ and $L_y$ symmetry requires hoppings satisfy

$$t_{23} = t_{14}G_m(2)G_m^*(3), \quad t_{43} = t_{12}G_y(4)G_y^*(3).$$

Thus, the flux surrounding the rectangle satisfies

$$
\begin{aligned}
e^{-i\Phi} &= e^{i\phi_{1\to2\to3\to4\to1}} = e^{i\phi_{12}}e^{i\phi_{23}}e^{i\phi_{34}}e^{i\phi_{41}} \\
&= e^{i\phi_{12}}e^{i\phi_{14}}e^{i\phi_{21}}e^{i\phi_{41}}G_m(2)G_m^*(3)G_y^*(4)G_y(3) \\
&= G_m^*(3)G_y^*(M(3))G_m(L_y^{-1}(3))G_y(3) = \eta^*.
\end{aligned}
\tag{145}
$$

The plaquette in Fig. 23(a) can also take a varied form. In particular, when we replace the reflection effectively by a rotation around a horizental axis, the plaquette will be folded into a three-dimensional plaquette such as that in Fig. 23(b). This also occurs in Fig. 26(b) and Fig. 30(c) in lattice models for example.

### v. Cohomology invariants between translation and glide-reflection

Finally, we look at the cohomology invariant between glide-reflection and translation. The relation $g_xL_xg_x^{-1} = \eta L_x$ implies

$$
\begin{aligned}
L_xg_xL_xg_x^{-1} &= G_xL_xG_yg_x(G_xL_x)(G_yg_x)^{-1} \\
&= G_x(L_xG_yL_x^{-1})L_xg_xL_xg_x^{-1}((L_xg_x)^{-1}G_xL_xg_x^{-1})G_y^{-1} \\
&= G_x(\boldsymbol{r})G_{g_x}(L_x(\boldsymbol{r}))G_x(L_xg_x^{-1}(\boldsymbol{r}))G_{g_x}^*(\boldsymbol{r}) = \tau.
\end{aligned}
\tag{146}
$$

Consider a rectangle plaquette spanned by $L_x$ and $g_x$ as illustrated in Fig. 23(b). The hoppings satisfy

$$t_{23} = t_{14}G_x(2)G_x^*(3), \quad t_{43} = t_{21}G_{g_x}(4)G_{g_x}^*(3).$$

Thus, the flux surrounding the rectangle satisfies

$$
\begin{aligned}
e^{-i\Phi} &= e^{i\phi_{1\to2\to3\to4\to1}} \\
&= e^{i\phi_{12}}e^{i\phi_{23}}e^{i\phi_{34}}e^{i\phi_{41}} \\
&= e^{i\phi_{12}}e^{i\phi_{14}}e^{i\phi_{12}}e^{i\phi_{41}}\mathsf{G}_x(2)\mathsf{G}_x^*(3)\mathsf{G}_{g_x}^*(4)\mathsf{G}_{g_x}(3) \\
&= e^{i2\phi_{12}}\mathsf{G}_x^*(3)\mathsf{G}_{g_x}^*(L_x(3))\mathsf{G}_x(L_x g_x^{-1}(3))\mathsf{G}_{g_x}(3) \\
&= \mathsf{G}_x^*(3)\mathsf{G}_{g_x}^*(L_x(3))\mathsf{G}_x^*(L_x g_x^{-1}(3))\mathsf{G}_{g_x}(3) = \tau^*.
\end{aligned}
\tag{147}
$$

where we use the constrain of $\mathbb{Z}_2$ gauge fields, $e^{i\phi_{12}}$, $\mathsf{G}_{g_x}(\boldsymbol{r}) \in \{\pm 1\}$.

In the following, we will construct lattice models to realize all the $\mathbb{Z}_2$ projective algebras for the 17 wallpaper groups.

### b. P1

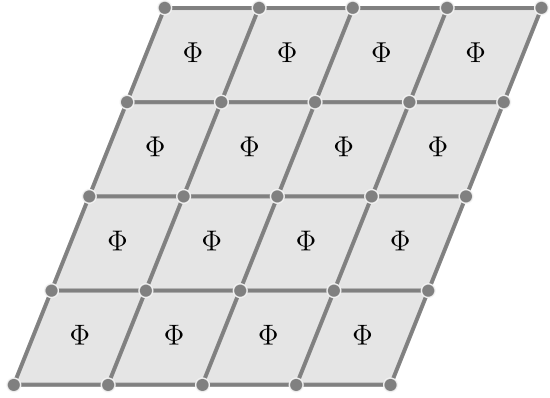

Supplementary Fig.24: Lattice with $\mathbb{Z}_2$-projective P1 symmetry. When each plaquette has $\Phi = \pi$ flux, the two translation operators become anti-commute. When $\Phi = 0$, this lattice has ordinary P1 symmetry.

The factor systems of $P1$ are labelled by the cohomology invariant $\sigma = \mathsf{L}_a\mathsf{L}_b\mathsf{L}_a^{-1}\mathsf{L}_b^{-1}$. According to what we analyzed in last section, we only need to add $\pi$-flux to each unit translation plaquette to realize the nontrivial cohomology invariant $\sigma = -1$, as shown in Fig. 24.

### c. P2

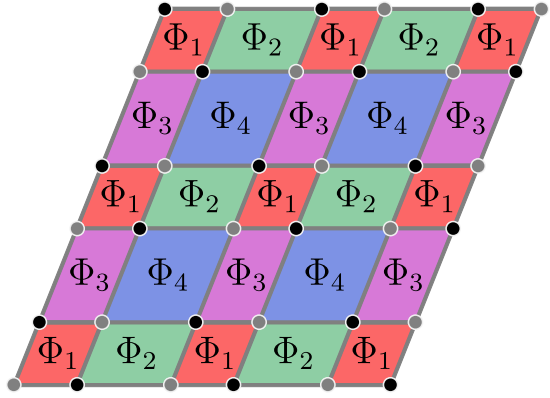

Supplementary Fig.25: An example of flux lattice with $\mathbb{Z}_2$-projective $P2$ symmetry. The fluxes corresponding to cohomology invariants $\alpha_1, \alpha_2, \alpha_3, \alpha_4$ are $\Phi_1, \Phi_2, \Phi_3, \Phi_4$.

According to Eq. (58), the $\mathbb{Z}_2^4$ classes of projective algebra of $P2$ are labelled by cohomology invariants $\alpha_1, \alpha_2, \alpha_3, \alpha_4$ of the four different rotation centers. So we construct a lattice that each class of rotation centers is surrounded by an independent plaquette, as Fig. 25 shows. Comparing with the distribution of cohomology invariants in Fig. 5, we attach flux $\Phi_1, \Phi_2, \Phi_3, \Phi_4$ to these plaquettes so that

$$e^{i\Phi_i} = \alpha_i, \quad i = 1, 2, 3, 4. \tag{148}$$

### d. Pm

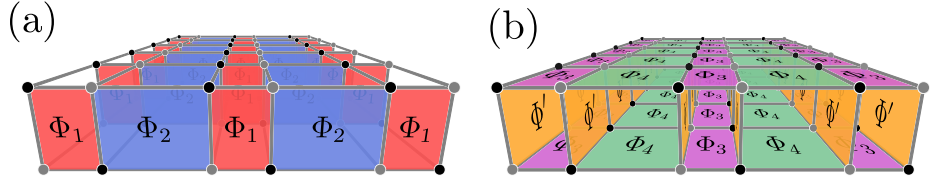

Supplementary Fig.26: An example of flux lattice with $\mathbb{Z}_2$-projective $Pm$ symmetry. (a) The fluxes $\Phi_1, \Phi_2$ correspond to cohomology invariants $\beta_1, \beta_2$ of the two classes of reflections. (b) The fluxes $\Phi_3 + \Phi', \Phi_4 + \Phi'$ correspond to cohomology invariants $\eta_1, \eta_2$ between reflection and translations.

Since the group $Pm$ contains reflections, to realize all the possible $\mathbb{Z}_2$- projective algebras by nearest hopping models, we need a bilayer lattice as that in Fig. 26. The reflection $M_x$ is replaced by a rotation $R_\pi$. We will keep the notation of $M_x$, since the two groups are isomorphic.

According to Eq. (65), the cohomology invariants determinate $\mathbb{Z}_2^4$ classes of projective algebras of $Pm$. There are two kinds of conboundary invariants.

The first kind contains the cohomology invariants $\alpha_1, \alpha_2$ of reflections (Eq. (65a), Eq. (65b)), which can be realized by adding fluxes $\Phi_1, \Phi_2$ as Fig. 26(a) shows.

The second kind contains the cohomology invariants $\eta_1, \eta_2$, which are commutators between reflections and translations (Eq. (65c), Eq. (65d)). As discussed in Sec. Supplementary Note 3 a iv, they can be realized by adding flux $\Phi_3 + \Phi', \Phi_4 + \Phi'$ as Fig. 26(b) shows.

In summary, relations between flux distribution and cohomology invariants are given by

$$e^{i\Phi_1} = \alpha_1, \tag{149a}$$

$$e^{i\Phi_2} = \alpha_2, \tag{149b}$$

$$e^{i(\Phi_3+\Phi')} = \eta_1, \tag{149c}$$

$$e^{i(\Phi_4+\Phi')} = \eta_2. \tag{149d}$$

### e. Pg

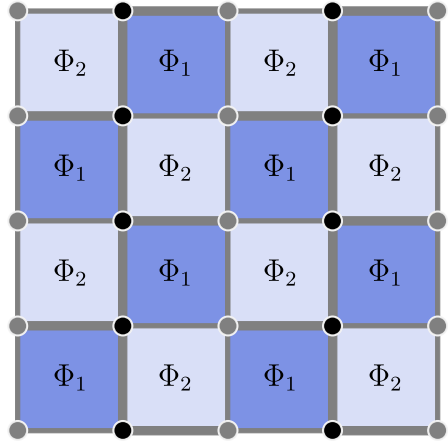

Supplementary Fig.27: An example of flux lattice with $\mathbb{Z}_2$-projective $Pg$ symmetry.

The group $Pg$ has only one cohomology invariant $\eta$ to label factor systems between translation and glide-reflection. As discussed in Sec. Supplementary Note 3 a iv, this cohomology invariant can be realized by adding flux $\Phi = \Phi_1 + \Phi_2$ as Fig. 27.(2) shows. The relation between flux and coboundary invariant is

$$e^{i(\Phi_1+\Phi_2)} = \tau. \tag{150}$$

### f. Cm

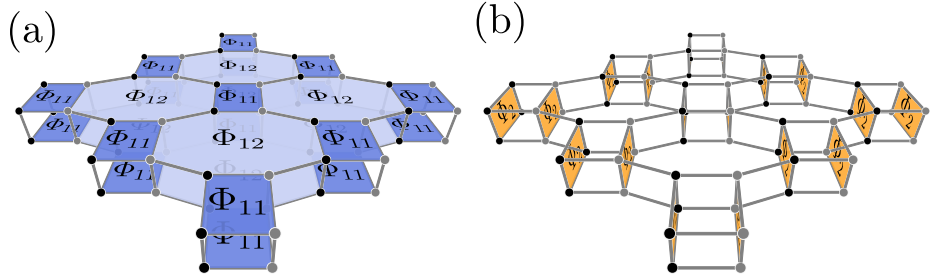

Supplementary Fig.28: An example of flux lattice with $\mathbb{Z}_2$-projective $Cm$ symmetry. (a) The flux $\Phi_1 = \Phi_{11} + \Phi_{12}$ corresponds to the cohomology invariant of translation. (b) The flux $\Phi_2$ correspond to the cohomology invariant of the reflection (two-fold rotation).

The group $Cm$ also contains reflections, so we need a bilayer lattice and replace the reflection by a two-fold rotation. According to Eq. (70), two independent cohomology invariants of $Cm$ are $\sigma$ and $\alpha$. The cohomology invariant $\sigma$ is determined by the flux $\Phi_1 = \Phi_{11} + \Phi_{12}$ in the unit translation plaquette as Fig. 28.(a) shows. And the cohomology invariant $\alpha$ is determined by the flux $\Phi_2$ in the plaquette around horizontal rotation axis as Fig. 28.(b) shows.

Relations between flux distribution and cohomology invariants are

$$e^{i(\Phi_{11}+\Phi_{12})} = e^{i\Phi_1} = \sigma, \tag{151a}$$

$$e^{i\Phi_2} = \beta. \tag{151b}$$

### g. Pmm

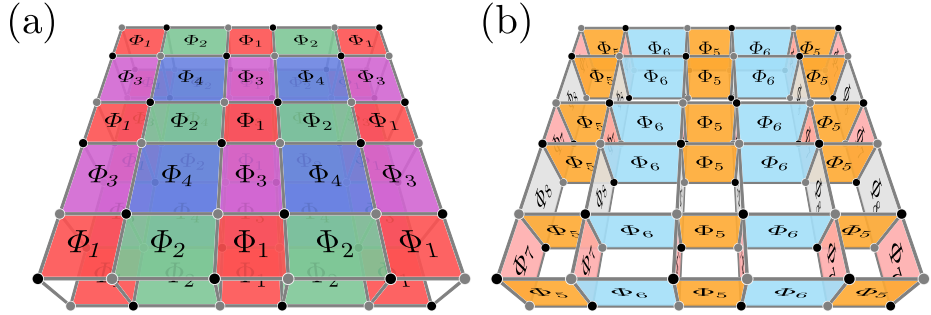

Supplementary Fig.29: An example of flux lattice with $\mathbb{Z}_2$-projective $Pmm$ symmetry. (a) The flux $\Phi_i, i = 1, 2, 3, 4$ correspond to cohomology invariants $\alpha_i, i = 1, 2, 3, 4$ of rotations. (b) The flux $\Phi_i, i = 5, 6, 7, 8$ correspond to cohomology invariants $\beta_i, i = 1, 2, 3, 4$ of reflections (two-fold rotations).

According to Eq. (77) and Fig. 9, the $\mathbb{Z}_2^8$ classes of projective algebras of $Pmm$ are labelled by cohomology invariants $\alpha_i, \beta_i, i = 1, 2, 3, 4$. In order to realize all the possible projective algebras we construct a lattice in which each rotation center and reflection axis is surrounded by an independent plaquette, and we attach fluxes $\Phi_i, i = 1, 2, 3, 4, 5, 6, 7, 8$ to it as shown in Fig. 29.

In summary, relations between flux distribution and cohomology invariants are

$$e^{i\Phi_i} = \alpha_i, \quad i = 1, 2, 3, 4, \tag{152}$$

$$e^{i\Phi_{i+4}} = \beta_i, \quad i = 1, 2, 3, 4. \tag{153}$$

### h. Pmg

The group $Pmg$ also contains reflections, so we need a bilayer lattice. According to Fig. 10 and Eq. (83), the $\mathbb{Z}_2^4$ classes of projective algebra of $Pmg$ are labelled by cohomology invariants $\alpha_i, i = 1, 2, \beta$ and $\eta$. Cohomology invariants $\alpha_1, \alpha_2$ are invariants of rotations, which can be realized by fluxes $\Phi_1, \Phi_2$ as illustrated in Fig. 30.(a). $\beta$ is the cohomology invariant of reflection, which can realized by flux $\Phi_3$ in the reflection invariant plaquette as Fig. 30.(b) shows. $\eta$ is the cohomology invariant between translation and reflection, which can realized by fluxes $\Phi_4 = \Phi_{41} + \Phi_{42} + \Phi_{43} + \Phi_{44}$ as Fig. 30.(c) shows.

In summary, relations between flux distribution and cohomology invariants are

$$e^{i\Phi_1} = \alpha_1, \tag{154a}$$

$$e^{i\Phi_2} = \alpha_2, \tag{154b}$$

$$e^{i\Phi_3} = \beta, \tag{154c}$$

$$e^{i(\Phi_{41}+\Phi_{42}+\Phi_{43}+\Phi_{44})} = \eta. \tag{154d}$$

### i. Pgg

According to Fig. 11 and Eq. (86), factor systems of $Pgg$ can be labelled by cohomology invariants $\alpha_1, \alpha_2$ of rotation centers. Thus we attach fluxes $\Phi_1, \Phi_2$ to the plaquettes around these two rotation centers, as Fig. 31 shows.

In summary, relations between flux distribution and cohomology invariants are

$$e^{i\Phi_i} = \alpha_i, \quad i = 1, 2. \tag{155a}$$

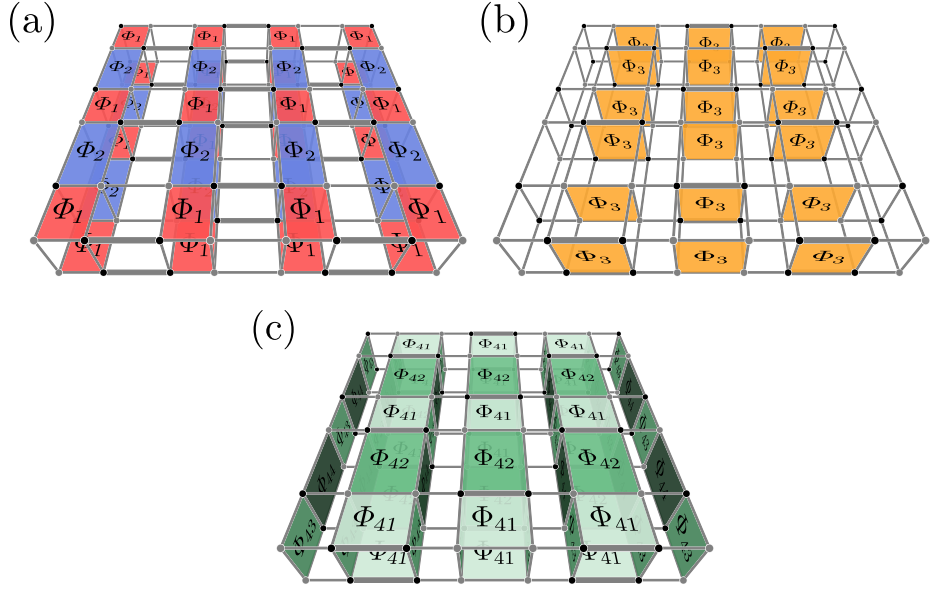

Supplementary Fig.30: An example of flux lattice with $\mathbb{Z}_2$-projective $Pmg$ symmetry. (a) The fluxes $\Phi_1, \Phi_2$ correspond to cohomology invariants $\alpha_1, \alpha_2$ of rotations. (b) The flux $\Phi_3$ corresponds to the cohomology invariant of reflection $\beta$. (c) The flux $\Phi_4 = \Phi_{41} + \Phi_{42} + \Phi_{43} + \Phi_{44}$ corresponds to cohomology invariant $\eta$ between reflection and translation.

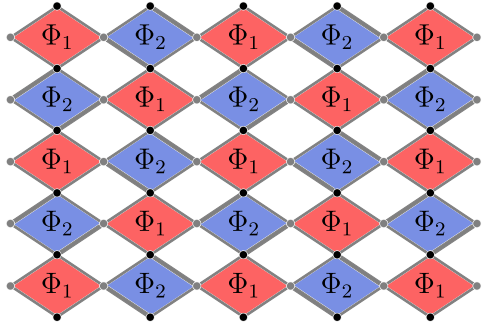

Supplementary Fig.31: An example of flux lattice with $\mathbb{Z}_2$-projective $Pgg$ symmetry. The flux $\Phi_1, \Phi_2$ correspond to cohomology invariants $\alpha_1, \alpha_2$ of rotations.

Flux distributions are shown in Fig. 31.

### j. Cmm

The group $Cmm$ contains reflections, so we need a bilayer lattice.

According to Fig. 12 and Eq. (93), factor systems of $Cmm$ can be labelled by cohomology invariants $\alpha_i, i = 1, 2, 3$ and $\beta_i, i = 1, 2$. To realize these cohomology invariants, we attach fluxes $\Phi_1, \Phi_2, \Phi_3$ to the plaquettes around these three rotation centers as Fig. 32.(a) shows, and attach fluxes $\Phi_4, \Phi_5$ to the plaquettes around these two reflection axes as Fig. 32.(b) shows.

In summary, relations between flux distribution and cohomology invariants are

$$e^{i\Phi_i} = \alpha_i, i = 1, 2, 3, \tag{156}$$
$$e^{i\Phi_{i+3}} = \beta_i, i = 1, 2. \tag{157}$$

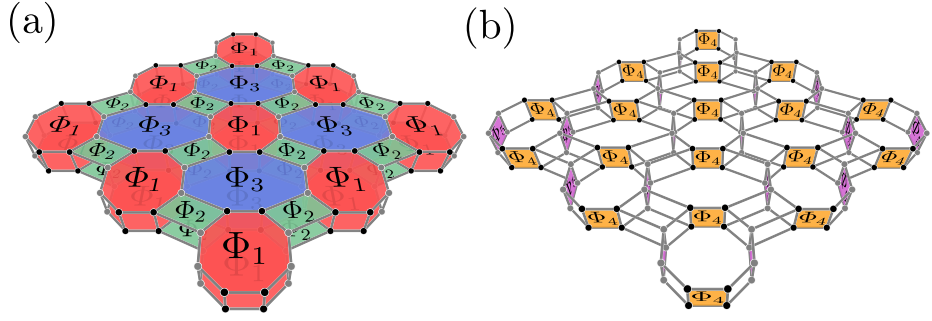

Supplementary Fig.32: An example of flux lattice with $\mathbb{Z}_2$-projective $Cmm$ symmetry. (a) The fluxes $\Phi_i, i = 1, 2, 3$ correspond to cohomology invariants $\alpha_i, i = 1, 2, 3$ of rotations. (b) The fluxes $\Phi_i, i = 4, 5$ correspond to cohomology invariants $\alpha_i, i = 4, 5$ of reflections.

### k. P4

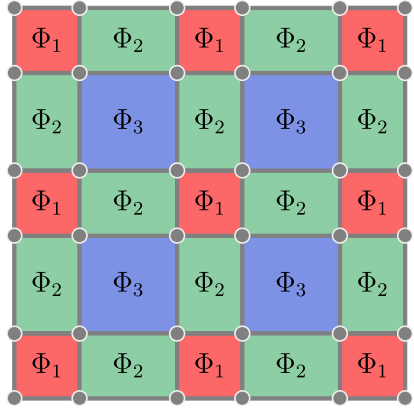

Supplementary Fig.33: An example of flux lattice with $\mathbb{Z}_2$-projective $P4$ symmetry.

According to Fig. 13 and Eq. (99), factor systems of $P4$ can be labelled by cohomology invariants $\alpha_i, i = 1, 2, 3$ of rotations. Thus we need lattice in which each rotation center is surrounded by an independent plaquette, as Fig. 33 shows.

To realize these invariants, we attach fluxes $\Phi_1, \Phi_2, \Phi_3$ to the plaquettes around the three rotation centers respectively, as Fig. 33 shows.

In summary, relations between flux distribution and cohomology invariants are

$$e^{i\Phi_i} = \alpha_i, i = 1, 2, 3. \tag{158}$$

### l. P4m

The group $P4m$ contains reflections, so we need a bilayer lattice.

According to Fig. 14 and Eq. (106), factor systems of $P4m$ can be labelled by cohomology invariants $\alpha_i, i = 1, 2, 3$ and $\beta_i, i = 1, 2, 3$. To realize these cohomology invariants, we attach fluxes $\Phi_1, \Phi_2, \Phi_3$ to the plaquettes around the three rotation centers as Fig. 34.(a) shows, and attach fluxes $\Phi_4, \Phi_5, \Phi_6$ to the plaquettes around the three reflection axes as Fig. 34.(b) shows.

In summary, relations between flux distribution and cohomology invariants are

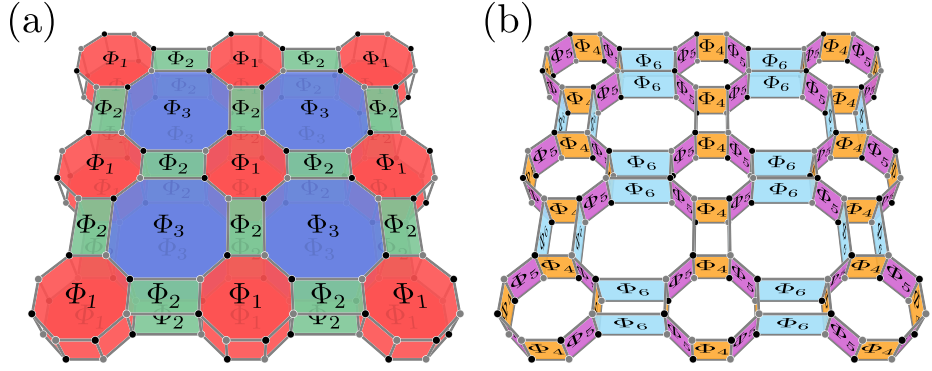

Supplementary Fig.34: An example of flux lattice with $\mathbb{Z}_2$-projective $P4m$ symmetry. (a) The fluxes $\Phi_i, i = 1, 2, 3$ correspond to cohomology invariants $\alpha_i, i = 1, 2, 3$ of rotations. (b) The fluxes $\Phi_i, i = 4, 5, 6$ correspond to cohomology invariants $\beta_i, i = 1, 2, 3$ of reflections.

$$e^{i\Phi_i} = \alpha_i, \quad i = 1, 2, 3, \tag{159}$$

$$e^{i\Phi_{i+3}} = \beta_i, \quad i = 1, 2, 3. \tag{160}$$

### m. P4g

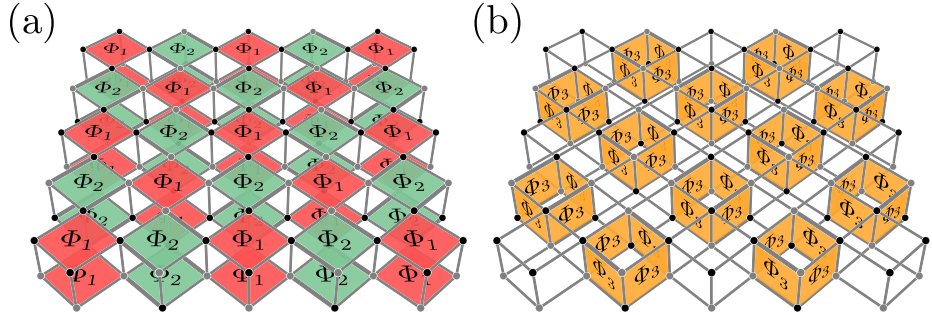

Supplementary Fig.35: An example of flux lattice with $\mathbb{Z}_2$-projective $P4g$ symmetry. (a) The fluxes $\Phi_i, i = 1, 2$ correspond to cohomology invariants $\alpha_i, i = 1, 2$ of rotations. (b) The flux $\Phi_3$ corresponds to cohomology invariant $\beta$ of reflection.

The group $P4g$ contains reflections, so we need a bilayer lattice.

According to Fig. 15 and Eq. (108), factor systems of $P4m$ can be labelled by cohomology invariants $\alpha_i, i = 1, 2$ and $\beta$. To realize these cohomology invariants, we attach fluxes $\Phi_1, \Phi_2$ to the plaquettes around these two rotation centers as Fig. 35.(a) shows, and attach fluxes $\Phi_3$ to the plaquettes around the reflection axis as Fig. 35.(b) shows.

In summary, relations between flux distribution and cohomology invariants are

$$e^{i\Phi_i} = \alpha_i, \quad i = 1, 2, \tag{161}$$

$$e^{i\Phi_3} = \beta. \tag{162}$$

### n. P3

The projective algebra of $P3$ is only depend on the cohomology invariant $\sigma$ of translation. Thus, if we attach flux $\Phi$ to each plaquette as Fig. 36 shows, we can realize the projective relation Eq. (114). The relation of the cohomology invariant and flux is

$$e^{i\Phi} = \sigma. \tag{163}$$

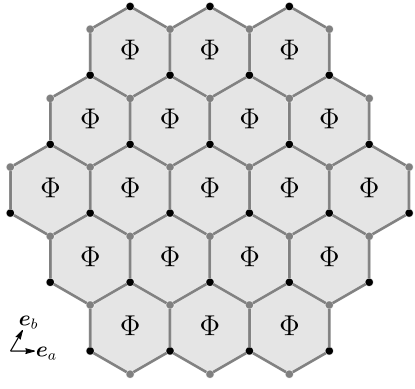

Supplementary Fig.36: An example of flux lattice with $\mathbb{Z}_2$-projective $P3$ symmetry.

### o. P3m1

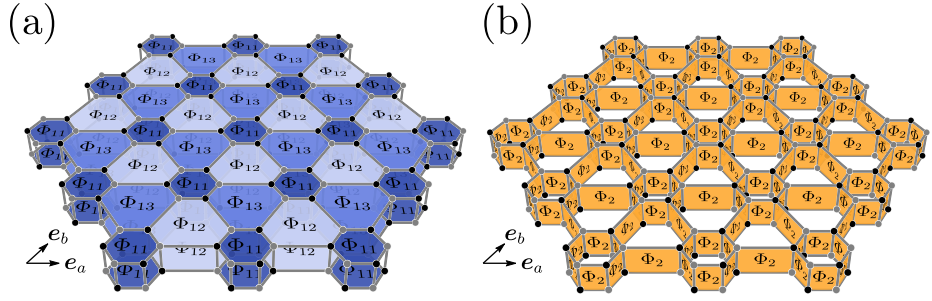

Supplementary Fig.37: An example of flux lattice with $\mathbb{Z}_2$-projective $P3m1$ symmetry. (a) The flux $\Phi_1 = \Phi_{11} + \Phi_{12} + \Phi_{13}$ in the unit translation area corresponds to the cohomology invariant $\sigma$ of translation. (b) The flux $\Phi_2$ corresponds to cohomology invariant $\beta$ of reflection.

The group $P3m1$ contains reflections, so we need a bilayer lattice.

According to Eq. (117), factor systems of $P3m1$ are labelled by cohomology invariant $\sigma$ and $\alpha$. We attach flux $\Phi_1$ to each unit translation plaquette as Fig. 37.(c) shows, and attach flux $\Phi_2$ to each plaquette around the reflection axis as Fig. 37.(d) shows.

In summary, relations between flux distribution and cohomology invariants are

$$e^{i\Phi_1} = e^{i(\Phi_{11}+\Phi_{12}+\Phi_{13})} = \sigma, \tag{164a}$$
$$e^{i\Phi_2} = \beta. \tag{164b}$$

### p. P31m

The group $P31m$ contains reflections, so we need a bilayer lattice.

According to Eq. (121) and Fig. 18, factor systems of $P31m$ are labelled by cohomology invariant $\sigma$ and $\alpha$. We attach flux $\Phi_1$ to each unit translation plaquette as Fig. 38.(c) shows, and attach flux $\Phi_2$ to each plaquette around the reflection axis as Fig. 37.(d) shows.

In summary, relations between flux distribution and cohomology invariants are

$$e^{i\Phi_1} = \sigma, \tag{165a}$$
$$e^{i\Phi_2} = \beta. \tag{165b}$$

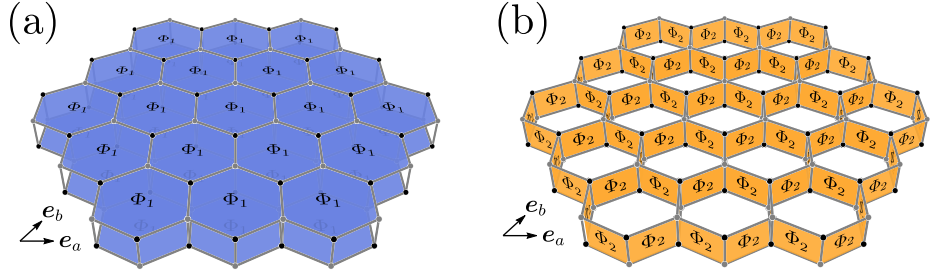

Supplementary Fig.38: An example of flux lattice with $\mathbb{Z}_2$-projective $P31m$ symmetry. (a) The flux $\Phi_1$ in the unit translation area corresponds to coboundary the invariant $\sigma$ of translation. (b) The flux $\Phi_2$ corresponds to cohomology invariant $\beta$ of reflection.

### q. P6

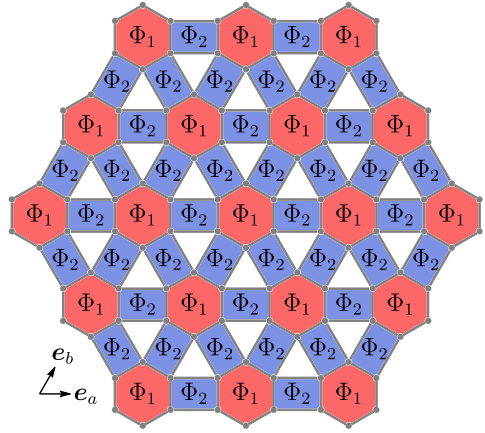

Supplementary Fig.39: An example of flux lattice with $\mathbb{Z}_2$-projective $P6$ symmetry. The fluxes $\Phi_1, \Phi_2$ correspond to cohomology invariants $\alpha_1, \alpha_2$ of rotations.

According to Eq. (127) and Fig. 19, factor systems of $P6$ can be labelled by two cohomology invariants $\alpha_1, \alpha_2$ of even-fold rotations. Thus we need a lattice in which each rotation center is surrounded by an independent plaquette. To realize the cohomology invariants, we attach fluxes $\Phi_1, \Phi_2$ to the plaquettes around these rotation centers, as Fig. 39 shows.

In summary, relations between flux distribution and cohomology invariants are

$$e^{i\Phi_i} = \alpha_i, \quad i = 1, 2. \tag{166}$$

### r. P6m

The group $P6m$ contains reflections, so we need a bilayer lattice.

According to Eq. (132) and Fig. 20, factor systems of $P6m$ can be labelled by two cohomology invariants $\alpha_1, \alpha_2$ of even-fold rotations and two cohomology invariants $\beta_1, \beta_2$ of reflections. To realize these factors, we attach fluxes $\Phi_1, \Phi_2$ to the plaquettes around these rotation centers as Fig. 40.(a) shows, and attach fluxes $\Phi_3, \Phi_4$ to the plaquettes around these reflection axes as Fig. 40.(b) shows.

In summary, relations between flux distribution and cohomology invariants are

$$e^{i\Phi_i} = \alpha_i, \quad i = 1, 2, e^{i\Phi_{i+2}} = \beta_i, \quad i = 1, 2. \tag{167}$$

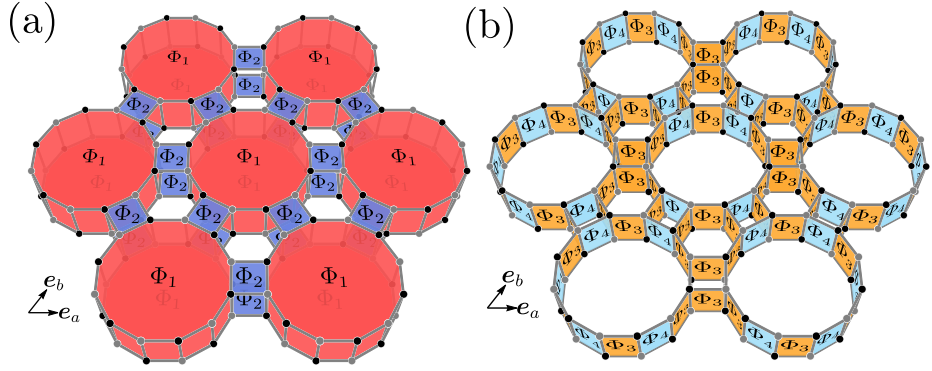

Supplementary Fig.40: An example of flux lattice with $\mathbb{Z}_2$-projective $P6m$ symmetry. (a) The fluxes $\Phi_i, i = 1, 2$ correspond to cohomology invariants $\alpha_i, i = 1, 2$ of rotations. (b) The fluxes $\Phi_i, i = 3, 4$ correspond to cohomology invariants $\beta_i, i = 1, 2$ of reflections.

## Supplementary Note 4. Other technical details for results in the main text

### a. Energy bands of flux lattices with all classes of $\mathbb{Z}_2$-projective P2 symmetries

In this section, we will analyze a model with projective $P2$ symmetry to show the consequences of the flux. For simplicity, we choose $\mathsf{L}_a, \mathsf{L}_b$ to be perpendicular to each other, so that $\mathsf{L}_a = \mathsf{L}_x, \mathsf{L}_b = \mathsf{L}_y$. The model is shown in Fig. 41. The parameters take values $t_1^x = 1, t_2^x = 2, t_1^y = 1.5, t_2^y = 2, m = 1$. We keep the notation that $\mathsf{R}$ is the rotation operator corresponding to cohomology invariant $\alpha_1$.

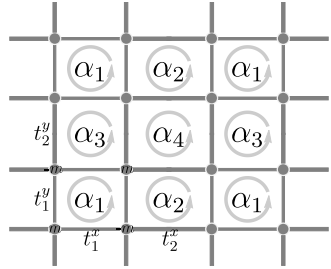

Supplementary Fig.41: Flux lattice with projective $P2$ symmetry. There are undetermined cohomology invariants $\alpha_1, \alpha_2, \alpha_3, \alpha_4$. $t_1^x, t_2^x, t_1^y, t_2^y$ are hopping amptitudes and $m$ is the onsite energy.

In the following, we will disucss the symmetry properties of energy bands for all classes of $\mathbb{Z}_2$-projective algebras.

(i) When cohomology invariants $(\alpha_1, \alpha_2, \alpha_3, \alpha_4) = (1, 1, 1, 1)$, the group $P2$ is linearly represented in momentum space. The time-reversal $T$ and rotation $\mathsf{R}$ share the same high symmetry points $\Gamma, X, Y, M$. The energy bands are shown in Fig. 42.

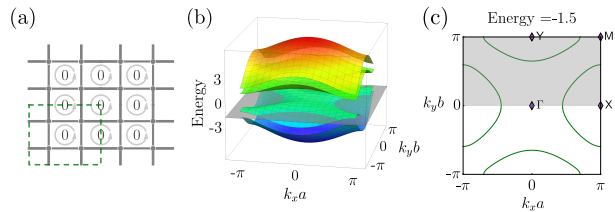

Supplementary Fig.42: Flux distribution and energy bands corresponds to $(\alpha_1, \alpha_2, \alpha_3, \alpha_4) = (1, 1, 1, 1)$.(a) Flux distribution. (b)(c) Energy bands and a constant-energy section.

(ii) When cohomology invariants $(\alpha_1, \alpha_2, \alpha_3, \alpha_4) = (-1, -1, -1, -1)$, the relations between $\mathsf{R}$ and $\mathsf{L}_x$, $\mathsf{L}_y$ is maintained, so the distribution of high symmetry points in momentum space is the same as that in the case $(\alpha_1, \alpha_2, \alpha_3, \alpha_4) = (1, 1, 1, 1)$. However, at the general point , the little cogroup generator $\mathsf{R}T$ satisfies $(\mathsf{R}T)^2 = -1$. Thus the energy bands is two-fold degenerate, see Fig. 43.

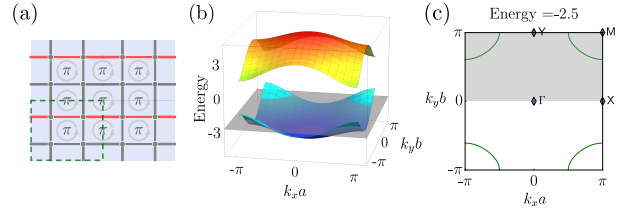

Supplementary Fig.43: Flux distribution and energy bands correspond to $(\alpha_1, \alpha_2, \alpha_3, \alpha_4) = (-1, -1, -1, -1)$.(a) Flux distributions. (b)(c) Energy bands and a constant-energy section.

(iii) When two of the cohomology invariants are negative, there are six possibilities: $(\alpha_1, \alpha_2, \alpha_3, \alpha_4) = (1, -1, 1, -1)$, $(-1, 1, -1, 1)$, $(1, 1, -1, -1)$, $(-1, -1, 1, 1)$, $(1, -1, -1, 1)$, $(-1, 1, 1, -1)$. They have the same algebra. The high symmetry points of rotation $\mathsf{R}$ are translated by 1/4 reciprocal lattice vector compared with the high symmetry points of $T$. The energy bands are shown in Fig. 44.

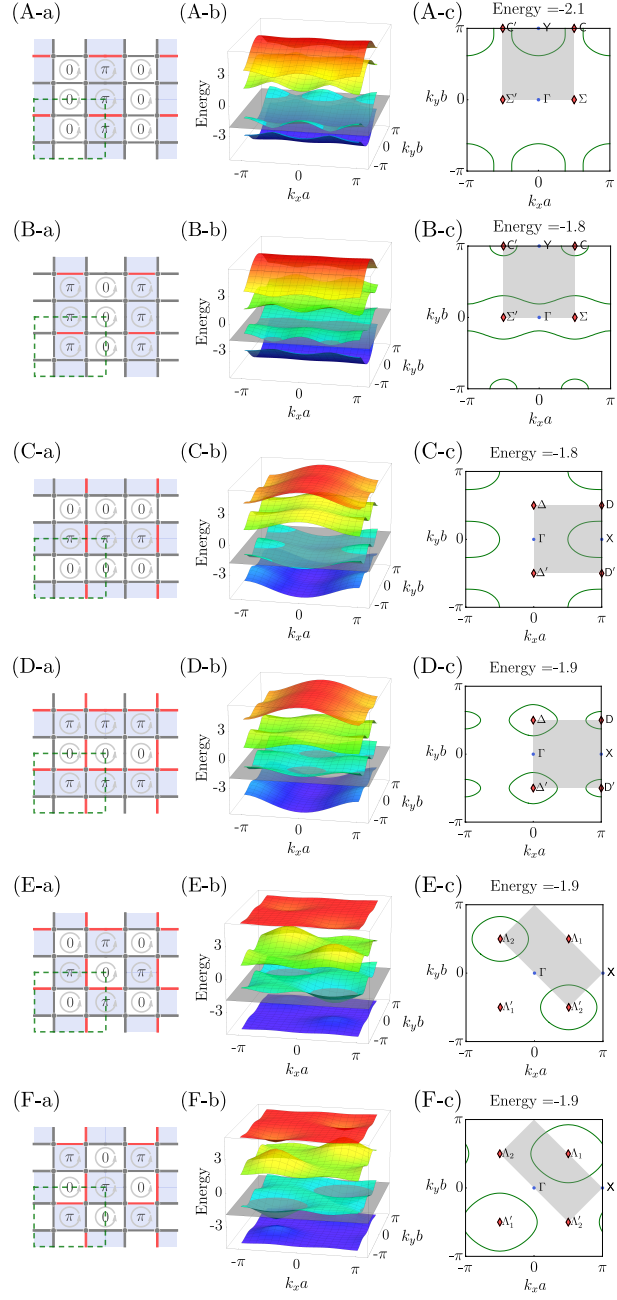

Supplementary Fig.44: Flux distributions and energy bands when two of the cohomology invariants are negative. Flux distribution corresponds to $(\alpha_1, \alpha_2, \alpha_3, \alpha_4) =$ (A-a)$(1, 1, -1, -1)$, (B-a)$(-1, -1, 1, 1)$, (C-a)$(1, 1, -1, -1)$, (D-a)$(-1, -1, 1, 1)$, (E-a)$(1, -1, -1, 1)$, (F-a)$(-1, 1, 1, -1)$. (b)(c) Energy bands and constant-energy sections.

(iv)   When one of the cohomology invariants are negative, there are four possibilities: $(\alpha_1, \alpha_2, \alpha_3, \alpha_4) = (1, -1, 1, -1)$, $(-1, 1, -1, 1)$, $(1, 1, -1, -1)$, $(-1, -1, 1, 1)$, which have the same algebra. Since the original unit cell contains total $\pi$-flux, we have to enlarge the unit cell to maintain the translation symmetry of the gauge connection, as Fig. 45 shows. There are additional high symmetry points of rotation $\mathsf{L}_x\mathsf{R}$ and time-reversal $\mathsf{L}_x T$, which are translated by $\boldsymbol{G}_y/4$ from the high symmetry point of $\mathsf{R}$ and $T$. The energy bands are two-fold degenerate at $D, D'$.

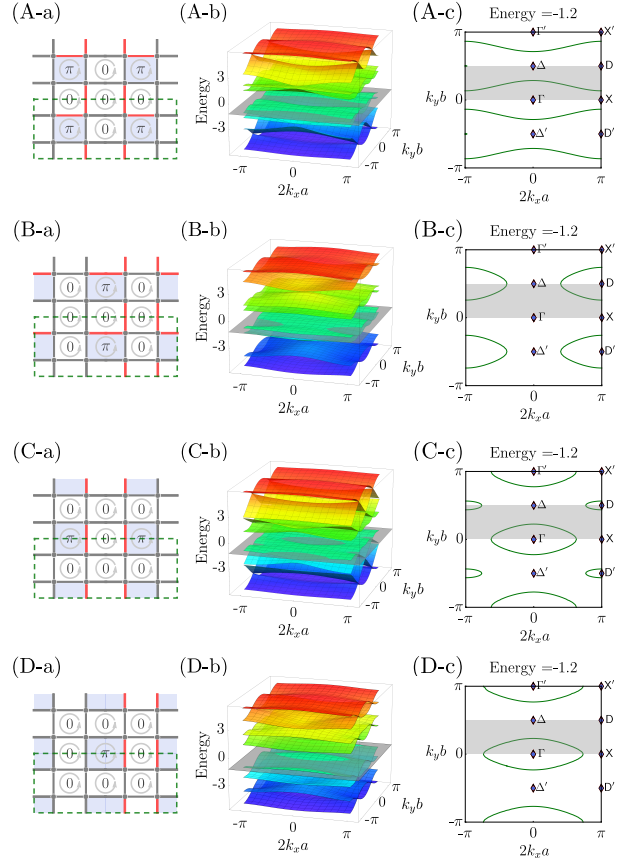

Supplementary Fig.45: Flux distributions and energy bands when one of the cohomology invariant is negative. Flux distribution corresponds to $(\alpha_1, \alpha_2, \alpha_3, \alpha_4) =$ (A-a)$(-1, 1, 1, 1)$, (B-a)$(1, -1, 1, 1)$, (C-a)$(1, 1, -1, 1)$, (D-a)$(1, 1, 1, -1)$. (b)(c) Energy bands and constant-energy sections.

(v) When three of the cohomology invariants are negative, there are four possibilities: $(\alpha_1, \alpha_2, \alpha_3, \alpha_4) =$ $(1, -1, -1, -1)$, $(-1, 1, -1, 1)$, $(-1, -1, 1, -1)$, $(-1, -1, -1, 1)$, which has the same algebra. Since the original unit cell countains total $\pi$-flux, we have to enlarge the unit cell to maintain the translation symmetry of the gauge connection, as Fig. 46 shows. There are additional high symmetry points of rotation $\mathsf{L}_x\mathsf{R}$ and time-reversal $\mathsf{L}_x T$, which are translated by $\boldsymbol{G}_y/4$ from the high symmetry points of $R$ and $T$. The energy bands are two-fold degenerate at every point.

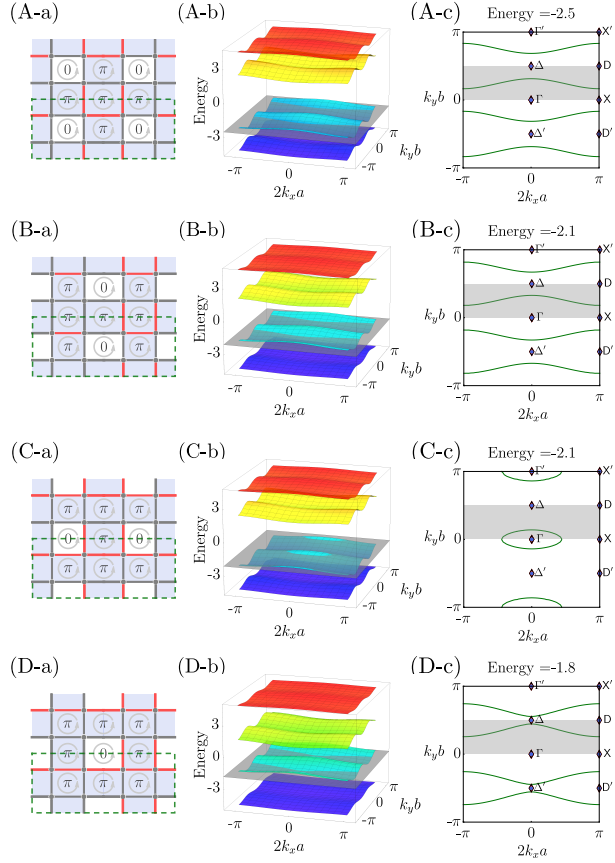

Supplementary Fig.46: Flux distributions and energy bands when three of the cohomology invariants are negative. Flux distribution corresponds to $(\alpha_1, \alpha_2, \alpha_3, \alpha_4) = $ (A-a)$(1, -1, -1, -1)$, (B-a)$(-1, 1, -1, -1)$, (C-a)$(-1, -1, 1, -1)$, (D-a)$(-1, -1, -1, 1)$. (b)(c) Energy bands and constant-energy sections.

### b. Projective symmetry enforced Berry phase

Projective symmetry could enforce Berry phase along some directions to be quantized value. Consider projective $P2$ symmetr with coboundary invariants $(\alpha_1, \alpha_2, \alpha_3, \alpha_4) = (\alpha_1, \alpha_1, -\alpha_1, -\alpha_1)$, i.e., $(\sigma, \eta_a, \eta_b, \alpha) = (1, 1, -1, \alpha_1)$. As analyzed in the last paragraph, the high-symmetry points redistribute as in Fig. 47.(a).

In this case, anti-unitary operator $\hat{R}\hat{T}$ will shift momentum $\boldsymbol{k}$ to $\boldsymbol{k} + \boldsymbol{G}_y/2$ because of the nontrivial $\eta_b$. $\hat{R}\hat{T}$ acts on the eigenstate of valence bands $|\psi_{\boldsymbol{k}}^i\rangle$ as

$$\hat{R}\hat{T} |\psi_{\boldsymbol{k}}^a\rangle = \sum_b U^{ba}(\boldsymbol{k}) |\psi_{\boldsymbol{k}+\frac{1}{2}\boldsymbol{G}_y}^b\rangle . \tag{168}$$

Where $U$ is a unitary matrix of $N_{\text{occ}} \times N_{\text{occ}}$. Using this relation twice we get

$$(\hat{R}\hat{T})^2 |\psi_{\boldsymbol{k}}^a\rangle = \sum_{bc} U^{cb}(\boldsymbol{k} + \boldsymbol{G}_y/2)(U^*)^{ba}(\boldsymbol{k}) |\psi_{\boldsymbol{k}}^c\rangle = \alpha_1 |\psi_{\boldsymbol{k}}^a\rangle . \tag{169}$$

Thus,

$$U^*(\boldsymbol{k} + \boldsymbol{G}_y/2)U(\boldsymbol{k}) = \alpha I_n \tag{170}$$

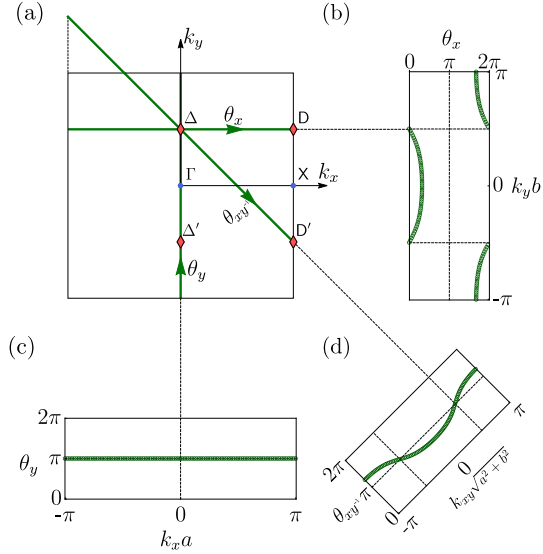

Supplementary Fig.47: Projective symmetry enforced Berry phase. (a) The Brillouin zone and high symmetry points. (b) Berry phase along $\boldsymbol{G}_x$ direction, it is quantized at $k_y b = \pm \pi/2$. (c) Berry phase along $\boldsymbol{G}_y$ direction, it is quantized at all $k_x$ and the value is enfored to be $-i\log(\alpha_1)$. (d) Berry phase along $\boldsymbol{G}_x - \boldsymbol{G}_y$ direction, it is quantized at $k_{xy}\sqrt{a^2 + b^2} = \pm \pi/2$, where $k_{xy}$ is the projection of $\boldsymbol{k}$ in the $\boldsymbol{a} + \boldsymbol{b}$ direction.

The symmetry requires the berry connection to satisfy

$$A_y^{ab}(\boldsymbol{k}) = i \langle \psi_{\boldsymbol{k}}^a | \frac{\partial}{\partial k_y} |\psi_{\boldsymbol{k}}^b \rangle = i \langle \hat{R}\hat{T} \frac{\partial}{\partial k_y} \psi_{\boldsymbol{k}}^b | \hat{R}\hat{T}\psi_{\boldsymbol{k}}^a \rangle = -i \langle \hat{R}\hat{T}\psi_{\boldsymbol{k}}^a | \frac{\partial}{\partial k_y} | \hat{R}\hat{T}\psi_{\boldsymbol{k}}^b \rangle$$

$$= -i \sum_{cd} \langle U^{ca}(\boldsymbol{k})\psi_{\boldsymbol{k}+\boldsymbol{G}_y/2}^c | \frac{\partial}{\partial k_y} |U^{db}(\boldsymbol{k})\psi_{\boldsymbol{k}+\boldsymbol{G}_y/2}^d \rangle = -\sum_{cd} (U^\dagger)^{ac}(\boldsymbol{k}) A_y^{cd}(\boldsymbol{k}+\boldsymbol{G}_y/2) U^{db}(\boldsymbol{k}) + \sum_c (U^\dagger)^{ac}(\boldsymbol{k}) \frac{\partial}{\partial k_y} U^{cb}(\boldsymbol{k})$$

(171)

Take trace of both sides we can have

$$\text{tr}\left(A_y(\boldsymbol{k}) + A_y(\boldsymbol{k}+\boldsymbol{G}_y/2)\right) = \text{tr}\left(U^\dagger(\boldsymbol{k})\frac{\partial}{\partial k_y} U(\boldsymbol{k})\right)$$

(172)

Thus, the Berry phase along $\boldsymbol{G}_y$ direction as Fig.47(a) is

$$\theta_y = \int_0^{2\pi} dk_y \text{tr} A_y(\boldsymbol{k}) = \int_0^\pi dk_y \text{tr}\left(A_y(\boldsymbol{k}) + A_y(\boldsymbol{k}+\boldsymbol{G}_y/2)\right)$$

$$= \int_0^\pi dk_y \text{tr}\left(U^\dagger(\boldsymbol{k})\frac{\partial}{\partial k_y} U(\boldsymbol{k})\right) = \ln \det U(\boldsymbol{k})|_{k_y=0}^{k_y=\pi}$$

$$= -i N_{\text{occ}} \ln \alpha \mod 2\pi.$$

(173)

We see that $\theta_y$ is quantized to $\pi$ or $0$ according to $\alpha_1 = -1$ or $1$ when we consider odd number of valence bands. The Berry phase along other directions will not be restricted, but will be quantized to $0$ or $\pi$ when they pass $\hat{R}$ invariant points. One example for Berry phases is shown in Fig. 47.(b)(c)(d).

We see that although the projective algebra with $(1, 1, -1, -1)$ and that with $(-1, -1, 1, 1)$ are isomorphic, their Berry phases $\theta_y$ are different, because Berry phase is an unit-cell dependent quantity and the unit-cell convention will break the isomorphism of the algebras. However, for the two cases, the projective symmetry groups are isomorphic, so they have the same classification of band topology. This provides us an example that isomorphic projective algebras have the same topological classification of bands but with different physical meanings.

### c. High degeneracy of $P3m1$ at $\Gamma$ point when $\sigma = -1$

In this section, we give the irreducible representation matrix of $P3m1$ at $\Gamma$ point when it take cohomology invariant $\sigma = -1$. Since the $\Gamma$ point is the highest symmetry point, the little group is the group $P3m1$ itself.

Without considering the time-reversal symmetry, there are three irreducible representations at $\Gamma$, which are given by $\Gamma_{i=1,2,3}$ in Table. II.

| Irrep | Rep by generators | | | |
|---|---|---|---|---|
| | $\hat{L}_a$ | $\hat{L}_b$ | $\hat{R}$ | $\hat{M}$ |
| $\Gamma_1$ | $i\sigma_1$ | $i\sigma_3$ | $U_R$ | $U_M$ |
| $\Gamma_2$ | $i\sigma_1$ | $i\sigma_3$ | $U_R$ | $-U_M$ |
| $\Gamma_3$ | $i\sigma_1 \otimes \sigma_0$ | $i\sigma_3 \otimes \sigma_0$ | $U_R \otimes D_R^3$ | $U_M \otimes D_M^3$ |

Supplementary Table.II: Representation of $\mathbb{Z}_2^2 \rtimes D_3$ without time-reversal symmetry.

In Table. II,

$$U_R = \frac{1}{2} \begin{pmatrix} -1-i & -1+i \\ 1+i & -1+i \end{pmatrix}, \tag{174a}$$

$$U_M = \frac{i}{\sqrt{2}} \begin{pmatrix} -1 & -i \\ i & 1 \end{pmatrix}, \tag{174b}$$

$$D_R^3 = \begin{pmatrix} \cos\frac{2\pi}{3} & -\sin\frac{2\pi}{3} \\ \sin\frac{2\pi}{3} & \cos\frac{2\pi}{3} \end{pmatrix}, \tag{174c}$$

$$D_M^3 = \begin{pmatrix} 1 & 0 \\ 0 & -1 \end{pmatrix}. \tag{174d}$$

When we consider time reversal symmetry, if $\beta = 1$, irreducible representations $\Gamma_1,\Gamma_2$ stick together to form $\Gamma_1^T$, while the representation $\Gamma_3$ is time-reversal invariant. If $\beta = -1$, three irreducible representations $\Gamma_{i=1,2,3}$ need to be doubled to $\Gamma_{i=1,2,3}^T$ as in Table. III. The irreducible representation $\Gamma_3^T$ is 8 dimensional.

| Irrep | Rep by generators | | | | | $\alpha$ |
|---|---|---|---|---|---|---|
| | $\hat{L}_a$ | $\hat{L}_b$ | $\hat{R}$ | $\hat{M}$ | $\hat{T}$ | |
| $\Gamma_1^T$ | $i\sigma_1 \otimes \sigma_0$ | $i\sigma_3 \otimes \sigma_0$ | $U_R \otimes \sigma_0$ | $U_M \otimes \sigma_3$ | $\sigma_2 \otimes \sigma_2 K$ | 1 |
| $\Gamma_3^T$ | $i\sigma_1 \otimes \sigma_0$ | $i\sigma_3 \otimes \sigma_0$ | $U_R \otimes D_R^3$ | $U_M \otimes D_M^3$ | $\sigma_2 \otimes \sigma_2 K$ | |
| $\Gamma_1^T$ | $i\sigma_1 \otimes \sigma_0$ | $i\sigma_3 \otimes \sigma_0$ | $U_R \otimes \sigma_0$ | $U_M \otimes \sigma_0$ | $\sigma_2 \otimes \sigma_2 K$ | -1 |
| $\Gamma_2^T$ | $i\sigma_1 \otimes \sigma_0$ | $i\sigma_3 \otimes \sigma_0$ | $U_R \otimes \sigma_0$ | $-U_M \otimes \sigma_0$ | $\sigma_2 \otimes \sigma_2 K$ | |
| $\Gamma_3^T$ | $i\sigma_1 \otimes \sigma_0 \otimes \sigma_0$ | $i\sigma_3 \otimes \sigma_0 \otimes \sigma_0$ | $U_R \otimes D_R^3 \otimes \sigma_0$ | $U_M \otimes D_M^3 \otimes \sigma_3$ | $\sigma_2 \otimes \sigma_2 \otimes \sigma_1 K$ | |

Supplementary Table.III: Representation of $\mathbb{Z}_2^2 \rtimes D_3$ with time-reversal symmetry.

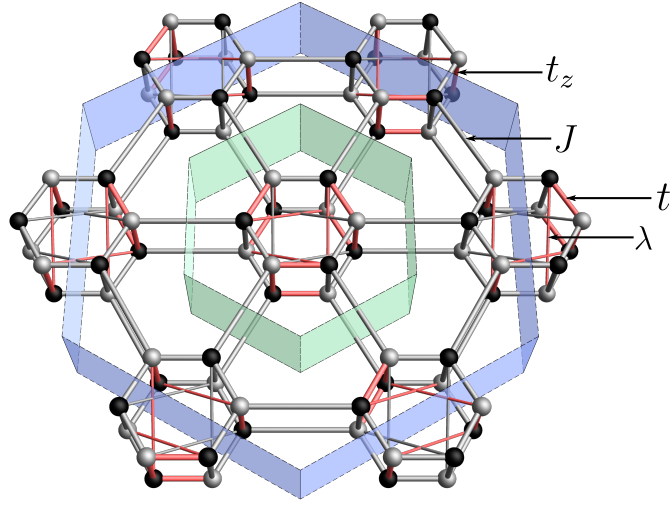

Supplementary Fig.48: Gauge convention of the lattice model of $p3m1$ correspond to cohomology invariants $\sigma = -1, \alpha = -1$. The gray(red) bonds denotes positive(negative) hoppings. When we add gauge to the model, the original unit cell(region within green boundary) is enlarged to the region within the blue boundary.

The 8 dimensional irreducible representation can be found at the lowest 8 bands of the model in Fig.38 when we take the parameters as $|t| = 2, |J| = 10, |t_z| = 5$, we also introduce a inter-layer hopping $|\lambda| = 5$ to lift the degeneracy at the general point(otherwise the bands will be four-folds degenerated at the general point).

## Supplementary Note 5. Engineering gauge fluxes in artificial crystals

In this section, we first present a general mechanism, namely the so-called "dark-bright" mechanism, for generating $\mathbb{Z}_2$ gauge fields on artificial crystals, and then briefly survey on emergent $\mathbb{Z}_2$ gauge fields in various crystal systems.

### a. The Dark-Bright Mechanism for Engineering $\mathbb{Z}_2$ gauge field

We would like to emphasize an important fact, i.e., $\mathbb{Z}_2$ gauge fields preserve the time-reversal symmetry, which are essentially different from other $U(1)$ gauge fields. Thus, $\mathbb{Z}_2$ gauge fields can be realized without introducing magnetism or magnetic fields. As such, $\mathbb{Z}_2$ gauge fields can be realized in low energies in a large class of lattice structures preserving time reversal symmetry. Here, we introduce the so-called dark-bright mechanism to achieve $\mathbb{Z}_2$ gauge fields.

Consider two sites with the hopping and onsite energies as $t > 0$ and $\epsilon$ in Fig. 49. The Hamiltonian of this system is written as

$$H = \begin{pmatrix} \epsilon & t \\ t & \epsilon \end{pmatrix}.$$  (175)

The eigen state and eigen energy can be obtained as

$$\begin{aligned} E_+ &= \epsilon + t, & |+\rangle &= |a\rangle + |b\rangle, \\ E_- &= \epsilon - t, & |-\rangle &= |a\rangle - |b\rangle, \end{aligned}$$  (176)

where $|a\rangle, |b\rangle$ are the local wave functions, or Wannier wave functions.

For $t > 0$, the ground state is the anti-bonding state and the excitation is the bonding state. If the sign of $t$ is reversed as $-t$, the configuration is exchanged. By inserting an ancillary site between them with onsite energy $\Delta$ as

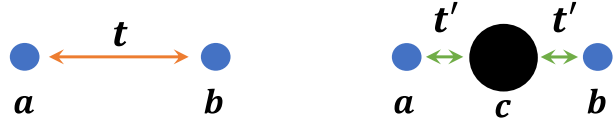

Supplementary Fig.49: Left one denotes the hopping between two sites. The onsite energy is $\epsilon$. For the right one, an ancillary site with onsite energy as $\Delta$ and $\Delta \gg \epsilon$ is inserted between the two original sites. The hopping energy between original and inserted sites is $t'$.

shown in Fig. 49, the Hamiltonian is written as

$$H' = \begin{pmatrix} \epsilon & 0 & t' \\ 0 & \epsilon & t' \\ t' & t' & \Delta \end{pmatrix}. \tag{177}$$

In the limit of $\Delta \gg \epsilon, t'$, we have the eigen values and vectors as

$$
\begin{aligned}
E &= \epsilon, & |-\rangle &= (|a\rangle - |b\rangle)/\sqrt{2}, \\
E &\approx \epsilon - \frac{2t'^2}{\Delta - \epsilon}, & |+\rangle &\approx \left(|a\rangle + |b\rangle - \frac{2t'}{\Delta - \epsilon}|c\rangle\right)/\sqrt{2}, \\
E &\approx \Delta + \frac{2t'^2}{\Delta - \epsilon}, & |e\rangle &\approx \frac{t'}{\Delta - \epsilon}|a\rangle + \frac{t'}{\Delta - \epsilon}|b\rangle + |c\rangle.
\end{aligned}
\tag{178}
$$

Since $\Delta \gg \epsilon, t'$, we can take $|e\rangle$ as high-energy excitation state, which is irrelevant to the energy scale of interest. The state $|-\rangle$, which is called "dark state", is decoupled with the inserted site. The state $|+\rangle$ is called "bright state". Due to $\Delta \gg \epsilon, t'$, the occupation on the inserted site can be ignored. Then, in the subspace of dark and bright states as $\{|-\rangle, |+\rangle\}$, we have the Hamiltonian as

$$H'' = \epsilon |-\rangle \langle-| + \left(\epsilon - \frac{2t'^2}{\Delta - \epsilon}\right)|+\rangle \langle+| \tag{179}$$

By taking the approximation $|+\rangle \approx (|a\rangle + |b\rangle)/\sqrt{2}$ since $|\frac{2t'}{\Delta - \epsilon}| \ll 1$, we have the effective Hamiltonian in the subspace of $\{|a\rangle, |b\rangle\}$ as

$$H_{\text{eff}} = \begin{pmatrix} \epsilon - \frac{t'^2}{\Delta - \epsilon} & -\frac{t'^2}{\Delta - \epsilon} \\ -\frac{t'^2}{\Delta - \epsilon} & \epsilon - \frac{t'^2}{\Delta - \epsilon} \end{pmatrix}, \tag{180}$$

which mimics the $\pi$ hopping phase with the hopping amplitude as $\frac{t'^2}{\Delta - \epsilon}$. If we set $t'^2 = t(\Delta - \epsilon)$, we have the effective hopping coefficient between the sites $a$ and $b$ as $-t$. Note that the loss of the fidelity comes from the occupation on the inserted site. The higher $\Delta$ is, the better fidelity the system has.

### b. A brief survey on emergent $\mathbb{Z}_2$ gauge fields in various crystal systems

In this section, we give a brief review about $\mathbb{Z}_2$ gauge fields in artificial systems, including cold atoms in optical lattices, photonic/acoustic crystals, periodic mechanical systems, electric circuit arrays, and condensed matter systems.

- In photonic crystals, the gauge field can be generated by modulation of the resonant frequencies, e.g., by adjusting the gap between site ring and link-ring wave guides.

- In acoustic crystals, $\mathbb{Z}_2$ hopping phases can be readily realized by coupling the resonators with wave guides on different sides.

- For cold atoms in optical lattices, we introduce two methods: rotating the optical lattice and laser-assisted tunneling. i) Rotating optical lattice can introduce weak and uniform effective magnetic field and the side effect of Coriolis force should be compensated. ii) For the laser-assisted tunneling, the atomic hopping with desired gauge potentials can be engineered by coupling internal levels of atoms with laser beams. Different kinds of gauge fields can be induced, even the nonabelian ones.

- For periodic mechanical systems, effective $\mathbb{Z}_2$ gauge field can be generated by tuning the stiffness coefficients of the spring connections.

- For electric circuit arrays, $\mathbb{Z}_2$ gauge fields can be realized by suitably choosing the capacitances and inductances.

- For strongly correlated systems, there are emergent gauge fields in the low-energy effective theories. The $\mathbb{Z}_2$ gauge field, which defines the $\mathbb{Z}_2$ spin liquid, can naturally emerge in quantum spin liquid. In the mean-field theory of quantum spin liquid, close to the ground states the spinors are coupled to gauge field, particularly a $\mathbb{Z}_2$ gauge field as demonstrated in several works. Actually, perhaps it was the first time that physicists noticed the importance of the projective representations of space groups with a given gauge configuration. Another example is the Kitaev-type exactly solvable model, where non-dynamical $\mathbb{Z}_2$ gauge fields are coupled with Majorana fermions.



\end{document}